\documentclass[prb,twocolumn,notitlepage,superscriptaddress]{revtex4-2}
\usepackage{amsmath,amsfonts,amssymb,amsthm,epsfig,array}
\usepackage{dsfont}
\usepackage{slashed}
\usepackage{graphics}
\usepackage{float}
\usepackage{verbatim}
\usepackage{color}
\usepackage[dvipsnames]{xcolor}
\usepackage[hidelinks,colorlinks,linkcolor=blue,
citecolor=blue,urlcolor=blue]{hyperref}
\usepackage[titletoc,title]{appendix}
\usepackage{multirow}
\usepackage{bibentry}
\usepackage{tabularx}
\usepackage[mathscr]{euscript}
\usepackage{mathtools}

\newcommand{\bsl}[1]{\boldsymbol{#1}}

\newcommand{\shpa}{{\mkern3mu\vphantom{\perp}\vrule depth 0pt\mkern2mu\vrule depth 0pt\mkern3mu}}
\newcommand{\shpacap}{\raisebox{0.5pt}{$\scriptscriptstyle\parallel$}}

\newcommand{\bra}[1]{\langle #1|}
\newcommand{\ket}[1]{|#1 \rangle}
\newcommand{\braket}[2]{\left\langle #1 | #2  \right\rangle}

\newcommand{\ii}{\mathrm{i}}

\newcommand{\dsZ}{\mathbb{Z}}
\newcommand{\dsN}{\mathbb{N}}
\newcommand{\dsR}{\mathbb{R}}

\newcommand{\Tr}{\mathop{\mathrm{Tr}}}
\renewcommand{\Re}{\mathop{\mathrm{Re}}}
\renewcommand{\Im}{\mathop{\mathrm{Im}}}
\renewcommand{\O}{\mathop{\mathrm{O}}}
\newcommand{\SO}{\mathrm{SO}}
\newcommand{\SU}{\mathrm{SU}}
\newcommand{\U}{\mathrm{U}}

\newcommand{\eqnref}[1]{Eq.\,\eqref{#1}}
\newcommand{\figref}[1]{Fig.\,\ref{#1}}

\newcommand{\appref}[1]{Appendix.\,\ref{#1}}
\newcommand{\refcite}[1]{Ref.\,\cite{#1}}

\newcommand{\mat}[1]{\left(\begin{matrix}#1\end{matrix}\right)}

\newcommand{\eq}[1]{\begin{equation} #1 \end{equation}}

\newcommand{\eqa}[1]{\begin{align}\begin{split} #1 \end{split}\end{align}}

\usepackage{environ}
\NewEnviron{eqs}{%
\begin{equation}\begin{split}
    \BODY
\end{split}\end{equation}
}

\let\oldAA\AA
\renewcommand{\AA}{\text{\normalfont\oldAA}}

\newcommand{\ie}{{\emph{i.e.}}}
\newcommand{\eg}{{\emph{e.g.}}}
\newcommand{\TR}{\mathcal{T}}
\newcommand{\cc}{\mathcal{K}}
\newcommand{\s}{\mathrm{s}}

\newcommand{\W}{\mathcal{W}}

\newcommand{\N}{\mathcal{N}}
\renewcommand{\H}{\mathcal{H}}

\newcommand{\BZ}{\text{1BZ}}
\newcommand{\MBZ}{\text{MBZ}}

\newtheorem{definition}{Definition}

\newtheorem{assumption}{Assumption}
\newcommand{\asmref}[1]{Asm.\,\ref{#1}}

\newtheorem{proposition}{Proposition}
\newcommand{\propref}[1]{Prop.\,\ref{#1}}

\begin{document}
\title{Euler Obstructed Cooper Pairing in Twisted Bilayer Graphene: Nematic Nodal Superconductivity and Bounded Superfluid Weight}
\author{Jiabin Yu}
\email{jiabinyu@umd.edu}
\affiliation{Condensed Matter Theory Center and Joint Quantum Institute, Department of Physics, University of Maryland, College Park, MD 20742, USA}
\author{Ming Xie}
\affiliation{Condensed Matter Theory Center and Joint Quantum Institute, Department of Physics, University of Maryland, College Park, MD 20742, USA}
\author{Fengcheng Wu}
\affiliation{School of Physics and Technology, Wuhan University, Wuhan 430072, China}
\affiliation{Wuhan Institute of Quantum Technology, Wuhan 430206, China}
\author{Sankar Das Sarma}
\affiliation{Condensed Matter Theory Center and Joint Quantum Institute, Department of Physics, University of Maryland, College Park, MD 20742, USA}

\begin{abstract}
Magic-angle twisted bilayer graphene (MATBG) hosts normal-state nearly-flat bands with nonzero Euler numbers and shows superconductivity.
In this work, we study the effects of the nontrivial
normal-state band topology on the intervalley $C_{2z}\TR$-invariant mean-field Cooper pairing order parameter in MATBG.
We show that the pairing order parameter can always be split into a trivial channel and an Euler obstructed channel in all gauges for the normal-state basis, generalizing the previously-studied channel splitting in the Chern gauge.
The nonzero normal-state Euler numbers require the pairing gap function of the Euler obstructed channel to have zeros, while the trivial channel can have a nonvanishing pairing gap function.
When the pairing is spontaneously nematic, we find that a sufficiently-dominant Euler obstructed channel with two zeros typically leads to nodal superconductivity.
Under the approximation of exactly-flat bands, we find that the mean-field zero-temperature superfluid weight is generally bounded from below, no matter whether the Euler obstructed channel is dominant or not, generalizing the previously-derived bound for the uniform s-wave pairing.
We numerically verify these statements for pairings derived from a local attractive interaction.
Our work suggests that Euler obstructed Cooper pairing may play an essential role in the superconducting MATBG.
\end{abstract}

\maketitle

\section{Introduction}

The normal state of MATBG (\ie, twisted bilayer graphene with twist angle near $1.1^{\circ}$~\cite{Cao2018TBGMott,Cao2018TBGSC,Bistritzer2011BMModel}) was theoretically shown to host topologically nontrivial nearly-flat bands near the charge neutrality, based on the Bistritzer-MacDonald (BM) model~\cite{Bistritzer2011BMModel,Po2018TBGMottSC,Song2019TBGFragile,Po2019TBGFragile, Ahn2019TBGFragile}. 
The nontrivial band topology is characterized by the $C_{2z}\TR$-protected nonzero Euler numbers~\cite{Ahn2019TBGFragile} (or equivalently Wilson loop winding numbers~\cite{Song2019TBGFragile}), where $C_{nj}$ is the spinless part of the $n$-fold rotation about the $j$ axis (with $j=z$ out of plane) and $\TR$ is the spinful time-reversal symmetry.
When the nearly-flat bands are partially filled, superconductivity was observed in MATBG~\cite{Cao2018TBGSC,Yankowitz2019SCMATBG,Lu2019SCMATBG,Stepanov2019SCCorrTBG,Saito2020SCCorrTBG,Cao2021NematicSCMATBG,deVries2021JosephsonMATBG,Oh2021NodalSCMATBG,Dibattista2021NodalSCMATBG,Tian2021ExpTBGSW}, and the superconductivity may be nematic~\cite{Cao2021NematicSCMATBG} (despite some debate~\cite{Yu2021Jan5NematicSC}) and nodal~\cite{Oh2021NodalSCMATBG,Dibattista2021NodalSCMATBG} when there are 2$\sim$3 holes per moir\'e unit cell.
Here being nematic means breaking $C_{3z}$.
Various theoretical mechanisms~\cite{Xu2018SCMATBG,Guo2018SCMATBG,Liu2018SCMATBG,Isobe2018SCMATBG,Kennes2018SCMATBG,You2019SCMATBG,Wu2018PhononSCMATBG,Lian2019PhononSCMATBG,Wu2019PhononResSCMATBG,Wu2019TSCMATBG,Chichinadze2019Oct16NemSCTBG,Khalaf2021SkyrSCMATBG,Wang2021SCMATBG,Christos2020SCWZWMATBG,Classen2020HOVH,Kozii2020May26KohnLuttingerMATBG,Chatterjee2020SkyrSCMATBG,Khalaf2020SymConsSCMATBG,Cea2021SCTBG,Shavit2021Jul18TBGCISC,Huang2021SCMATBG,Kwan2021SkyrmionsTBG} have been proposed to explain the observed superconductivity.
Yet, it is still unclear whether the potential nematic nodal feature of the superconducting MATBG has any relation to the normal-state Euler numbers.

In this work, we reveal the relation by studying the intervalley $C_{2z}\TR$-invariant mean-field pairing order parameter that is either spin-singlet or spin-triplet with a momentum-independent spin direction, which is energetically possible in MATBG~\cite{Wu2018PhononSCMATBG,Wu2019PhononResSCMATBG,Wu2019TSCMATBG}.
By generalizing the theory for 3D semimetals in \refcite{Yu2021EOCP}, we find that regardless of the gauge for the normal-state basis, the pairing order parameter can always be split into one trivial channel and one nontrivial channel.
The nonzero normal-state Euler numbers require the pairing gap function of the nontrivial channel to have zeros, and determine the total winding number of the zeros, whereas the trivial channel is allowed to have a nonvanishing pairing gap function.
Then, the nontrivial channel is called the Euler obstructed pairing channel.
Our gauge-independent formalism of the Euler obstructed Cooper pairing is a generalization of the known channel splitting in the Chern gauge~\cite{Zaletel2019Nov5MATBGIntegerFilling,Kang2020Feb24DiracNodeMATBG,Zaletel2020Sep4NematicTSMMATBG,Zaletel2020Sep30SoftModesTBG,Khalaf2021SkyrSCMATBG} (or for the Chern bands~\cite{Li2018WSMObstructedPairing,Murakami2003BerryPhaseMSC,Zaletel2020AHTBG}); our gauge-independent formalism is more convenient for numerical calculations as it saves us from explicit gauge fixing.

When the considered $C_{2z}\TR$-invariant pairing is spontaneously nematic, we find that a sufficiently-dominant Euler obstructed channel with two zeros typically leads to nodal superconductivity.
This serves as a mechanism that connects the nematic pairing to nodal superconductivity, although a nematic pairing in general does not necessarily lead to nodal superconductivity~\cite{Fu2014TSCNematic} (especially in multi-band cases like MATBG).
Our mechanism roots in the normal-state Euler numbers, and our mechanism is general in the sense that it is independent of the specific interaction that accounts for the considered pairing form.
We would like to mention that the role of nematicity in the Euler-obstructed-pairing-induced nodal superconductivity revealed in the current work is absent in \refcite{Yu2021EOCP}.

We further provide analytic and numerical supports for the potential existence of a spontaneously-nematic dominant Euler obstructed pairing in MATBG.
First, we use the formalism of Euler obstructed pairing to analytically derive a lower bound of the mean-field zero-temperature superfluid weight for the  considered  $C_{2z}\TR$-invariant pairing, under the exact-flat-band approximation.
Our bound generalizes the previously-derived bound for time-reversal invariant uniform s-wave pairing in \refcite{Xie2020TopologyBoundSCTBG}.
Thus, under the exact-flat-band approximation, the superfluid weight is bounded from below even for pairings with a dominant Euler obstructed channel, regardless of the specific interaction that accounts for the pairing form.
Second, we numerically verify the above statements for the pairings given by a local attractive interaction; the interaction has a similar form as that mediated by acoustic phonons~\cite{Wu2018PhononSCMATBG,Wu2019PhononResSCMATBG,Wu2019TSCMATBG}.
In particular, we find that a spontaneously-nematic pairing with a dominant Euler obstructed channel can arise from the local interaction, which, together with the bounded superfluid weight, implies the potential existence of a spontaneously-nematic dominant Euler obstructed pairing in MATBG.

\section{Euler Obstructed Cooper Pairing in MATBG}

We start by introducing the Euler obstructed Cooper pairing in MATBG.
The BM model contains two decoupled valley $\pm$ related by the $C_{2z}$ or $\TR$ symmetries, and within each valley, the model has $C_{2z}\TR$, $C_{3z}$ and spin-charge $\U(2)$ symmetries.
Because of the normal-state global spin $\SU(2)$ symmetry, we only need to consider the spinless parts for $C_{nz}$, as mentioned above.
The model has other exact and approximate symmetries~\cite{Song2019TBGFragile,Song2020TBGII}, but they are not required for the discussion below.
With the twist angle $\theta$ near $1.1^{\circ}$, BM model captures the normal state of MATBG (that is not aligned with the hBN substrate~\cite{Zaletel2020AHTBG}), and has two nearly-flat bands with additional spin degeneracy near the charge neutrality in each valley.
We use $\ket{u_{\pm,\bsl{k},a}}\otimes\ket{s}$ to label the periodic parts of the Bloch basis for the nearly-flat bands, where $a=1,2$ labels the spinless basis of the two nearly-flat bands in one valley, and $s=\uparrow,\downarrow$ is the spin index.
Defining $\ket{u_{\pm,\bsl{k}}}=(\ket{u_{\pm,\bsl{k},1}},\ket{u_{\pm,\bsl{k},2}})$, the nontrivial topology of $\ket{u_{\pm,\bsl{k}}}$ is manifested by the nonzero Euler number or Wilson loop winding number $\N_\pm=1$~\cite{Song2019TBGFragile,Ahn2019TBGFragile,Yu2021EOCP}.

For the superconductivity in MATBG, we only consider the pairing between the nearly-flat bands, owing to the large normal-state band gaps ($\sim 20$meV) above and below the nearly-flat bands.
We consider the following mean-field Cooper pairing operator
\eq{
\label{eq:H_pairing}
H_{pairing}=\sum_{\bsl{k}\in \text{MBZ}} c^{\dagger}_{+,\bsl{k}} \Delta(\bsl{k})\otimes \Pi (c^{\dagger}_{-,-\bsl{k}})^T +\ h.c.\ ,
}
where $c^{\dagger}_{\pm,\bsl{k}}= (..., c^{\dagger}_{\pm,\bsl{k},a ,s} , ...)$ and $c^{\dagger}_{\pm,\bsl{k},a ,s}$ is the creation operator for the Bloch state of $\ket{u_{\pm,\bsl{k},a}}\otimes\ket{s}$, and MBZ is short for moir\'e Brillouin zone.
We have chosen and will always choose the pairing to be intervalley, since only the intervalley pairing can couple electrons with exactly the same energy and opposite momenta.
Throughout the work, we also choose the pairing to be $C_{2z}\TR$-invariant and to have a momentum-independent spin part $\Pi$.
In particular, we consider two cases for $\Pi$, (i) spin-singlet $\Pi=\ii s_y$ and (ii) spin-triplet $\Pi=\ii (\hat{n}\cdot\bsl{s}) \ii s_y$ with $\hat{n}$ any real momentum-independent unit vector, where $s_{x,y,z}$ are Pauli matrices for the spin index.
For spin-triplet, we can always choose the spin index of the basis to keep $\hat{n}=(0,-1,0)$, \ie, $\Pi=s_0$.
The chosen pairing form is satisfied by certain solutions of the mean-field linearized gap equation owing to the $C_{2z}\TR$ and spin $\SU(2)$ symmetries in the normal state~\cite{Wu2018PhononSCMATBG,Wu2019PhononResSCMATBG,Wu2019TSCMATBG}, but remains an assumption at zero temperature.
$\Delta(\bsl{k})$ in \eqnref{eq:H_pairing} is the spinless part of the pairing gap function, which is the focus of our work.

\begin{figure}
    \centering
    \includegraphics[width=\columnwidth]{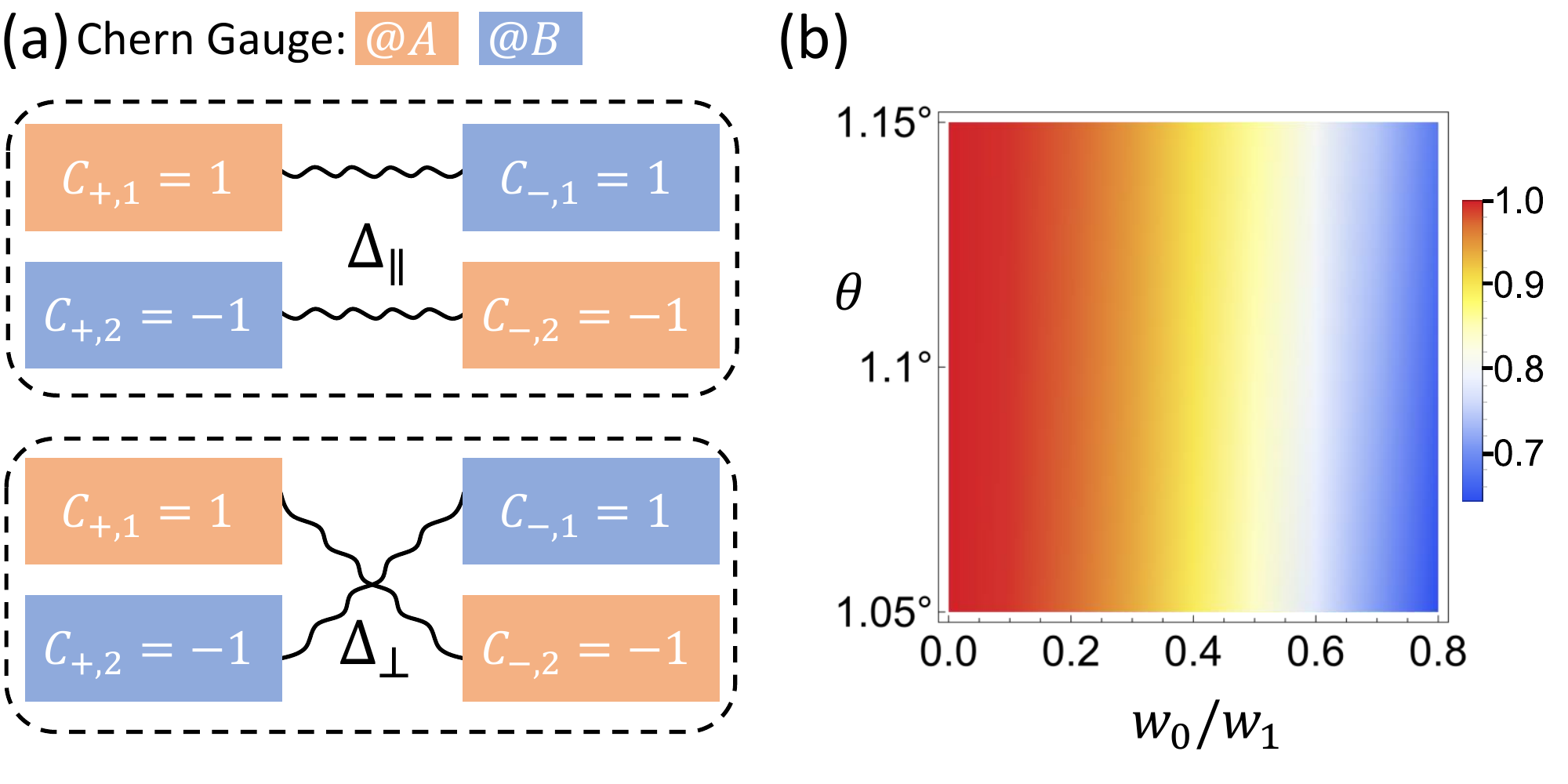}
    \caption{(a) Schematic illustration of the two channels $\Delta_{\shpacap}$ and $\Delta_{\perp}$ (\eqnref{eq:Delta_para_perp_Ch}) in the Chern gauge.
    The blocks stand for the spinless basis for the nearly-flat bands in the Chern gauge, where $\pm$ stand for the valleys.
    The orange and purple blocks stand for the normal states that are polarized to the sublattice $A$ and $B$~\cite{Tarnopolsky2019MagicAngleChiralLimit}, respectively, though the polarization may not be complete~\cite{Zaletel2019Nov5MATBGIntegerFilling}.
    (b) Plot of the probability of $\ket{u_{+,1}(\bsl{k})}$ in the Chern gauge at sublattice $A$ averaged over the MBZ, showing the sublattice polarization discussed in \refcite{Tarnopolsky2019MagicAngleChiralLimit,Zaletel2019Nov5MATBGIntegerFilling}.
    $w_0$ and $w_1$ are the interlayer AB and AA tunneling strengths in the BM model, respectively.    }
    \label{fig:EOCP_Ch}
\end{figure}

Before our work, there were related discussions~\cite{Zaletel2020AHTBG,Zaletel2020SkyrmionTBG,Zaletel2019Nov5MATBGIntegerFilling,Kang2020Feb24DiracNodeMATBG,Zaletel2020Sep4NematicTSMMATBG,Zaletel2020Sep30SoftModesTBG,Khalaf2021SkyrSCMATBG} on how the normal-state band topology affects $\Delta(\bsl{k})$ in the Chern gauge~\cite{Slager2020EulerOptical,Xie2020TopologyBoundSCTBG,Bouhon2020WeylNonabelian,Xie2021TBGVI} for $\ket{u_{\pm,\bsl{k}}}$, which we specify below for our chosen pairing form.
In the Chern gauge, $\ket{u_{\pm,\bsl{k},a}}$ has well-defined Chern number $C_{\pm,a}$; we henceforth choose $C_{\pm, 1} = -C_{\pm, 2} = \N_{\pm}=1$ and choose the following symmetry representations for the Chern gauge
\eqa{
\label{eq:sym_rep_Ch}
& (C_{2z}\TR)c^\dagger_{\pm,\bsl{k}}(C_{2z}\TR)^{-1} = c^\dagger_{\pm,\bsl{k}}\tau_x\otimes \ii s_y\\
& C_{2z} c^\dagger_{+,\bsl{k}} C_{2z}^{-1} =  c^\dagger_{-,-\bsl{k}}
\ ,
}
where $\tau$'s are the Pauli matrices for the spinless basis.
Based on the Chern numbers of the paired Chern states~\cite{Zaletel2020Sep30SoftModesTBG}, we can split $\Delta(\bsl{k})$ into two channels as
\eq{
\label{eq:Delta_split}
\Delta(\bsl{k})= \Delta_{\shpa}(\bsl{k}) + \Delta_{\perp}(\bsl{k}), 
}
where $\Delta_{\shpa}$ ($\Delta_{\perp}$) contains the pairings between Chern states with the same (opposite) Chern numbers (\figref{fig:EOCP_Ch}(a)).~\footnote{Channel splitting based on a different band topology in a different symmetry class was studied in \refcite{Sun2020Z2PairingObstruction}.}
Owing to $C_{2}\TR$ symmetry, we have
\eq{
\label{eq:Delta_para_perp_Ch}
\Delta_{\shpa}(\bsl{k})= \mat{ d_{\shpa}^*(\bsl{k})  & \\ & d_{\shpa}(\bsl{k}) },\  \Delta_{\perp}(\bsl{k})= \mat{ & d_{\perp}(\bsl{k})  \\ d_{\perp}^*(\bsl{k})  & }\ ,
}
where $d_{b}(\bsl{k})=|\Delta_{b}(\bsl{k})|e^{\ii \theta_{b}(\bsl{k})}$ with $b=\perp,\shpa$, and $|\Delta_{b}(\bsl{k})|=\sqrt{\Tr[\Delta_{b}(\bsl{k}) \Delta_{b}^\dagger(\bsl{k})]/2}$.
If $\Delta_b$ has zeros (\ie, $|\Delta_{b}|$ has zeros) but is not everywhere-vanishing, an integer winding number can naturally be defined for each isolated zero $i$ of $\Delta_b$ as
\eq{
\label{eq:W_b_i_Ch}
\W_{b,i}=-\frac{(-1)^b}{2\pi}\int_{\gamma_{b,i}}d\bsl{k}\cdot \nabla_{\bsl{k}} \theta_{b}(\bsl{k})
}
where $(-1)^{\perp}=1$, $(-1)^{\shpa}=-1$, and $\gamma_{b,i}$ is a circle around the zero $i$ of $\Delta_b$.
Then, \refcite{Li2018WSMObstructedPairing,Murakami2003BerryPhaseMSC} (which studied the pairing between Chern states) suggests that
\eqa{
\label{eq:total_winding_Ch}
& \sum_{i} \W_{\perp,i} = C_{+,1}+C_{-,2} = 0 \ , \\
& \sum_{i} \W_{\shpa,i} = -C_{+,2}-C_{-,2} = 2\ .
}
(See SM for details.)
As the total winding number $\sum_{i} \W_{b,i}$ is by definition zero if $\Delta_b$ has no zeros, \eqnref{eq:total_winding_Ch} suggests that $\Delta_{\shpa}$ must have zeros, while $\Delta_{\perp}$ can be nonvanishing~\cite{Li2018WSMObstructedPairing,Murakami2003BerryPhaseMSC,Zaletel2020Sep30SoftModesTBG}.
According to the terminology defined in \refcite{Li2018WSMObstructedPairing}, \eqnref{eq:Delta_para_perp_Ch} and \eqnref{eq:total_winding_Ch} suggest that each element of $\Delta_{\shpa}$ in the Chern gauge is a monopole Cooper pairing, since the nonzero total winding number indicates that the monopole Harmonics~\cite{Wu1976MonopoleHarmonics} are required for the full description of $\Delta_{\shpa}$ in the Chern gauge.
Thus, $\Delta_{\shpa}$ in the Chern gauge can be viewed as a $C_{2z}\TR$-protected double version of monopole Cooper pairing.

The relation between $\Delta_{\shpa}$ and the monopole Cooper pairing relies on the Chern gauge, because the monopole Cooper pairing is only defined between Chern states.
Nevertheless, as a generalization of \refcite{Yu2021EOCP} (generalization from sphere-like Fermi surfaces in \refcite{Yu2021EOCP} to torus-like MBZ here), we find that the channel splitting into trivial $\Delta_{\perp}$ and nontirival $\Delta_{\shpa}$ can be done for all gauges (even beyond the Chern gauge) by using the Wilson line and the gauge-invariant operator $P_{\Delta}(\bsl{k})=\ket{u_{+,\bsl{k}}}\Delta(\bsl{k})\bra{u_{-,-\bsl{k}}^{C_{2z}\TR}}$, where $\ket{u_{\pm,\bsl{k}}^{C_{2z}\TR}}=C_{2z}\TR\ket{u_{\pm,\bsl{k}}}$.
The gauge transformations of the generally defined $\Delta_{\shpa}$ and $\Delta_{\perp}$ are the same as the guage transformation of $\Delta$, meaning that $|\Delta_b(\bsl{k})|$ and the zeros of $\Delta_{b}(\bsl{k})$ are gauge invariant.
Then, we can define the gauge-invariant winding number $W_{b,i}$ for the $i$th zero of $\Delta_{b}(\bsl{k})$, and have 
\eq{
\label{eq:W_b_N_short}
\sum_{i} \W_{b,i} = \N_+ - (-1)^b \N_- = 1 - (-1)^b\ .
}
(See SM for details.)
It means that the zeros of $\Delta_{\shpa}$ are generally enforced by the Euler numbers $\N_\pm$ for any gauges of the normal-state basis, even when the normal-state gauges do not have well-defined Chern numbers.
In other words, \eqnref{eq:total_winding_Ch} in Chern gauge is just a special case of the gauge-independent \eqnref{eq:W_b_N_short}.
Therefore, $\Delta_{\shpa}$ is called the Euler obstructed pairing channel.

Typically, the parity-even inter-sublattice pairing tends to have a dominant $\Delta_{\shpa}$, where the parity is equal (opposite) to the $C_{2z}$ eigenvalue for the spin-singlet (spin-triplet) pairings.
To show this, we can use the Chern gauge since $|\Delta_{b}|$ is gauge-invariant.
Based on \eqnref{eq:sym_rep_Ch} and \eqnref{eq:Delta_para_perp_Ch}, we find that $|\Delta_{\shpa}|=0$ for parity-odd pairing, and thus only the parity-even pairing can have a dominant $\Delta_{\shpa}$.
Then, since the states in the Chern gauge are polarized to the sublattice $A$ or $B$ of the BM model~\cite{Tarnopolsky2019MagicAngleChiralLimit,Zaletel2019Nov5MATBGIntegerFilling} (see also \figref{fig:EOCP_Ch}(b)), the parity-even $\Delta_{\shpa}$ ($\Delta_{\perp}$) mainly corresponds to inter-sublattice (intra-sublattice) pairing.

\section{Nematic Nodal Superconductivity in MATBG}

Next we consider the case where the Euler obstructed pairing channel is sufficiently dominant, implying that $|\Delta_{\perp}|$ is perturbatively small compared to $|\Delta_\shpa|$ and the pairing is parity-even, and discuss the resultant nodal superconductivity.
We only need to study the gapless nodes of the spin-up block of the Bogoliubov-de Gennes (BdG) Hamiltonian in $+$ valley, whose matrix representation is labelled as $\H(\bsl{k})$ for basis $(c^\dagger_{+,\bsl{k},\uparrow},c^T_{-,-\bsl{k},\downarrow})$ with $c^\dagger_{\pm,\bsl{k},s}=(c^\dagger_{\pm,\bsl{k},1,s},c^\dagger_{\pm,\bsl{k},2,s})$; it is because the BdG gapless nodes are the same for the spin-down block owing to the normal-state spin SU(2) symmetry and the pairing form \eqnref{eq:H_pairing}, and the BdG gapless nodes for the $-$ valley can be obtained from the particle-hole symmetry.
As the presence or absence of BdG nodes is gauge-independent, we use the Chern gauge for convenience, resulting in
\eq{
\label{eq:h_BdG}
\H(\bsl{k})=\mat{ h_+(\bsl{k})-\mu & \Delta_\perp(\bsl{k}) + \Delta_{\parallel}(\bsl{k}) \\  [\Delta_\perp(\bsl{k}) + \Delta_{\parallel}(\bsl{k})]^\dagger & -h_+^T(\bsl{k})+\mu}\ ,
}
where $\mu$ is the chemical potential, $h_+(\bsl{k})=\epsilon(\bsl{k}) + \Re[f (\bsl{k})]\tau_x + \Im[f (\bsl{k})]\tau_y$ 
describes the normal-state nearly-flat bands in valley $+$, the form of $\Delta_b(\bsl{k})$ is in \eqnref{eq:Delta_para_perp_Ch}, and we choose the zero-point energy such that $\epsilon(K_M)=0$.

Owing to the parity-even nature of the pairing, $\H$ has an effective spinless $C_{2z}\TR$ symmetry as $\rho_0\tau_x\cc$ and a chiral symmetry $\ii \rho_y\tau_x$, belonging to the nodal class CI which can support stable zero-energy BdG gapless points protected by nonzero chiralities~\cite{Bzdusek2017AZInversionNodal}.
Here, $\cc$ is the complex conjugate, and $\rho$'s are the Pauli matrices for the particle-hole index. (See SM for details.)
In the following, we will discuss the $\Delta_\shpa$-guaranteed nodal superconductivity based on $\H$ for both $C_{3z}$-invariant and spontaneously nematic pairings.
We choose $\mu\in [\epsilon(\Gamma_M)-|f(\Gamma_M)|,\epsilon(\Gamma_M)+|f(\Gamma_M)| ]$, which is typically true for 2$\sim$3 holes per moir\'e unit cell since the bottom and top of the set of nearly-flat bands are typically at $\Gamma_M$ for realistic parameter values. (See SM for details.)
We also choose the Euler obstructed $\Delta_{\shpa}$ (or equivalently $d_{\shpa}(\bsl{k})$) to only have two zeros with winding 1, since more zeros typically require more complex pairing structure which tends to be physically unfavored.

A sufficiently-dominant $\Delta_\shpa$ guarantees $\H$ to be gapless only if $\H^{(0)}$ (which is $\H$ with $|\Delta_\perp|=0$) is gapless.
By diagonalizing $\H^{(0)}$, we find that $\H^{(0)}$ is gapless if and only if $\mu\in E(\Sigma)$, where $\Sigma$ and $E(\Sigma)$ are defined in the following.
Let us consider the following deformation
\eq{
\label{eq:deformation_d_f}
d_\shpa(\bsl{k}) \pm \lambda \ii f(\bsl{k})\ ,
} 
where $\lambda$ is gradually increased from $0$ to $1$.
Owing to the normal-state Euler numbers, \eqnref{eq:deformation_d_f} must have zeros for all $\lambda\in[0,1]$, since the deformation cannot merge the initial two zeros of $d_\shpa(\bsl{k})$ that have the same winding.
Then, the zeros of \eqnref{eq:deformation_d_f} for all $\lambda\in[0,1]$ constitute $\Sigma$, and $E(\Sigma)$ consists of the values of $\epsilon(\bsl{k})\pm\sqrt{|f(\bsl{k})|^2 - |d_{\shpa}(\bsl{k})|^2}$ for all $\bsl{k}$ in $\Sigma$.  (See SM for details.)

\begin{figure}
    \centering
    \includegraphics[width=\columnwidth]{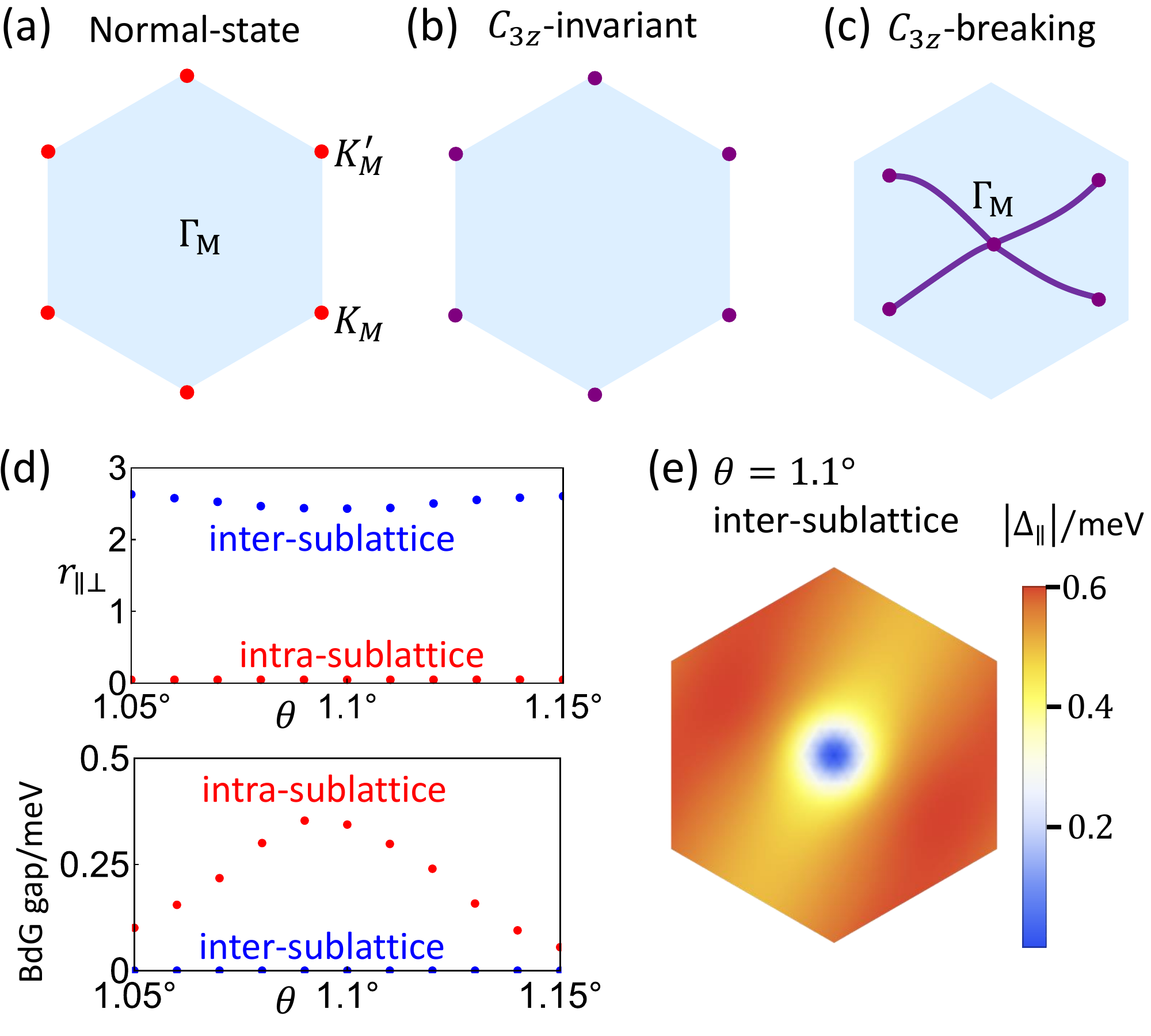}
    \caption{(a) The normal-state Dirac cones (red dots, zeros of $f(\bsl{k})$ in \eqnref{eq:h_BdG}) are typically located at $K_M$ and $K'_M$ in the MBZ (light blue).
    (b) Smallest $\Sigma$ (defined below \eqnref{eq:deformation_d_f}, purple) for $C_{3z}$-invariant pairing.
    (c) Illustrative $\Sigma$ (purple) for spontaneously nematic pairing, when pining both zeros of $\Delta_{\shpacap}(\bsl{k})$ at $\Gamma_M$.
    (d) Plots of the ratio $r_{\shpacap\perp}=\langle|\Delta_{\shpacap}|\rangle/\max(|\Delta_{\perp}|)$ and the BdG gap for the intra-sublattice and inter-sublattice pairings induced by the local attractive interaction at zero temperature.
    $\theta$ is the twist angle, and $\langle|\Delta_{b}|\rangle$ and $\max(|\Delta_{b}|)$ are the averaged and maximum values of $|\Delta_{b}(\bsl{k})|$ in the MBZ, respectively.
    (e) Plot of $|\Delta_{\shpacap}|$ of the inter-sublattice pairing in the MBZ for $\theta=1.1^{\circ}$ at zero temperature.
    $\Delta_{\shpacap}(\bsl{k})$ has two winding-$1$ zeros (or equivalently a winding-$2$ zero) at $\Gamma_M$, agreeing with the analogous discussion on the inter-Chern modes in \refcite{Zaletel2020Sep30SoftModesTBG}.
    }
    \label{fig:EOCP_NodalSC}
\end{figure}

The difference between $C_{3z}$-invariant and spontaneously nematic pairings lies in the different shapes of $\Sigma$.
$f(\bsl{k})$ typically has two zeros at $K_M$ and $K_M'$ (\figref{fig:EOCP_NodalSC}(a)).
For $C_{3z}$-invariant pairing, the two zeros of $d_{\shpa}(\bsl{k})$ are also pinned at $K_M$ and $K_M'$ by the $C_{3z}$ symmetry.
Then, \eqnref{eq:deformation_d_f} is typically zero at $K_M$ and $K_M'$, meaning that the initial two zeros of $d_{\shpa}(\bsl{k})$ typically does not move during the deformation.
As a result, $\Sigma$ is typically localized in the neighborhood of $K_M$ and $K_M'$ (the simplest case shown in \figref{fig:EOCP_NodalSC}(b)), and $E(\Sigma)$ only contains energies close to zero, leading to gapped $\H^{(0)}$ for considerably large $\mu$.
Therefore, a sufficiently-dominant $\Delta_{\shpa}$ cannot always guarantee nodal superconductivity when the pairing is $C_{3z}$-invariant, even if fine-tuning cases are ruled out. 
(See SM for details.)

On the other hand, for spontaneously nematic pairing, only one of the two zeros of $d_{\shpa}(\bsl{k})$ is constrained by the $C_{3z}$ eigenvalues, and is pinned at $\Gamma_{M}$.
Then, without invoking fine tuning, there must be continuous paths connecting $\Gamma_{M}$ to zeros of $d_\shpa(\bsl{k}) \pm \ii f(\bsl{k})$ (\figref{fig:EOCP_NodalSC}(c)), resulting that $\mu\in[\epsilon(\Gamma_M)-|f(\Gamma_M)|,\epsilon(\Gamma_M)+|f(\Gamma_M)| ]\subset E(\Sigma)$ and then $\H^{(0)}$ has gapless nodes with nonzero chiralties.
Therefore, when the pairing is spontaneously nematic, a sufficiently-dominant $\Delta_{\shpa}$ can always guarantee nodal superconductivity unless invoking fine tuning.
(See SM for details.)

The above mechanism for nematic nodal superconductivity is different from that discussed in \refcite{Wu2019TSCMATBG} since the latter does not involve any normal-state band topology.
More importantly, the mechanism in \refcite{Wu2019TSCMATBG} relies on a scalar pairing, which is not required in our work. (See SM for details.)

The statements in the above discussion are independent of the specific form of the interaction that accounts for the pairing form \eqnref{eq:H_pairing}.
Nevertheless, we use a local attractive interaction, which has a similar form as that mediated by the acoustic phonons~\cite{Wu2018PhononSCMATBG,Wu2019PhononResSCMATBG,Wu2019TSCMATBG}, to verify these general statements.
According to \refcite{Wu2019PhononResSCMATBG}, by tuning the interaction strength, we can get two types of $C_{2z}\TR$-invariant intervalley parity-even pairings: $C_{3z}$-invariant intra-sublattice pairings and spontaneously-nematic inter-sublattice pairings.
We obtain spin-triplet pairings of both types for 2.5 holes per moir\'e unit cell and $w_0/w_1=0.8$ by numerically solving the self-consistent equation, and find both the resultant intra-sublattice and inter-sublattice pairings have the form in \eqnref{eq:H_pairing}.
By using the gauge-invariant formlism, we find that the intra-sublattice and inter-sublattice pairings have dominant $\Delta_\perp$ and $\Delta_\shpa$ channels (\figref{fig:EOCP_NodalSC}(d)), respectively, agreeing with the above argument.
Moreover, since the inter-sublattice pairing has two winding-1 zeros for $\Delta_{\shpa}$ (as exemplified in \figref{fig:EOCP_NodalSC}(e)), the corresponding BdG Hamiltonian must be nodal, which is also verified in \figref{fig:EOCP_NodalSC}(d). (See details in SM.)
In short, the inter-sublattice pairing that we get from the local interaction is a spontaneously-nematic pairing that has an Euler obstructed channel dominant enough to guarantee nodal superconductivity.

Nodal superconductivity for the inter-valley inter-sublattice pairing was also shown in \refcite{Po20212DNodalSCBy1DAIII}.
The 2D nodal superconductivity in \refcite{Po20212DNodalSCBy1DAIII} is enforced by the normal-state chiral-symmetry-protected winding numbers, but the normal-state chiral symmetry is not exact in the BM model with realistic parameter values.
In contrast, our mechanism relies on the normal-state $C_{2z}\TR$-protected Euler numbers, which are exactly well-defined in the BM model with realistic parameter values.

\section{Bounded Superfluid Weight}

We now discuss lower bound of superfluid weight within the mean-field approximation.
We adopt the exact-flat-band approximation~\cite{Peotta2015SWGeometric,Liang2017SWBandGeo,Xie2020TopologyBoundSCTBG,Herzogarbeitman2021SWBound,Torma2021SFReview}, where we choose the normal-state flat bands to be exactly-flat.
By using the formalism of Euler obstructed Cooper pairing, we obtain a lower bound for the trace of the zero-temperature superfluid weight for the $C_{2z}\TR$-invariant pairing in \eqnref{eq:H_pairing}, which reads
\eq{
\label{eq:SW_bound}
\Tr[D_{SF}]\geq \left\langle \frac{\left[|\Delta_{\perp}(\bsl{k})|-|\Delta_{\shpa}(\bsl{k})|\right]^2}{\sqrt{\left[|\Delta_{\perp}(\bsl{k})|-|\Delta_{\shpa}(\bsl{k})|\right]^2+\mu^2}} \right\rangle_g \frac{4 e^2}{\pi}\N_+\ ,
}
where we have chosen the unit system in which $\hbar=c=1$, $e$ is the elementary charge,
\eq{
\left\langle x(\bsl{k}) \right\rangle_g = \frac{ \int_{\text{MBZ}} d^2 k\ x(\bsl{k}) \Tr[g(\bsl{k})] }{ \int_{\text{MBZ}} d^2 k\ \Tr[g(\bsl{k})] }\ ,
}
and $g_{ij}(\bsl{k})=\frac{1}{2}\Tr[\partial_{k_i} P_+(\bsl{k}) \partial_{k_j} P_+(\bsl{k})]$ is the Fubini-Study metric for $P_+(\bsl{k})=\ket{u_{+,\bsl{k}}}\bra{u_{+,\bsl{k}}}$.
If we choose the time-reversal-invariant uniform s-wave pairing used in \refcite{Xie2020TopologyBoundSCTBG}, \eqnref{eq:SW_bound} reproduces the lower bound presented in \refcite{Xie2020TopologyBoundSCTBG}; however our \eqnref{eq:SW_bound} holds for any pairing of the form \eqnref{eq:H_pairing}, even if the pairing is not uniform s-wave (like the inter-sublattice pairing in \figref{fig:EOCP_NodalSC}).
For MATBG with $\theta$ very close to $1.1^\circ$ and with pairings derived from the local attractive interaction mentioned above, the exact-flat-band approximation is valid for the study of the superfluid weight, and $\Tr[D_{SF}]$ estimated from the bound in \eqnref{eq:SW_bound} is roughly $ 10^8$ H$^{-1}$ for both intra-sublattice and inter-sublattice pairings, similar to the values theoretically estimated in \refcite{Xie2020TopologyBoundSCTBG} and reported in \refcite{Tian2021ExpTBGSW}, meaning that \eqnref{eq:SW_bound} is reasonably tight as a lower bound.
(See details in SM.)

\section{Discussion}

In summary, we have identified MATBG as a realistic superconductor that potentially hosts Euler obstructed Cooper pairing.
In this work, we allow several symmetries (like $C_{2x}$) of the BM model to be broken either spontaneously or externally in the normal state.
An interesting direction is to study the interplay between these symmetries and the Euler obstructed Cooper pairing.

\section{Acknowledgments}
J.Y. thanks B. Andrei Bernevig, Yang-Zhi Chou, Zhi-Da Song, and in particular Jie Wang for helpful discussions.
This work is supported by Laboratory of Physical Science.
F.W. is supported by start-up funds of Wuhan University.

\bibliography{bibfile_references.bib}

\appendix

\onecolumngrid
\tableofcontents

\section{Conventions}

We use $C_{nj}$ to label the spinless part of a $n$-fold rotation along axis $j$, and use $\overline{C}_{nj}$ to label the corresponding spinful operation.

We only consider the spinless part of any operation $g$ when acting $g$ on any spinless state.

We use the unit system in which $\hbar=1$, unless specified otherwise.

We always imply $\bsl{k}\in\dsR^2$ unless specifying $\bsl{k}\in${\BZ}, where ${\BZ}$ is short for the first Brillouin zone.

\section{Euler Obstructed Cooper Pairing in 2D Systems with $C_{2z}\TR$ Symmetry}
\label{app:general_formalism}

In this section, we will follow \refcite{Yu2021EOCP} to introduce the Euler obstructed Cooper pairing in 2D systems with $C_{2z}\TR$ symmetry.
We start with the general formalism and then focus on the Chern gauge.

\subsection{General Formalism}

In this part, we will introduce the gauge-invariant formalism for the Euler obstructed Cooper pairing in 2D systems.
In most the discussions of this part, we will not choose any specific gauge; but in certain cases, we might use choose convenient gauges to prove a gauge-independent statement.

\subsubsection{Normal State}

The normal-state that we consider is a 2D (effectively) noninteracting fermionic Hamiltonian $\widetilde{H}$ that satisfies the following assumptions.
\begin{assumption}
\label{asm:normal_state_C2T}
    $\widetilde{H}$ has 2D lattice translation symmetries and $\overline{C}_{2z}\TR$ symmetry, with $z$ perpendicular to the 2D system.
\end{assumption} 
\begin{assumption}
    \eq{\widetilde{H}=\widetilde{H}_+ + \widetilde{H}_-}
    The basis of $\widetilde{H}_\alpha$ is created by $\psi^\dagger_{\alpha,\bsl{k},i,s}$ with $\alpha =\pm$ the valley index, $\bsl{k}$ the momentum, $s$ the spin index, and $i$ labelling all other degrees of freedom. 
    $\{ \psi^\dagger_{\alpha,\bsl{k},i,s} , \psi_{\alpha',\bsl{k}',i',s'} \} = \delta_{\alpha\alpha'} \delta_{ii'} \delta_{ss'}\delta_{\bsl{k}\bsl{k}'}$.  
   $e^{-\ii\bsl{k}\cdot\bsl{r}}\psi^\dagger_{\alpha,\bsl{k}} e^{\ii \bsl{k}\cdot\bsl{r}} = \psi^\dagger_{\alpha,0}$ $\forall\bsl{k}\in${\BZ}, and $e^{-\ii\bsl{G}\cdot\bsl{r}}\psi^\dagger_{\alpha,0} e^{\ii \bsl{G}\cdot\bsl{r}} = \psi^\dagger_{\alpha,0} S_{\alpha,\bsl{G}}\otimes s_0$, where $s_{0,x,y,z}$ are the Pauli matrices for the spin index, $\psi^\dagger_{\alpha,\bsl{k}}=(..., \psi^\dagger_{\alpha,\bsl{k},i,s},...)$, $\bsl{G}$ is an arbitrary reciprocal lattice vector, and $S_{\alpha,\bsl{G}}$ is a unitary matrix with $S_{\alpha,\bsl{G}+\bsl{G}'}=S_{\alpha,\bsl{G}} S_{\alpha,\bsl{G}'}$.
\end{assumption}
\begin{assumption}
$\overline{C}_{2z}\TR$ does not change the valley index $\alpha$. $\widetilde{H}_{\alpha}$ has its onw spin-charge $\U(2)$ symmetry for $\alpha\in\{+,-\}$.
\end{assumption}
Then, we have  $\overline{C}_{2z}\TR\psi^{\dagger}_{\pm,\bsl{k}} (\overline{C}_{2z}\TR)^{-1}= \psi^{\dagger}_{\pm,\bsl{k}}\widetilde{U}_{\pm}\otimes(-\ii s_x)\ \forall\ \bsl{k}\in\text{{\BZ}}$, meaning that $[\overline{C}_{2z}\TR, \widetilde{H}_{\alpha}] = 0$.
\eq{
\label{eq:htilde_gen}
\widetilde{H}_{\pm} = \sum_{\bsl{k}\in \text{{\BZ}}} \psi^{\dagger}_{\pm,\bsl{k}} \widetilde{h}_{\pm}(\bsl{k})\otimes s_0 \psi_{\pm,\bsl{k}}\ ,
}
meaning that $[C_{2z}\TR, \widetilde{H}_{\alpha}] = 0$.
In the following, we will use $C_{2z}\TR$ more often.

We can extend the domain from of $\psi^{\dagger}_{\pm,\bsl{k}}$ from {\BZ} to $\dsR^2$ by $\psi^{\dagger}_{\pm,\bsl{k}} = e^{\ii \bsl{k}\cdot\bsl{r}}\psi^{\dagger}_{\pm,0}  e^{-\ii \bsl{k}\cdot\bsl{r}}$ $\forall\bsl{k}\in\dsR^2$, and we have $S^\dagger_{\alpha,\bsl{G}} \widetilde{h}_{\alpha}(\bsl{k}+\bsl{G})S_{\alpha,\bsl{G}} = \widetilde{h}_{\alpha}(\bsl{k})$.

\begin{assumption}
\label{asm:isolated_set_bands_low-energy}
Each $\widetilde{H}_\alpha$ (with $\alpha=\pm$) has an isolated set of two spin doubly degenerated bands at low energy.
\end{assumption}
We use Bloch basis $\ket{\psi_{\alpha,\bsl{k},a}}\otimes \ket{s} $ to label the Block basis of the two spin doubly degenerated bands in valley $\alpha$, where $a=1,2$ labelling the two spinless states. 
By defining $V_{\alpha,a}(\bsl{k})$ as the orthonormal linear combinations of the eigenvectors of $\widetilde{h}_{\alpha}(\bsl{k})$ for the two bands in valley $\alpha$, we have 
\eq{
\label{eq:psi_c_relation}
\ket{\psi_{\alpha,\bsl{k},a}}\otimes \ket{s}  = c^\dagger_{\alpha,\bsl{k},a,s}\ket{\Omega}\ ,
}
where $c^\dagger_{\alpha,\bsl{k},a,s} = \sum_{i} \psi^\dagger_{\alpha,\bsl{k},i,s} [V_{\alpha,a}(\bsl{k})]_i$.
Then, the projected Hamiltonian for the isolated set of bands in valley $\alpha$ reads
\eq{
\label{eq:h_gen}
H_{\alpha}=\sum_{\bsl{k}\in {\BZ}} c^\dagger_{\alpha,\bsl{k}} h_{\alpha}(\bsl{k})\otimes s_0 \ c_{\alpha,\bsl{k}}\ ,
}
where 
\eq{
h_{\alpha}(\bsl{k}) = V_{\alpha}^\dagger(\bsl{k}) \widetilde{h}_{\alpha}(\bsl{k}) V_{\alpha}(\bsl{k}) \ ,
}
$c^\dagger_{\alpha,\bsl{k}} = (..., c^\dagger_{\alpha,\bsl{k}, a,s},...)$, and 
$V_{\alpha}(\bsl{k}) = \mat{ V_{\alpha,1}(\bsl{k})  &  V_{\alpha,2}(\bsl{k}) }$

We can choose $V_\alpha(\bsl{k}+\bsl{G})=S_{\alpha,\bsl{G}} V_{\alpha} (\bsl{k})$.
Then, we have $\ket{\psi_{\alpha,\bsl{k},a}}=\ket{\psi_{\alpha,\bsl{k}+\bsl{G},a}}$ and $c^\dagger_{\alpha,\bsl{k}+\bsl{G},a,s}=c^\dagger_{\alpha,\bsl{k},a,s}$.
We define $\ket{u_{\alpha,\bsl{k},a}}=e^{-\ii \bsl{k}\cdot\bsl{r}}\ket{\psi_{\alpha,\bsl{k},a}}$ with $\bsl{r}$ the postition operator, and define $\ket{u_{\alpha,\bsl{k}}}=(\ket{u_{\alpha,\bsl{k},1}},\ket{u_{\alpha,\bsl{k},2}})$, which satisfies
\eq{
\ket{u_{\alpha,\bsl{k}+\bsl{G}}}=e^{-\ii \bsl{G}\cdot\bsl{r}}\ket{u_{\alpha,\bsl{k}}}\ \forall\ \text{reciprocal lattice vector } \bsl{G}\ .
}
Then, we have
\eq{
C_{2z}\TR \ket{u_{\alpha,\bsl{k}}} = \ket{u_{\alpha,\bsl{k}}} U_{\alpha}(\bsl{k})\ ,
}
$\ket{u_{\alpha,\bsl{k}}}$ has a $\U(2)$ gauge freedom 
\eq{
\label{eq:gauge_trans_u}
\ket{u_{\alpha,\bsl{k}}}\rightarrow \ket{u_{\alpha,\bsl{k}}} R_{\alpha,\bsl{k}}
}
with $R_{\alpha,\bsl{k}}\in \U(2)$ and $R_{\alpha,\bsl{k}+\bsl{G}}=R_{\alpha,\bsl{k}}$.
By gauge invariant, we mean invariant under \eqnref{eq:gauge_trans_u}.

The Wilson line matrix~\cite{Dai2011Z2WilsonLoop} is crucial for our later discussions.
The Wilson line matrix is defined as
\eq{
W_{\alpha}(\bsl{k}_0\xrightarrow{\gamma}\bsl{k})=\lim_{L\rightarrow \infty} \bra{u_{\alpha,\bsl{k}_0}} P_{\alpha}(\bsl{k}_1)P_{\alpha}(\bsl{k}_2)...P_{\alpha}(\bsl{k}_L) \ket{u_{\alpha,\bsl{k}}}\ ,
}
where $\gamma$ is a continuous path from $\bsl{k}_0$ to $\bsl{k}$, $\bsl{k}_1$,...,$\bsl{k}_L$ are sequentially arranged on $\gamma$ with $|\bsl{k}_{i+1}-\bsl{k}_{i}|=O\left(\frac{\text{length of }\gamma}{L}\right)$, and $P_{\alpha}(\bsl{k})=\ket{u_{\alpha,\bsl{k}}}\bra{u_{\alpha,\bsl{k}}}$.
Moreover, for any $\bsl{k}_0,\bsl{k}\in \dsR^2$, $\det[W_{\pm}(\bsl{k}_0\xrightarrow{\gamma}\bsl{k})]=\det[W_{\pm}(\bsl{k}_0,\bsl{k})]$ is independent of the path $\gamma$.

In particular, we define the continuous $\gamma$ to be effectively-closed if and only if $\bsl{k}-\bsl{k}_0$ is a reciprocal lattice vector.
Then, owing to the $C_{2z}\TR$ symmetry, $\det[W_{\alpha}(\bsl{k}\xrightarrow{\gamma}\bsl{k}+\bsl{G})]=\pm 1$ for any effectively-closed $\gamma$, which can be derived based on $C_{2z}\TR P_{\alpha}(\bsl{k}) (C_{2z}\TR)^\dagger=P_{\alpha}(\bsl{k})$.
Then, we choose the following assumption for the normal-state Hamiltonian $H_{\pm}$.
\begin{assumption}
\label{asm:oritentable}
   $\exists\ \bsl{k}_{0,\beta}\in\dsR^2$\text{ such that }$\det[W_{\pm}(\gamma_\beta)]=1$, where $\gamma_\beta$ is the straight path from $\bsl{k}_{0,\beta}$ to $\bsl{k}_{0,\beta}+\bsl{b}_{\beta}$, $\beta=1,2$, and $\bsl{b}_{1,2}$ are the basis vectors of the reciprocal lattice.
\end{assumption} 
This assumption implies that $\det[W_{\pm}(\bsl{k}\xrightarrow{\gamma}\bsl{k}+\bsl{G})]=1$ for all effectively-closed $\gamma$, since we can always express $\det[W_{\pm}(\bsl{k}\xrightarrow{\gamma}\bsl{k}+\bsl{G})]$ in terms of the products of $\det[W_{\pm}(\gamma_1)]$ and $\det[W_{\pm}(\gamma_2)]$.

With the properties of the Wilson line, we can define the following path-independent $\eta$ factor \eq{
\eta_{\alpha,\bsl{k}_0}(\bsl{k})=\sqrt{\det\left(\braket{u_{\alpha,\bsl{k}_0}^{C_{2z}\TR}}{u_{\alpha,\bsl{k}_0}}\right)}\det[W_{\alpha}(\bsl{k}_0,\bsl{k})]\ ,
} 
where $\bsl{k}_0$ is treated as a base point, and $\ket{u_{\alpha,\bsl{k}}^{C_{2z}\TR}}=C_{2z}\TR \ket{u_{\alpha,\bsl{k}}}$.
Then, we can define
\eqa{
\label{eq:N_pm}
& Q_{\alpha,\bsl{k}_0}(\bsl{k})=\frac{\eta_{\alpha,\bsl{k}_0}^*(\bsl{k})}{\sqrt{2}} \ket{u_{\alpha,\bsl{k}}}(-\ii\tau_y)\bra{u_{\alpha,\bsl{k}}^{C_{2z}\TR}} \\
& \Phi_{\alpha,\bsl{k}_0}(\bsl{k})=\frac{1}{\sqrt{2}}Tr[Q_{\alpha,\bsl{k}_0}(\bsl{k})\partial_{k_x}P_{\alpha}(\bsl{k})\partial_{k_y}P_{\alpha}(\bsl{k})]-(\partial_{k_x}\leftrightarrow \partial_{k_y}) \\
& \N_{\alpha,\bsl{k}_0} = \frac{1}{2\pi} \int_{\text{{\BZ}}} d^2 k \Phi_{\alpha,\bsl{k}_0}(\bsl{k})\\
& \N_{\alpha}=|\N_{\alpha,\bsl{k}_0}|\ ,
}
where (and also for the rest of the work) the square root is always fixed in the principle branch as
\eq{
\sqrt{e^{\ii\theta}}= e^{\ii (\theta+2\pi n)/2}\ \text{for } n\in \dsZ \text{ such that } \theta+2\pi n \in (-\pi,\pi]\ .
}
Among the quantities defined above, $\N_{\alpha}$ is invariant under the gauge transformation \eqnref{eq:gauge_trans_u} and the shift of the base point $\bsl{k}_0$, since both the gauge transformation \eqnref{eq:gauge_trans_u} and the shift of the base point $\bsl{k}_0$ can only change $Q_{\alpha,\bsl{k}_0}(\bsl{k})$, $ \Phi_{\alpha,\bsl{k}_0}(\bsl{k})$ and $\N_{\alpha,\bsl{k}_0}$ by the same sign factor.

\begin{figure}
    \centering
    \includegraphics[width=0.8\columnwidth]{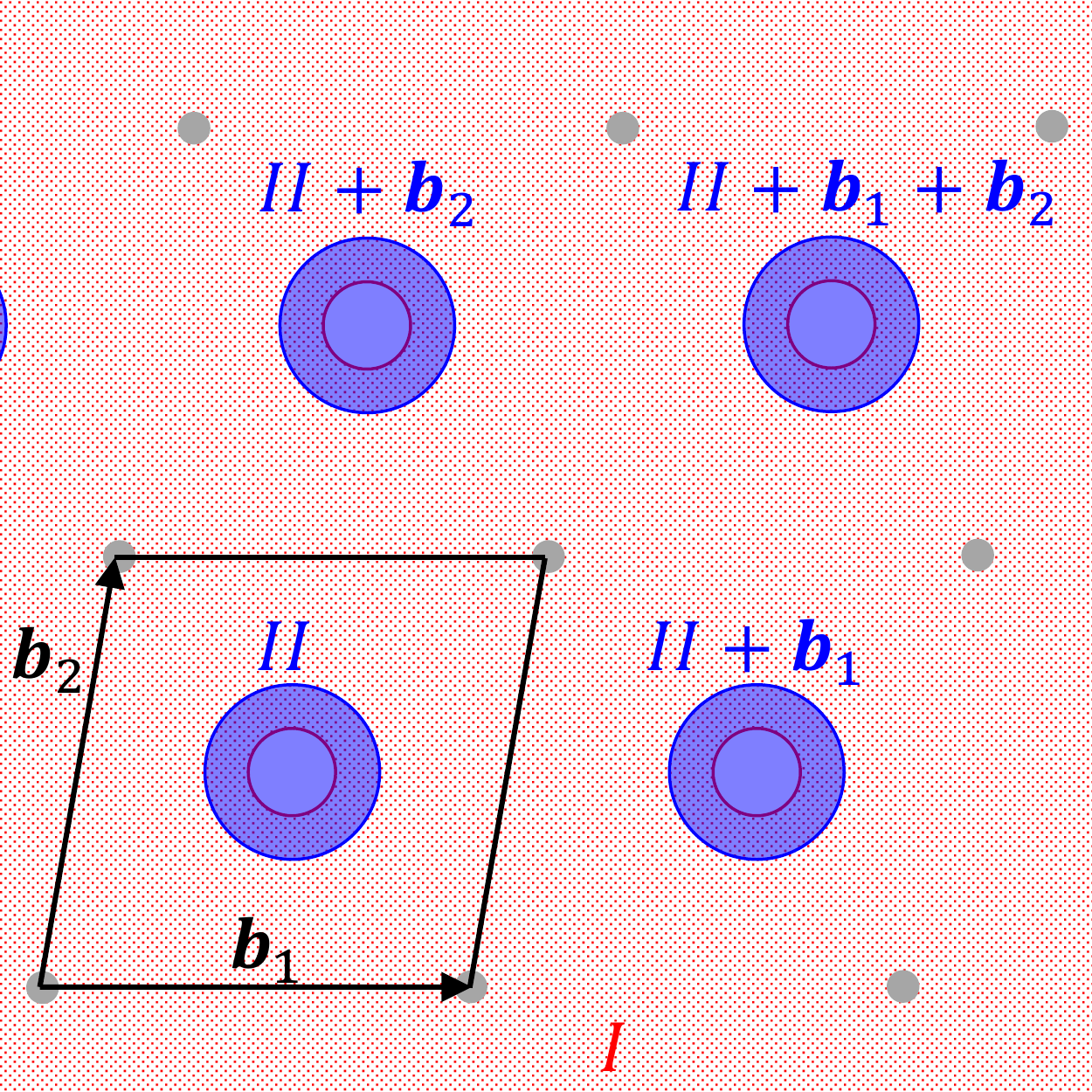}
    \caption{This shows the patches that we choose for the patchwise-smooth real oriented gauge.
    The gray dots are reciprocal lattice points, $\bsl{b}_{1,2}$ are the basis vectors of the reciprocal lattice, and the tetragon marked by the black lines is the unit cell of the reciprocal lattice.
    Patch $I$ (red doted) covers $\dsR^2$ except holes that form a reciprocal lattice, and the holes are covered by the patch $II$ (blue) and its copies given by shifts along the reciprocal lattice vectors.
    This choice of patches is allowed by \asmref{asm:oritentable}, which makes the corresponding real vector bundle orientable.
    }
    \label{fig:patch}
\end{figure}

Now we show $\N_{\alpha,\bsl{k}_0}\in \dsZ$ and $\N_{\alpha}\in\dsN$ for any gauges and base points.
All we need to show is that $\N_{\alpha,\bsl{k}_0}\in\dsZ$ holds for one gauge and one choice of base point.
It is because if it is true, then $\N_{\alpha,\bsl{k}_0}\in\dsZ$ and $\N_{\alpha}\in\dsN$ holds for all gauges and all choices of base points, since the gauge transformation and the base point shift can only change $\N_{\alpha,\bsl{k}_0}$ by a sign.
Then, we will use a special gauge to show $\N_{\alpha,\bsl{k}_0}\in\dsZ$, which is the real oriented gauge~\cite{Ahn2018MonopoleNLSM,Ahn2019TBGFragile} as described in the following.
First, there exists a gauge 
\eq{
\label{eq:real_gauge}
\ket{\widetilde{u}_{\alpha,\bsl{k}}}=\ket{u_{\alpha,\bsl{k}}} \widetilde{R}_\alpha(\bsl{k})\sqrt{\Lambda_{\alpha}(\bsl{k})}\ ,
}
which satisfies $C_{2z}\TR\ket{\widetilde{u}_{\alpha,\bsl{k}}}=\ket{\widetilde{u}_{\alpha,\bsl{k}}}$.
Here $\widetilde{R}_\alpha(\bsl{k})$ is an orthogonal real matrix that satisfies 
\eq{
[\widetilde{R}_{\alpha}(\bsl{k})]^T U_{\alpha}(\bsl{k}) \widetilde{R}_{\alpha}(\bsl{k}) = \Lambda_{\alpha}(\bsl{k})
}
with $\Lambda_{\alpha}(\bsl{k})$ diagonal, and the square roots in $\sqrt{\Lambda_{\alpha}(\bsl{k})}$ act on the diagonal elements~\cite{Bouhon2020WeylNonabelian}.
Furthermore, we ensure $\ket{\widetilde{u}_{\alpha,\bsl{k}+\bsl{G}}}=e^{-\ii \bsl{G}\cdot \bsl{r}}\ket{\widetilde{u}_{\alpha,\bsl{k}}}$.
$\ket{\widetilde{u}_{\alpha,\bsl{k}}}$ is called a real gauge.
The real gauge is not unique: $\ket{\widetilde{u}_{\alpha,\bsl{k}}}\rightarrow \ket{\widetilde{u}_{\alpha,\bsl{k}}}R_{\alpha,\bsl{k}}$ with $R_{\alpha,\bsl{k}}\in \O(2)$ and $R_{\alpha,\bsl{k}+\bsl{G}}=R_{\alpha,\bsl{k}}$ gives another real gauge.
If the $C_{2z}\TR$-protected topology of the nearly flat bands is nontrivial, it is impossible to make $\ket{\widetilde{u}_{\alpha,\bsl{k}}}$ smooth everywhere in $\dsR^2$.
However, owing to \asmref{asm:oritentable}, we can choose patches that cover $\dsR^2$ as \figref{fig:patch}, and we can make $\ket{\widetilde{u}_{\alpha,\bsl{k}}}$ to be smooth in each patch, noted as $\ket{\widetilde{u}_{\alpha,\bsl{k}}^{A}}$ where $A=I$, $II$, $II+\bsl{b}_1$, $II+\bsl{b}_2$, $II+\bsl{b}_1+\bsl{b}_2$, $...$ labels the patches. 
Note that for any reciprocal lattice vector $\bsl{G}$, $\ket{\widetilde{u}_{\alpha,\bsl{k}+\bsl{G}}^{I}}=e^{-\ii \bsl{G}\cdot\bsl{r}}\ket{\widetilde{u}_{\alpha,\bsl{k}}^{I}}$ if $\bsl{k}+\bsl{G},\bsl{k}\in I$, and $\ket{\widetilde{u}_{\alpha,\bsl{k}+\bsl{G}}^{II+\bsl{G}}}=e^{-\ii \bsl{G}\cdot\bsl{r}}\ket{\widetilde{u}_{\alpha,\bsl{k}}^{II}}$ if $\bsl{k}\in II$.
For any $\bsl{k}\in A\cap A'$, $\braket{\widetilde{u}_{\alpha,\bsl{k}}^{A}}{\widetilde{u}_{\alpha,\bsl{k}}^{A'}}\in \O(2)$.
As shown in \figref{fig:patch}, the intersection only occurs between $I$ and copies of $II$, and then we further choose the proper ${\widetilde{u}_{\alpha,\bsl{k}}^{A}}$ with $A=II,II+\bsl{b}_1,II+\bsl{b}_2,...$ to realize 
\eq{
\braket{\widetilde{u}_{\alpha,\bsl{k}}^{A}}{\widetilde{u}_{\alpha,\bsl{k}}^{A'}}\in \SO(2)\ \forall\ \text{any $\bsl{k}\in A\cap A'$}\ .
}
As a result, $\ket{\widetilde{u}_{\alpha,\bsl{k}}^{A}}$ is a patchwise-smooth real oriented gauge, and the real oriented gauge for $\ket{u_{\alpha,\bsl{k}}}$ is $\ket{u_{\alpha,\bsl{k}}^{RO}}=\ket{\widetilde{u}_{\alpha,\bsl{k}}^{A_{\bsl{k}}}}$, where $A_{\bsl{k}}$ is the unique patch chosen for $\bsl{k}$.
We emphasize that the existence of the smooth real $\ket{\widetilde{u}_{\alpha,\bsl{k}}^{I}}$ in $I$, as well as the existence of both the patchwise-smooth real oriented gauge $\ket{\widetilde{u}_{\alpha,\bsl{k}}^{A}}$ and the real oriented gauge $\ket{u_{\alpha,\bsl{k}}^{RO}}$, relies on \asmref{asm:oritentable}, which makes the real vector bundle given by $\ket{\widetilde{u}_{\alpha,\bsl{k}}}$ orientable.

For the real oriented gauge $\ket{u_{\alpha,\bsl{k}}^{RO}}$, we have 
\eqa{
& \eta_{\alpha,\bsl{k}_0}(\bsl{k})=1\ ,\\
& Q_{\alpha,\bsl{k}_0}(\bsl{k})=\frac{1}{\sqrt{2}} \ket{u_{\alpha,\bsl{k}}^{RO}}(-\ii\tau_y)\bra{u_{\alpha,\bsl{k}}^{RO}}\ ,
}
which leads to 
\eq{
\Phi_{\alpha,\bsl{k}_0}(\bsl{k})=f_\alpha(\bsl{k})=\braket{\partial_{k_x} u_{\alpha,\bsl{k},1}^{RO}}{\partial_{k_y} u_{\alpha,\bsl{k},2}^{RO}}-(\partial_{k_x}\leftrightarrow \partial_{k_y})\ ,
}
where $f_\alpha(\bsl{k})$ is the real curvature for the real oriented gauge $\ket{u_{\alpha,\bsl{k}}^{RO}}$~\cite{Ahn2019TBGFragile}.
Then, 
\eq{
\N_{\alpha,\bsl{k}_0}=\frac{1}{2\pi}\int_{\text{{\BZ}}} d^2 k\  \Phi_{\alpha,\bsl{k}_0}(\bsl{k}) =\frac{1}{2\pi}\int_{\text{{\BZ}}} d^2 k\ f_\alpha(\bsl{k}) = e_{2,\alpha}\in \dsZ\ ,
}
where $e_{2,\alpha}$ is the integer Euler class for the real oriented gauge $\ket{u_{\alpha,\bsl{k}}^{RO}}$.
As a result, we know $\N_{\alpha,\bsl{k}_0}\in \dsZ$ and $\N_{\alpha}\in\dsN$ for any gauges and base points.
Since $\N_{\alpha}=|e_{2,\alpha}|$ for any real oriented gauge and $\N_{\alpha}$ is gauge-independent, $\N_{\alpha}$ is called the Euler number~\cite{Yu2021EOCP}, which is distinguished from the Euler class $e_{2,\alpha}$~\cite{Ahn2019TBGFragile} that relies on the real oriented gauges.

In the following, we will introduce an extra assumption on the normal-state Hamiltonian.
\begin{assumption}
\label{asm:N_pm_nonzero}
    $\N_{\pm}\neq 0$
\end{assumption}
With this assumption, we can define 
\eqa{
\label{eq:Q_Phi}
&  Q_{\alpha}(\bsl{k})= \frac{\N_{\alpha,\bsl{k}_0}}{\N_{\alpha}}  Q_{\alpha,\bsl{k}_0}(\bsl{k}) \\ 
&  \Phi_{\alpha}(\bsl{k})= \frac{\N_{\alpha,\bsl{k}_0}}{\N_{\alpha}}  \Phi_{\alpha,\bsl{k}_0}(\bsl{k})\ ,
}
which are independent of the gauge and the choice of the base point.
The $C_{2z}\TR$ symmetry requires
\eq{
\label{eq:C2T_Q_Phi}
C_{2z}\TR Q_{\alpha}(\bsl{k}) (C_{2z}\TR)^{-1}= Q_{\alpha}(\bsl{k})\ ,\ \Phi_{\alpha}(\bsl{k})\in \dsR\ .
}
Moreover, $Q_{\alpha}(\bsl{k})$ is smooth everywhere in $\dsR^2$ since it is smooth for real oriented gauges and is gauge invariant, which means $\Phi_{\alpha}(\bsl{k})$ is smooth everywhere in $\dsR^2$ since $P_{\alpha}(\bsl{k})$ must be globally smooth.
These properties are useful in the following discussion.

\subsubsection{Euler Obstructed Cooper Pairing}
Now we discuss the Euler obstructed Cooper pairing.
The normal state that we consider satisfies \asmref{asm:normal_state_C2T}-\ref{asm:N_pm_nonzero}.
We consider the zero-temperature $C_{2z}\TR$ intervalley mean-field pairing operator among the low-energy isolated sets of bands in \asmref{asm:isolated_set_bands_low-energy}, and we choose it to satisfy the following condition.
\begin{assumption}
\label{asm:pairing}
\eq{
H_{pairing}=\sum_{\bsl{k}\in\text{\text{{\BZ}}}}c^{\dagger}_{+,\bsl{k}}\Delta(\bsl{k})\otimes\Pi (c^{\dagger}_{-,-\bsl{k}})^T+ h.c.\ ,
}
where  $c^{\dagger}_{\pm,\bsl{k}}=(..., c^{\dagger}_{\pm,\bsl{k},a,s} ,...)$, $\Delta(\bsl{k})$ is for spinless index $a$,  $\Pi$ is for the spin index $s$, $s_y \Pi^* s_y = \Pi$, and $\Pi^\dagger \Pi = s_0$. 
Moreover, $[C_{2z}\TR,H_{pairing}]=0$, and 
\eq{
\sum_{a,a'} [\Delta(\bsl{k})]_{aa'} \ket{u_{+,\bsl{k},a}}\ket{u_{-,-\bsl{k},a'}}
}
is smooth everywhere in $\dsR^2$.
\end{assumption}

The spinless part of the pairing $\Delta(\bsl{k})$ is the focus of this work.
We first discuss how to split $\Delta(\bsl{k})$ into two channels in a gauge independent way.
To do so, we define an operator
\eq{
P_{\Delta}(\bsl{k})=\ket{u_{+,\bsl{k}}}\Delta(\bsl{k})\bra{u^{C_{2z}\TR}_{-,-\bsl{k}}}\ ,
}
which is invariant under the gauge transformation \eqnref{eq:gauge_trans_u}, since $\Delta(\bsl{k})$ transforms as $\Delta(\bsl{k})\rightarrow R_{+,\bsl{k}}^\dagger \Delta(\bsl{k}) R_{-,-\bsl{k}}^*$ under \eqnref{eq:gauge_trans_u}.
Then, $P_{\Delta}(\bsl{k})$ should be smooth everywhere in $\dsR^2$.
It is because the $C_{2z}\TR$ symmetry restricts the total Chern number of $\ket{u_{\alpha,\bsl{k}}}$ to zero, and thus $\ket{u_{\alpha,\bsl{k}}}$ has complex gauge that is smooth everywhere in $\dsR^2$~\cite{Brouder2007Wannier}.
Then, $\Delta(\bsl{k})$ is smooth for any globally smooth complex gauges of $\ket{u_{\alpha,\bsl{k}}}$, resulting that $P_{\Delta}(\bsl{k})$ is smooth everywhere in $\dsR^2$ for any complex smooth gauges of $\ket{u_{\alpha,\bsl{k}}}$.
As $P_{\Delta}(\bsl{k})$ is gauge invariant, $P_{\Delta}(\bsl{k})$ is smooth everywhere in $\dsR^2$ for all gauges.

The gauge-independent way of splitting $\Delta(\bsl{k})$ is achieved by defining
\eq{
\label{eq:P_b}
P_{b}(\bsl{k})=\frac{1}{2}P_{\Delta}(\bsl{k})-(-1)^b Q_{+,\bsl{k}} P_{\Delta}(\bsl{q})Q_{-,-\bsl{k}}=\ket{u_{+,\bsl{k}}}\Delta_b(\bsl{k})\bra{u_{-,-\bsl{k}}^{C_{2z}\TR}}\ ,
}
which is gauge invariant and is smooth everywhere in $\dsR^2$.
Here $b\in\{\perp,\shpa\}$, $(-1)^{\perp}=1$, and $(-1)^{\shpa}=-1$.
Then, clearly, 
\eq{
\label{eq:Delta_split_app}
\Delta(\bsl{k})=\Delta_\perp(\bsl{k})+\Delta_\shpa(\bsl{k})\ .
}
Next, we discuss the winding of the possible zeros of $\Delta_b(\bsl{k})$.
First,
\eq{
|\Delta_b(\bsl{k})|=\sqrt{\frac{1}{2}\Tr[\Delta_b(\bsl{k})\Delta_b^\dagger(\bsl{k})]}=\sqrt{\frac{1}{2}\Tr[P_b(\bsl{k})P_b^\dagger(\bsl{k})]}
}
is gauge invariant and is smooth at where $|\Delta_b(\bsl{k})|\neq 0$.
Then, for any $\bsl{k}$ such that $|\Delta_b(\bsl{k})|\neq 0$, we can define
\eq{
\label{eq:v_b}
\bsl{v}_b(\bsl{k})=\frac{1}{\sqrt{2}}\Tr[Q_+(\bsl{k}) \hat{P}_b(\bsl{k})\nabla_{\bsl{k}}  \hat{P}_b^\dagger(\bsl{k})]\ ,
}
where $\hat{P}_b(\bsl{k})=P_b(\bsl{k})/|\Delta_b(\bsl{k})|$.
Clearly, $\bsl{v}_b(\bsl{k})$ and $\hat{P}_b(\bsl{k})$ are gauge invariant and are smooth at any $\bsl{k}$ such that $|\Delta_b(\bsl{k})|\neq 0$.
Owing to the $C_{2z}\TR$ invariance of the pairing operator, we have $C_{2z}\TR P_{\Delta}(\bsl{k}) (C_{2z}\TR)^{-1} = P_{\Delta}(\bsl{k})$.
Combined with \eqnref{eq:C2T_Q_Phi}, we know that the $C_{2z}\TR$ symmetry requires $\bsl{v}_b(\bsl{k})$ to be real.

\begin{figure}
    \centering
    \includegraphics{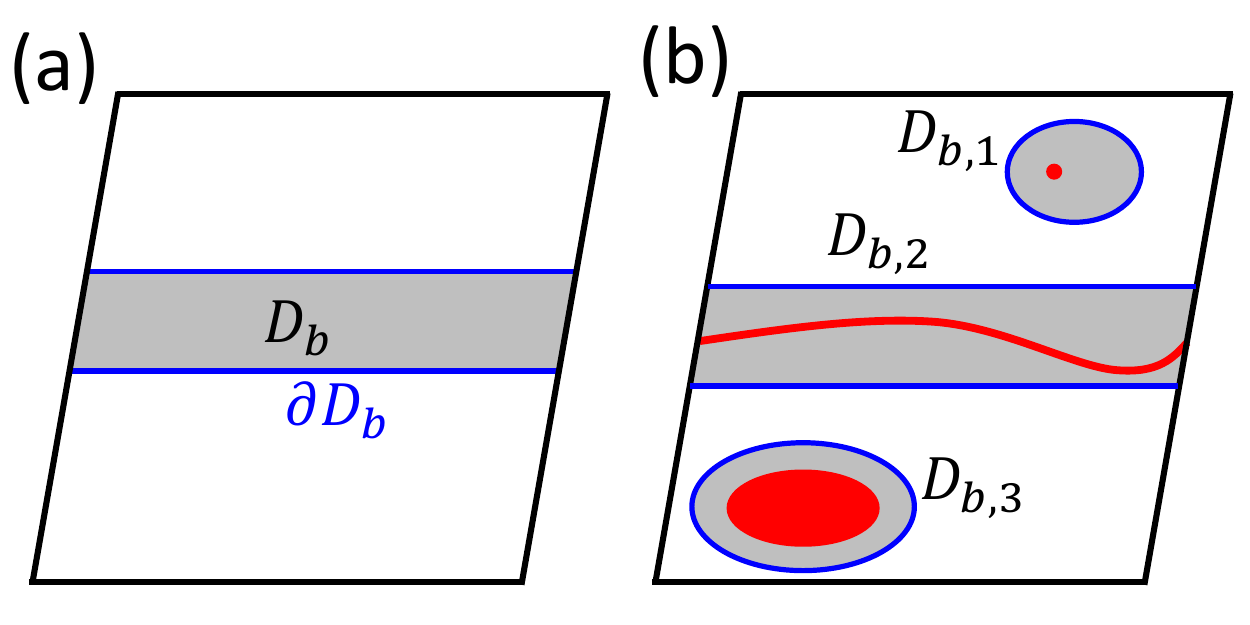}
    \caption{In this figure, we show the $D_{b}$  (\eqnref{eq:W_b}) or $D_{b,i}$  (\eqnref{eq:W_b_i}) regions in the reciprocal unit cell, whose boundary is marked by the black lines.
    In (a), we show a generic $D_b$ (gray area) that is a closed connected proper subset of the reciprocal unit cell.
    The boundary $\partial D_b$ (blue lines) is chosen as if the reciprocal unit cell is a torus with parallel opposite edges being identified.
    In (b), we choose three closed connected regions $D_{b,i}$ (gray areas) that include isolated zeros (red) of $\Delta_b$ in their interior. 
    $D_{b,1}$ contains a point zero, $D_{b,2}$ contains a line of zeros, and $D_{b,3}$ contains an area of zeros.
    Again, the boundaries (blue lines) of $D_{b,i}$'s are determined as if the reciprocal unit cell is a torus.
    }
    \label{fig:EOCP_D_Regions_App}
\end{figure}

The zeros of $\Delta_b(\bsl{k})$ are the zeros of $|\Delta_b(\bsl{k})|$, and the winding numbers of the possible zeros of $\Delta_b(\bsl{k})$ are defined by $\bsl{v}_b(\bsl{k})$ and $\Phi_{\alpha}(\bsl{k})$.
Since $\Delta_b(\bsl{k})$, $\bsl{v}_b(\bsl{k})$ and $\Phi_{\alpha}(\bsl{k})$ are periodic in $\bsl{G}$, \ie,
\eq{
\Delta_b(\bsl{k}+\bsl{G})=\Delta_b(\bsl{k})\ ,\ \bsl{v}_b(\bsl{k}+\bsl{G})=\bsl{v}_b(\bsl{k})\ ,\ \Phi_{\alpha}(\bsl{k}+\bsl{G})=\Phi_{\alpha}(\bsl{k})\ \forall\ \text{reciprocal lattice vector }\bsl{G}\ ,
}
we can focus on the reciprocal unit cell and treat it as a torus for the study of winding numbers of pairing zeros.
Before defining the winding numbers of the pairing zeros, let us first define a winding number for $\Delta_b(\bsl{k})$ associated with a closed connected region $D_b$ that is a subset of the reciprocal unit cell (with boundary), when $|\Delta_b(\bsl{k})|$ that is not everywhere zero in $\dsR^2$.
Specifically, we require the boundary $\partial D_b$ does not touch any zeros of $\Delta_b$, and then the winding number reads
\eq{
\label{eq:W_b}
\W_{b}(D_b)=\frac{1}{2\pi}\int_{\partial D_{b}} d\bsl{k}\cdot \bsl{v}_b(\bsl{k}) + \frac{1}{2\pi}\int_{D_b} d^2 k \left[\Phi_{+}(\bsl{k})-(-1)^b\Phi_{-}(-\bsl{k})\right]\ .
}
We emphasize that $\partial D_b$ is chosen as if the reciprocal unit cell is a torus as schematically shown in \figref{fig:EOCP_D_Regions_App}(a).
$\W_{b}(D_b)$ is gauge invariant.
More importantly, $\W_{b}(D_b)$ is an integer, as discussed in the following. 
First, we consider the case where $D_b$ is the same as the reciprocal unit cell, which makes $\partial D_b=\emptyset$, resulting in $\W_{b}(D_b)=\N_+-(-1)^b\N_-\in \dsZ$.
Then, we turn to the case where $D_b$ is a proper subset of the reciprocal unit cell.
In this case, we can choose the patch $I$ in \figref{fig:patch} to fully cover $D_b$, and then choose a real oriented gauge $\ket{u_{\pm,\bsl{k}}^{RO}}$ that is smooth in $I$ and has $e_{2,\pm}>0$.
For this real oriented gauge $\ket{u_{\pm,\bsl{k}}^{RO}}$, we have
\eq{
\hat{P}_{b}(\bsl{k})=\ket{u_{+,\bsl{k}}^{RO}}\mat{1 & \\ & (-1)^b}e^{\ii \theta_{b}(\bsl{k})\tau_y}\bra{u_{-,\bsl{k}}^{RO}}\ ,
}
resulting in 
\eq{
\bsl{v}_b=-(-1)^b \nabla_{\bsl{k}}\theta_b(\bsl{k})- (-1)^b \bra{u_{-,-\bsl{k},1}^{RO}}\nabla_{-\bsl{k}}\ket{u_{-,-\bsl{k},2}^{RO}}- \bra{u_{+,\bsl{k},1}^{RO}}\nabla_{\bsl{k}}\ket{u_{+,\bsl{k},2}^{RO}}\ .
}
Combined with 
\eq{
\Phi_{\pm,\bsl{k}}= \braket{\partial_{k_x} u_{\pm,\bsl{k},1}^{RO}}{\partial_{k_y} u_{\pm,\bsl{k},2}^{RO}}- (\partial_{k_x}\leftrightarrow \partial_{k_y})\ ,
}
we have 
\eq{
\W_{b}(D_b)=  -(-1)^b \frac{1}{2\pi}\int_{\partial D_b} d\bsl{k}\cdot \nabla_{\bsl{k}}\theta_b(\bsl{k})\in \dsZ\text{ for this real oriented gauge.}
}
Since $\W_{b}(D_b)$ is gauge-invariant, $\W_{b}(D_b)\in\dsZ$ is an integer for all gauges.

Now we are ready to define the winding number for the possible zeros of $|\Delta_b(\bsl{k})|$, given that $|\Delta_b(\bsl{k})|$ is not everywhere-vanishing in $\dsR^2$.
For each isolated connected region $i$ of zeros of $|\Delta_b(\bsl{k})|$ in the reciprocal unit cell, we choose a closed connected subset of the reciprocal unit cell, $D_{b,i}$, such that (i) $i$ is in the interior of $D_{b,i}$ and (ii) $D_{b,i}$ does not contain any other zeros of $|\Delta_b(\bsl{k})|$.
Then, since $D_{b,i}$ satisfies the requirements for $D_b$ in \eqnref{eq:W_b}, the winding number of the zero $i$ of $|\Delta_b(\bsl{k})|$ is defined as
\eq{
\label{eq:W_b_i}
\W_{b,i}=\W_{b}(D_{b,i})\ ,
}
which is an integer as discussed above.
Here we emphasize that the above discussion holds for $i$ being be a point zero (0D), a line of zeros (1D), or an area of zeros (2D), as schematically shown in \figref{fig:EOCP_D_Regions_App}(b).

At the end of this part, we show the main result for the Euler obstructed Cooper pairing.
If $|\Delta_{b}(\bsl{k})|$ is not everywhere zero in $\dsR^2$, we have
\eq{
\label{eq:W_b_i_N_app}
\sum_{i}\W_{b,i} = \N_+ - (-1)^b \N_-\ ,
}
where $i$ ranges over all isolated points, lines or areas of zeros of $\Delta_{b}(\bsl{k})$, and $\sum_{i}\W_{b,i}=0$ if $\Delta_{b}(\bsl{k})$ has no zeros.
To show this, we first consider the case where $\Delta_{b}(\bsl{k})$ has zeros.
In this case, we can continuously deform $D_{b,i}$ such that $\cup_i D_{b,i}$ covers the reciprocal unit cell and $D_{b,i}$'s at most intersect at boundaries.
After the deformation, we have $\sum_{i}\int_{D_{b,i}} d\bsl{k}\cdot \bsl{v}_{b}(\bsl{k}) = 0$, meaning that \eqnref{eq:W_b_i_N_app} holds after the deformation.
To show \eqnref{eq:W_b_i_N_app} holds for the generic case before the deformation, note that we can keep all $\partial D_{b,i}$'s not touching any zeros of $\Delta_b(\bsl{k})$ in the process, since all zeros of $\Delta_b(\bsl{k})$ are included inside $D_{b,i}$'s, meaning that $\int_{\partial D_{b,i}}d\bsl{k}\cdot \bsl{v}(\bsl{k})$ is continuously changing during the deformation.
Then, combined with the fact that $\Phi_\pm(\bsl{k})$ is smooth everywhere in $\dsR^2$, the deformation cannot cause any discontinuous change of $\W_{b,i}$, and thus cannot change $\W_{b,i}$.
Thus, \eqnref{eq:W_b_i_N_app} holds for the generic case before the deformation.

We now consider the case where $\Delta_{b}(\bsl{k})$ has no zeros.
In this case, we know $\sum_{i}\W_{b,i}=0$, and thus we only need to show $\N_+ - (-1)^b \N_-=0$.
Since $\Delta_{b}(\bsl{k})$ has no zeros, $\bsl{v}_b(\bsl{k})$ is smooth everywhere in $\dsR^2$.
Then, $\W_{b}(D_b)$ is the same for any closed connected subset $D_b$ of the reciprocal unit cell, since continuously deforming $D_b$ cannot cause any singular behavior of $\bsl{v}_b(\bsl{k})$ on $\partial D_b$ and thus  cannot cause any discontinuous change of  $\W_{b}(D_b)$.
As a result, we have $\W_{b}(D_b)=0$ for any closed subset $D_b$ of the reciprocal unit cell, since we can make $D_b$ infinitesimal and have $|\W_{b}(D_b)|\ll 1$.
Then, choosing $D_b$ to be the same as the reciprocal unit cell gives  $\N_+ - (-1)^b \N_-=0$.

In sum, \eqnref{eq:W_b_i_N_app} holds as long as $|\Delta_{b}(\bsl{k})|$ is not everywhere-vanishing in $\dsR^2$, no matter whether $\Delta_{b}(\bsl{k})$ has zeros or not.
In particular, we know if $\N_+ - (-1)^b \N_-\neq 0$, $\Delta_{b}(\bsl{k})$ must have zeros, since no zeros mean $\N_+ - (-1)^b \N_- = 0$.
In other words, when $\N_+ - (-1)^b \N_-\neq 0$ occurs, the nonzero normal-state Euler numbers prevent the corresponding pairing channel to have a nonvanishing pairing gap function $\Delta_b$.
Therefore, when $\N_+ - (-1)^b \N_-\neq 0$ occurs, $\Delta_b$ is called an Euler obstructed pairing channel.

\subsubsection{Comments on the Gauge Invariance}

In the last part of this section, we would like to comment on the gauge invariance of the formalism proposed above.
The key quantities $\N_\pm$ (\eqnref{eq:N_pm}), $\bsl{Q}_{\alpha}(\bsl{k})$ (\eqnref{eq:Q_Phi}), $\Phi_{\alpha}(\bsl{k})$ (\eqnref{eq:Q_Phi}), $P_b(\bsl{k})$ (\eqnref{eq:P_b}), $\bsl{v}_b(\bsl{k})$ (\eqnref{eq:v_b}) and $\W_{b,i}$ (\eqnref{eq:W_b_i}) in the above formalism are manifestly gauge-invariant, in the sense that (i) they take the same values for all gauges, and (ii) no special gauges are chosen in their expressions.
However, the sense of gauge invariance here is weaker than the gauge invairance of the Berry curvature, as discussed in the following.

To derive the Berry curvature at any momentum $\bsl{k}$, we only need to know the states near $\bsl{k}$, and the resultant Berry curvature is gauge invariant.
However, to get $\bsl{Q}_{\alpha}(\bsl{k})$, $\Phi_{\alpha}(\bsl{k})$, $P_b(\bsl{k})$ or $\bsl{v}_b(\bsl{k})$, knowing the states near $\bsl{k}$ is not enough, since there is a Wilson line matrix embedded in their definitions which connect $\bsl{k}$ to a base point $\bsl{k}_0$.
So $\bsl{Q}_{\alpha}(\bsl{k})$, $\Phi_{\alpha}(\bsl{k})$, $P_b(\bsl{k})$ or $\bsl{v}_b(\bsl{k})$ are not locally defined at $\bsl{k}$.
The Wilson line matrix is crucial for the gauge invariance of those quantities, and what it does is to align the gauge of $\ket{u_{\alpha,\bsl{k}}}(-\ii\tau_y)\bra{u_{\alpha,\bsl{k}}^{C_{2z}\TR}}$ at $\bsl{k}$ to that at $\bsl{k}_0$.
Therefore, although we have not chosen any special gauges in the expressions of $\N_\pm$, $\bsl{Q}_{\alpha}(\bsl{k})$, $\Phi_{\alpha}(\bsl{k})$, $P_b(\bsl{k})$, $\bsl{v}_b(\bsl{k})$ and $\W_{b,i}$, the gauges of $\ket{u_{\alpha,\bsl{k}}}(-\ii\tau_y)\bra{u_{\alpha,\bsl{k}}^{C_{2z}\TR}}$ have been implicitly aligned by the Wilson line embedded in their expression.
On the other hand, there are no gauge alignments in the expression of the Berry curvature, and thus the sense of gauge invariance here is weaker than the gauge invariance of the Berry curvature.

Nevertheless, no special gauges are chosen in the expressions of $\N_\pm$, $\bsl{Q}_{\alpha}(\bsl{k})$, $\Phi_{\alpha}(\bsl{k})$, $P_b(\bsl{k})$, $\bsl{v}_b(\bsl{k})$ and $\W_{b,i}$. 
Therefore, the numerical evaluation of them is very convenient, since we do not need to care about the random $\U(2)$ rotations or $\U(1)$ phases of the eigenvectors in numerical calculations. 

\subsection{Chern Gauge}

In the above, we have introduced the gauge-invariant formalism for the Euler obstructed Cooper pairing, which is convenient for numerical calculations. 
In some cases, choosing certain special gauges may help understand the physics or help prove a gauge-invariant conclusion.
In this part, we will discuss the Euler obstructed Cooper pairing in a class of special gauges, called Chern gauges, under \asmref{asm:normal_state_C2T}-\ref{asm:pairing}.
The Cooper pairings between Chern states (or in the Chern gauge) were discussed in \refcite{Li2018WSMObstructedPairing,Murakami2003BerryPhaseMSC,Zaletel2020Sep30SoftModesTBG,Khalaf2021SkyrSCMATBG}, and we will see the following discussion agrees with those previous discussions.

The Chern gauges for the bands with nonzero Euler numbers have been carefully discussed in \refcite{Slager2020EulerOptical,Xie2020TopologyBoundSCTBG,Bouhon2020WeylNonabelian}.
Here we briefly review it.
The Chern gauges for the set of two nearly flat bands in valley $\alpha$ are derived from the real oriented gauges.
Specifically, given a real oriented gauge $\ket{u^{RO}_{\alpha,\bsl{k}}}$ with real curvature $f_{\alpha}(\bsl{k})$ and Euler class $e_{2,\alpha}$, we can get a complex gauge as
\eq{
\label{eq:Chern_Gauge}
\ket{u_{\alpha,\bsl{k}}^{Ch}}=\ket{u_{\alpha,\bsl{k}}^{RO}}\frac{1}{\sqrt{2}}\mat{1 & 1 \\ \ii & -\ii}\ .
}
$P_{Ch,\alpha,a}(\bsl{k})=\ket{u_{\alpha,\bsl{k},a}^{Ch}}\bra{u_{\alpha,\bsl{k},a}^{Ch}}$ is smooth everywhere in $\dsR^2$.
It is because for any $\bsl{k}'\in\dsR^2$, $\ket{u^{RO}_{\alpha,\bsl{k}}}$ can be made smooth in a neighborhood of $\bsl{k}'$ by performing $\ket{u_{\alpha,\bsl{k}}^{RO}}\rightarrow \ket{u_{\alpha,\bsl{k}}^{RO}} e^{\ii \tau_y \phi_{\bsl{k}}}$ with $\phi_{\bsl{k}}\in \dsR$ around $\bsl{k}'$, which means we can perform $\ket{u_{\alpha,\bsl{k}}^{Ch}}\rightarrow \ket{u_{\alpha,\bsl{k}}^{Ch}} e^{\ii \tau_z \phi_{\bsl{k}}}$ around $\bsl{k}'$ to make $\ket{u_{\alpha,\bsl{k}}^{Ch}}$ smooth in a neighborhood of $\bsl{k}'$.
Since $P_{Ch,\alpha,a}(\bsl{k})$ is invariant under $\ket{u_{\alpha,\bsl{k}}^{Ch}}\rightarrow \ket{u_{\alpha,\bsl{k}}^{Ch}} e^{\ii \tau_z \phi_{\bsl{k}}}$, $P_{Ch,\alpha,a}(\bsl{k})$ must be smooth everywhere in $\dsR^2$.
Then, the globally smooth $P_{Ch,\alpha,a}(\bsl{k})$ means $\ket{u_{\alpha,\bsl{k},a}^{Ch}}$ is the basis of a rank-1 vector bunle, which has well-defined Berry curvature 
\eq{
F_{\alpha,a}(\bsl{k})=(-\ii)\Tr[P_{Ch,\alpha,a}(\bsl{k})\partial_{k_x} P_{Ch,\alpha,a}(\bsl{k})\partial_{k_y} P_{Ch,\alpha,a}(\bsl{k})  ] - (\partial_{k_x}\leftrightarrow \partial_{k_y})
}
and well-defined Chern number.
\eq{
C_{\alpha,a}=\frac{1}{2\pi}\int_{\text{{\BZ}}}d^2k\ F_{\alpha,a}(\bsl{k})\ .
}
Therefore, $\ket{u_{\alpha,\bsl{k}}^{Ch}}$ obtained from \eqnref{eq:Chern_Gauge} is called a Chern gauge.
In particular, owing to \eqnref{eq:Chern_Gauge}, the Berry curvatures and Chern numbers for the Chern gauges are related to the real curvature and the Euler class of the real oriented gauge as
\eqa{
& F_{\alpha,1}(\bsl{k})=-F_{\alpha,2}(\bsl{k})=f_{\alpha}(\bsl{k}) \\
& C_{\alpha,1}=-C_{\alpha,2}=e_{2, \alpha}\ ,
}
where the opposite Berry curvatures and Chern numbers for the two components of the Chern gauge are guaranteed by the $C_{2z}\TR$ symmetry, as 
\eq{
C_{2z}\TR\ket{u_{\pm,\bsl{k}}^{Ch}}=\ket{u_{\pm,\bsl{k}}^{Ch}}\tau_x\ .
}

Since we can always choose the real oriented gauge such that $e_{2,\pm}=\N_{\pm}$, we can always choose a Chern gauge $\ket{u^{Ch}_{\alpha,\bsl{k}}}$ such that 
\eqa{
\label{eq:C_N}
C_{\alpha,1}=-C_{\alpha,2}=\N_\alpha\ .
}
Then, we define an assumption for the Chern gauge as
\begin{assumption}
\label{asm:Chern_gauge}
Given any Chern gauge, we choose it to satisfy \eqnref{eq:C_N}.
\end{assumption}
With \asmref{asm:Chern_gauge}, we know $\eta_{\pm,\bsl{k}_0}^{Ch}(\bsl{k})=\ii$, and 
\eqa{
& Q_{\pm}(\bsl{k})=-\frac{\ii}{\sqrt{2}}\ket{u^{Ch}_{\pm,\bsl{k}}}\tau_z\bra{u^{Ch}_{\pm,\bsl{k}}} \\
& \Phi_{\pm}(\bsl{k})=F_{\pm,1}(\bsl{k})\ .
}
Combined with \eqnref{eq:P_b}, we have 
\eq{
\label{eq:Delta_split_Ch_app}
\Delta_{\perp}^{Ch}(\bsl{k})=\mat{ & d_\perp(\bsl{k})  \\ d_\perp^*(\bsl{k}) & }\ ,\ \Delta_{\shpa}^{Ch}(\bsl{k})=\mat{ d_\shpa^*(\bsl{k}) &  \\ & d_\shpa(\bsl{k}) }
}
with $d_b(\bsl{k})=|\Delta_b(\bsl{k})|e^{\ii \theta_b(\bsl{k})}$.
Furthermore, combined with \eqnref{eq:v_b} and \eqnref{eq:W_b_i}, we have
\eqa{
& \bsl{v}_{b}(\bsl{k})= -(-1)^b \nabla_{\bsl{k}}\theta_b(\bsl{k}) - \bsl{A}_{+,1}(\bsl{k})-(-1)^b\bsl{A}_{-,1}(-\bsl{k}) \\
& \W_{b,i}=- \frac{(-1)^b}{2\pi}\int_{\partial D_{b,i}}d\bsl{k}\cdot \nabla_{\bsl{k}}\theta_b(\bsl{k})\ ,
}
where $\bsl{A}_{\pm,a}(\bsl{k})=(-\ii)\braket{u_{\pm,\bsl{k},a}^{Ch}}{\nabla_{\bsl{k}}u_{\pm,\bsl{k},a}^{Ch}}$, and $\partial D_{b,i}$ encloses only the zero $i$ of $|\Delta_b(\bsl{k})|$ as defined in \eqnref{eq:W_b_i}.
Eventually, combining \eqnref{eq:C_N} and \eqnref{eq:W_b_i_N_app}, we have 
\eq{
\sum_{i} \W_{b,i}=-C_{+,2}-(-1)^b C_{-,1}=-C_{+,2}+(-1)^b C_{-,2}\ ,
}
which has the same form as the monopole Cooper pairing in \refcite{Li2018WSMObstructedPairing} if $\N_+-(-1)^b\N_-\neq 0$, and also agrees with the discussion in Chern gauge in \cite{Zaletel2020Sep30SoftModesTBG,Khalaf2021SkyrSCMATBG}.
Therefore, when choosing the Chern gauge for the normal-state basis, the Euler obstructed Cooper pairing can be treated as a $C_{2z}\TR$-protected double version of the monopole Cooper pairing.
However, the whole discussion in the Chern gauge, including the connection between the Euler obstructed Cooper pairing and the monopole Cooper pairing, is just a special case for the general formalism discussed in the last part.

At last, we would like to describe how we can numerically choose a Chern gauge that satisfies \asmref{asm:Chern_gauge}.
Numerically, it is straight forward to get a random gauge for $\ket{u_{\alpha}(\bsl{k})}$, \eg, by diagonalizing the Hamiltonian.
\begin{itemize}
    \item[1.] From $\ket{u_{\alpha}(\bsl{k})}$, we can get a real gauge $\ket{\widetilde{u}_{\alpha}(\bsl{k})}$ from \eqnref{eq:real_gauge}.
    \item[2.] Then, we pick a base point $\bsl{k}_0$ in {\BZ}, evaluate $\det(W_\alpha(\bsl{k}_0,\bsl{k}))$ for the real gauge $\ket{\widetilde{u}_{\alpha}(\bsl{k})}$ for every $\bsl{k}$; if $\det(W_\alpha(\bsl{k}_0,\bsl{k}))=-1$, perform $\ket{\widetilde{u}_{\alpha}(\bsl{k})}\rightarrow \ket{\widetilde{u}_{\alpha}(\bsl{k})}\tau_z$.
    After these operations, we have $\det(W_{\alpha}(\bsl{k},\bsl{k}'))=1$ for $\ket{\widetilde{u}_{\alpha}(\bsl{k})}$ and for any $\bsl{k}$, $\bsl{k}'$, and thus $\ket{\widetilde{u}_{\alpha}(\bsl{k})}$ becomes a real oriented gauge $\ket{u_{\alpha}^{RO}(\bsl{k})}$.
    \item[3.] Then, with \eqnref{eq:Chern_Gauge}, we can get the Chern gauge $\ket{u_{\alpha,a}^{Ch}(\bsl{k})}$ from $\ket{u_{\alpha,a}^{RO}(\bsl{k})}$.
    If $C_{\alpha,1}<0$, $\ket{u_{\alpha}^{Ch}(\bsl{k})}\rightarrow \ket{u_{\alpha}^{Ch}(\bsl{k})}\tau_x$, resulting in a Chern gauge that satisfies \asmref{asm:Chern_gauge}.
\end{itemize}

In the above process, we use the fact that if a real gauge $\ket{\widetilde{u}_{\alpha}(\bsl{k})}$ has  $\det(W_{\alpha}(\bsl{k}\xrightarrow{\gamma}\bsl{k}'))=1$ for any $\bsl{k}$, $\bsl{k}'$ and $\gamma$, then $\ket{\widetilde{u}_{\alpha}(\bsl{k})}$  is a real oriented gauge.
The reasoning is the following.
We can always have a patchwise-smooth real gauge $\ket{\widetilde{u}_{\alpha}^{A}(\bsl{k})}$ such that $\ket{\widetilde{u}_{\alpha}(\bsl{k})}=\ket{\widetilde{u}_{\alpha}^{A_{\bsl{k}}}(\bsl{k})}$.
Then, for any $\bsl{k}\in A\cap A'$, we have $\bsl{k}_{initial}\in A$ and $\bsl{k}_{final}\in A'$, and we can choose a path $\gamma$ from $\bsl{k}_{initial}$ through $\bsl{k}$ to $\bsl{k}_{final}$.
Then, $\det(W_{\alpha}(\bsl{k}_{initial}\xrightarrow{\gamma}\bsl{k}_{final}))=1$ for $\ket{\widetilde{u}_{\alpha}(\bsl{k})}$ gives
\eqa{
& 1=\lim_{L\rightarrow \infty}\det(\bra{\widetilde{u}_{\alpha}^{A}(\bsl{k}_{initial})}P_{\alpha}(\bsl{k}_1)...P_{\alpha}(\bsl{k}_L)\ket{\widetilde{u}_{\alpha}^{A}(\bsl{k})}) 
\det(\braket{\widetilde{u}_{\alpha}^{A}(\bsl{k})}{\widetilde{u}_{\alpha}^{A'}(\bsl{k})})
\det(\bra{\widetilde{u}_{\alpha}^{A'}(\bsl{k})}P_{\alpha}(\bsl{k}_{L+1})...P_{\alpha}(\bsl{k}_{2L})\ket{\widetilde{u}_{\alpha}^{A'}(\bsl{k}_{final})}) \\
& =\det(\braket{\widetilde{u}_{\alpha}^{A}(\bsl{k})}{\widetilde{u}_{\alpha}^{A'}(\bsl{k})})\ ,
}
meaning that $\ket{\widetilde{u}_{\alpha}^A(\bsl{k})}$ is a patchwise-smooth real oriented gauge and thus $\ket{\widetilde{u}_{\alpha}(\bsl{k})}$ is a real oriented gauge.

The above method of finding Chern gauges is equivalent to the method presented in \refcite{Xie2021TBGVI}.

\section{Superfluid Weight in 2D Systems with $C_{2z}\TR$ Symmetry}

In this section, we will discuss the superfluid weight.
We first review the general framework, and then focus on the case where \asmref{asm:normal_state_C2T}-\ref{asm:pairing} are satisfied.
In this section, we will adopt the Lorentz-Heaviside unit system with $\hbar=c=1$ unless specified otherwise.

\subsection{Review of General Formalism for Mean-field Superfluid Weight}

Let us first review the general formalism of the superfuild weight within the mean-field approximation following \refcite{Xie2020TopologyBoundSCTBG}.
Consider a general mean-field Hamiltonian $\widetilde{H}_{MF}$ for a generic superconductor.
The trick to derive superfluid wieght is to put the superconductors in a constant vector field $\bsl{A}$, resulting in the mean-field Hamiltonian 
\eq{
\widetilde{H}_{MF}(\bsl{A})= \widetilde{H}(\bsl{A}) -\mu \hat{N}_\psi +\widetilde{H}_{pairing}\ ,
}
where $\bsl{A}=e\bsl{A}^{phy}$ with $\bsl{A}^{phy}$ the physical $\U(1)$ vector field,
\eq{
\widetilde{H}(\bsl{A}) = \sum_{\bsl{k}\in \text{{\BZ}}}\psi^\dagger_{\bsl{k}} \widetilde{h}(\bsl{k}+\bsl{A}) \psi_{\bsl{k}}
}
represents the normal state,
\eq{
\widetilde{H}_{pairing} = \sum_{\bsl{k}\in \text{{\BZ}}}\psi^\dagger_{\bsl{k}} \widetilde{\Delta}(\bsl{k}) (\psi_{-\bsl{k}}^\dagger)^T + h.c.
}
represents the mean-field pairing operator, $\hat{N}_\psi=\sum_{\bsl{k}\in\text{{\BZ}}}\psi^\dagger_{\bsl{k}}\psi_{\bsl{k}}$, and $\psi^\dagger_{\bsl{k}}=(...,\psi^\dagger_{\bsl{k},l},...)$ are the creation operators for the atomic Bloch basis.
We can also interpret $\bsl{A}$ as a momentum shift caused by an in-homogeneous deformation of the pairing order parameter, since we treat $\bsl{A}$ to be independent of temporal and spatial coordinates.

Very often, we do not address the problem in the atomic Bloch basis, and instead, we need to project the system into a set of bands with creation operators
\eq{
c^\dagger_{\bsl{k},\bsl{A}}=\psi^\dagger_{\bsl{k}}V(\bsl{k}+\bsl{A})\ ,
}
where $V(\bsl{k})$ has orthonormal columns.
Then, we care about the projected mean-field Hamiltonian
\eq{
H_{MF}(\bsl{A})= H(\bsl{A}) -\mu \hat{N}(\bsl{A}) + H_{pairing}(\bsl{A})\ ,
}
where
$\hat{N}(\bsl{A})=\sum_{\bsl{k}\in {\BZ}} c^\dagger_{\bsl{k},\bsl{A}} c_{\bsl{k},\bsl{A}}$, 
\eq{
 H(\bsl{A})=\sum_{\bsl{k}\in \text{{\BZ}}}c^\dagger_{\bsl{k},\bsl{A}}h(\bsl{k}+\bsl{A}) c_{\bsl{k},\bsl{A}}
}
with $h(\bsl{k})=V^\dagger(\bsl{k}) \widetilde{h}(\bsl{k}) V(\bsl{k})$, and 
\eq{
 H_{pairing}(\bsl{A})=\sum_{\bsl{k}\in \text{{\BZ}}}c^\dagger_{\bsl{k},\bsl{A}}D(\bsl{k},\bsl{A}) (c^\dagger_{-\bsl{k},\bsl{A}})^T + h.c.
}
with 
\eq{
D(\bsl{k},\bsl{A})= V^\dagger(\bsl{k}+\bsl{A}) \widetilde{\Delta}(\bsl{k}) V^*(-\bsl{k}+\bsl{A})\ .
}
Using $E_{\bsl{k},m}(\bsl{A})$ to label the eigenvalues of 
\eq{
\mat{
h(\bsl{k}+\bsl{A}) -\mu & D(\bsl{k},\bsl{A}) \\
D^\dagger(\bsl{k},\bsl{A}) & -[h(-\bsl{k}+\bsl{A}) -\mu]^T
}\ ,
}
we arrive at the mean-field Free energy of the system as
\eq{
\label{eq:Free_en_gen}
\Omega(\bsl{A})= \sum_{\bsl{k}\in {\BZ}} \left\{ \frac{1}{2}\Tr[h(\bsl{k}+\bsl{A})-\mu] -\frac{1}{\beta}\sum_{m}^{E_{\bsl{k},m}(\bsl{A})>0}\log\left[ 2\cosh(\frac{\beta E_{\bsl{k},m}(\bsl{A})}{2}) \right] \right\}\ ,
}
where $\beta=1/(k_B T)$ with $T$ the temperature and $k_B$ the Boltzmann constant.
At zero temperature, the free energy equals to the ground state energy of $H_{MF}(\bsl{A})$, which reads
\eq{
\Omega_0(\bsl{A})= \lim_{T\rightarrow 0} \Omega(\bsl{A}) =-\frac{1}{4}\sum_{\bsl{k}\in\text{{\BZ}},m} |E_{\bsl{k},m}(\bsl{A})| + \frac{1}{2}\sum_{\bsl{k}\in\text{{\BZ}}}\Tr[h(\bsl{k}+ \bsl{A})-\mu]\ .
}
The superfluid weight then reads
\eq{
\label{eq:SFweight_gen}
[D_{SF}(T)]_{ij}= \left.\frac{e^2}{\mathcal{V}} \frac{\partial^2\Omega(\bsl{A})}{\partial A_i \partial A_j }\right|_{\bsl{A}\rightarrow 0}
}
with $\mathcal{V}$ the volume of the system.
Detailed derivations can be found in the supplementary materials of \refcite{Xie2020TopologyBoundSCTBG}.

\subsection{Superfluid Weight in 2D systems that satisfy \asmref{asm:normal_state_C2T}-\ref{asm:pairing}}

Now we derive the expression for the superfluid weight of the 2D systems that satisfy \asmref{asm:normal_state_C2T}-\ref{asm:pairing}, generalizing the derivation for the uniform pairing in \refcite{Xie2020TopologyBoundSCTBG}.

Owing to \asmref{asm:normal_state_C2T}-\ref{asm:pairing}, we have $\psi^\dagger_{\bsl{k}}=(\psi^\dagger_{+,\bsl{k}}, \psi^\dagger_{-,\bsl{k}})$, 
\eq{
\widetilde{h}(\bsl{k}) = 
\mat{\widetilde{h}_{+}(\bsl{k})\otimes s_0 & \\ & \widetilde{h}_{-}(\bsl{k})\otimes s_0 }
}
$\widetilde{h}_{\pm}(\bsl{k})$ defined in \eqnref{eq:htilde_gen},
\eq{
V(\bsl{k})=\mat{V_{+}(\bsl{k})\otimes s_0 & \\ & V_{-}(\bsl{k})\otimes s_0 }
}
with $V_\pm$ defined above \eqnref{eq:psi_c_relation},  and
\eq{
\widetilde{\Delta}(\bsl{k})
=
V(\bsl{k}) 
\mat{ & \Delta(\bsl{k})\otimes \Pi \\ 
-\Delta^T(-\bsl{k})\otimes \Pi^T & }
V^T(-\bsl{k})
}
with $\Delta(\bsl{k})\otimes \Pi$ defined in \asmref{asm:pairing}.
Then, we have
\eq{
h(\bsl{k})=
\mat{h_{+}(\bsl{k})\otimes s_0 & \\ & h_{-}(\bsl{k})\otimes s_0 }
}
with $h_{\pm}(\bsl{k})=V^\dagger_{\pm}(\bsl{k}) \widetilde{h}_{\pm}(\bsl{k})V_{\pm}(\bsl{k})$,
and 
\eq{
D(\bsl{k},\bsl{A}) = 
\mat{
 & D_{+,\bsl{k}}(\bsl{A}) \otimes \Pi \\
 -D_{+,-\bsl{k}}^T(\bsl{A}) \otimes \Pi^T & 
}
}
with 
\eq{
D_{+,\bsl{k}}(\bsl{A})=V^\dagger_+(\bsl{k}+\bsl{A}) V_+(\bsl{k}) \Delta(\bsl{k}) V_-^T(-\bsl{k}) V^*_-(-\bsl{k}+\bsl{A})\ .
}

Finally, by using \eqnref{eq:Free_en_gen} and the fact that $\Pi^\dagger \Pi = s_0$, we arrive at 
\eq{
\label{eq:free_energy_TBG}
\Omega(\bsl{A}) = \sum_{\bsl{k}\in {\BZ}} \left\{ \sum_\alpha \Tr[h_{\alpha}(\bsl{k}+\bsl{A})-\mu]  - \frac{2}{\beta} \sum_{n}^{E_{+,\bsl{k},n}(\bsl{A})>0}\log\left[ 2\cosh(\frac{\beta E_{+,\bsl{k},n}(\bsl{A})}{2}) \right] \right\}\ ,
}
where $E_{+,\bsl{k},n}(\bsl{A})$ eigenvalues of 
\eq{
\label{eq:Hcal_kA}
\H(\bsl{k},\bsl{A}) = 
\mat{ 
h_+(\bsl{k}+\bsl{A})-\mu & D_{+,\bsl{k}}(\bsl{A}) \\ 
D^\dagger_{+,\bsl{k}}(\bsl{A}) & - h^T_{-}(-\bsl{k}+\bsl{A})+\mu
}\ .
}
The superfluid weight can be derived by substituting \eqnref{eq:free_energy_TBG} into \eqnref{eq:SFweight_gen},

In 2D, the superconductivity transition is typically Berezinskii–Kosterlitz–Thouless (BKT) transition~\cite{Xie2020TopologyBoundSCTBG}, and the corresponding BKT transition temperature can be estimated as
\eq{
\label{eq:TBKT_DSF}
k_B T_{BKT} \approx \frac{\pi }{8 e^2} \frac{\Tr[D_{SF}(T_{BKT})]}{2}\ .
}

\subsection{Bounded zero-temperature superfluid weight in 2D systems that satisfy \asmref{asm:normal_state_C2T}-\ref{asm:pairing}}

In the the remaining discussion, we will discuss the zero-temperature sueprfluid weight for superconductors that satisfies 
\asmref{asm:normal_state_C2T}-\ref{asm:pairing}.
Besides \asmref{asm:normal_state_C2T}-\ref{asm:pairing}, we further impose the following extra assumption
\begin{assumption}
\label{asm:ForSW}
The normal-state bands described by \eqnref{eq:h_gen} are exactly flat with zero energies, $\mu\neq 0$, and the normal state has $C_{2z}$ symmetry with $C_{2z} c^\dagger_{+,\bsl{k}} C_{2z}^{-1} = c^\dagger_{-,-\bsl{k}} U_{C_{2z}}(\bsl{k})\otimes s_0$.
\end{assumption}

Under \asmref{asm:ForSW}, we can choose a Chern gauge that satisfies \asmref{asm:Chern_gauge} and $C_{2z} c^\dagger_{+,\bsl{k}} C_{2z}^{-1} = c^\dagger_{-,-\bsl{k}}$.
Let us first use this gauge to derive the bound for the superfluid weight. 
With this gauge, we have 
\eq{
D_{+,\bsl{k}}(\bsl{A}) = V_{+}^{\dagger}(\bsl{k}+\bsl{A}) V_{+}(\bsl{k}) \Delta(\bsl{k}) V_+^T(\bsl{k}) V_+^*(\bsl{k}-\bsl{A})\ ,
}
resulting in
\eq{
D_{+,\bsl{k}}^*(\bsl{A}) = \tau_x D_{+,\bsl{k}}(\bsl{A}) \tau_x \Rightarrow D_{+,\bsl{k}}(\bsl{A}) = 
 \mat{ 
 d_{\shpa}^*(\bsl{k},\bsl{A}) & d_{\perp}(\bsl{k},\bsl{A}) \\
 d_{\perp}^*(\bsl{k},\bsl{A}) & d_{\shpa}(\bsl{k},\bsl{A}) 
 }\ .
}
Combined with $h_{\alpha}(\bsl{k})=0$ required by \asmref{asm:ForSW}, we have $E_{+,\bsl{k},n}(\bsl{A})$ taking values in $\{ \pm \sqrt{\mu^2+\lambda_{+,\bsl{k},a}(\bsl{A})} | a=1,2 \}$, resulting in the zero-temperature superfluid weight as 
\eq{
\label{eq:D_SF_efb_sim}
[D_{SF}]_{ij} = -\frac{2 e^2}{\mathcal{V}}\sum_{\bsl{k}\in\text{{\BZ}}} \sum_{a} \left.\partial_{A_i} \partial_{A_j}\sqrt{\mu^2 + \lambda_{+,\bsl{k},a}(\bsl{A})} \right|_{\bsl{A}\rightarrow 0}\ ,
}
where 
\eq{
\lambda_{+,\bsl{k},a}(\bsl{A})= (|d_{\perp}(\bsl{k},\bsl{A})| + (-1)^{a-1} |d_{\shpa}(\bsl{k},\bsl{A})|)^2\ .
}
Here both $d_{\perp}(\bsl{k},\bsl{A})$ and $d_{\shpa}(\bsl{k},\bsl{A})$ are complex, and $d_{b}(\bsl{k},0)=d_{b}(\bsl{k})$ with $d_{b}(\bsl{k})$ in \eqnref{eq:Delta_split_Ch_app}.

In particular, $\sum_{a}\sqrt{\mu^2+\lambda_{+,\bsl{k},a}(\bsl{A})}$ is a smooth function of $(\bsl{k},\bsl{A})$, because
\eq{
\sum_{a}\sqrt{\mu^2+\lambda_{+,\bsl{k},a}(\bsl{A})} = \sqrt{2(\mu+ \delta_+(\bsl{k},\bsl{A}))+ 2 \sqrt{\mu^4 + 2 \mu^2 \delta_+(\bsl{k},\bsl{A}) + \delta_-^2(\bsl{k},\bsl{A})}}
}
with $\delta_\pm(\bsl{k},\bsl{A}) = |d_{\perp}(\bsl{k},\bsl{A})|^2 \pm |d_{\shpa}(\bsl{k},\bsl{A})|^2$, and $|d_{b}(\bsl{k},\bsl{A})|^2$ is a smooth function of  $(\bsl{k},\bsl{A})$ since it is independent of the choice of Chern gauge (that satisfies the requirements) and we can always choose the Chern gauge to make $d_{b}(\bsl{k},\bsl{A})$ in a neighborhood of any $(\bsl{k},\bsl{A})$.
Furthermore, from the above equation, $\Tr[D_{SF}]$ is solely determined by $\left.\partial_{A_i}^{1,2} \delta_{\pm}(\bsl{k},\bsl{A})\right|_{\bsl{A}\rightarrow 0}$.
Straightforward derivations give 
\eq{
\label{eq:dA_delta}
\left.\partial_{A_i} \delta_{\pm}(\bsl{k},\bsl{A})\right|_{\bsl{A}\rightarrow 0} =0\ ,
}
and
\eqa{
\label{eq:dA2_delta}
& \left.\partial_{A_i}^2 \delta_+(\bsl{k},\bsl{A})\right|_{\bsl{A}\rightarrow 0} =-\left[ 2  \delta_+(\bsl{k},0) g_{+,ii}(\bsl{k}) + 4 z_i(\bsl{k}) \right]\\
& \left.\partial_{A_i}^2 \delta_-(\bsl{k},\bsl{A})\right|_{\bsl{A}\rightarrow 0} =-\left[ 2  \delta_-(\bsl{k},0) g_{+,ii}(\bsl{k}) \right]\ ,
}
where 
\eqa{
\label{eq:g_z}
& g_{+,ij}(\bsl{k})=\frac{1}{2}\Tr[\partial_{k_i}P_{+,\bsl{k}}\partial_{k_j}P_{+,\bsl{k}}] \\
& z_i(\bsl{k})=\frac{1}{2}\Tr[P_{\perp}(\bsl{k})C_2]\Tr[P_{\shpa}(\bsl{k})C_2 \partial_{k_i}P_{+,\bsl{k}}\partial_{k_i}P_{+,\bsl{k}}]\ ,
}
$P_{+,\bsl{k}}=\ket{u_{+,\bsl{k}}}\bra{u_{+,\bsl{k}}}$, and $P_{b}(\bsl{k})$ is defined in \eqnref{eq:P_b}.
$g_{+}(\bsl{k})$ is called the Fubini-Study metric~\cite{Ma2013FSMetric} for the two normal-state bands in valley $+$, which is a positive semi-definite matrix.

By exploiting \eqnref{eq:dA_delta} and \eqnref{eq:dA2_delta}, we can simplify the expression of $\Tr[D_{SF}]$ to 
\eq{
\label{eq:D_SF_sim}
\Tr[D_{SF}] = 4 e^2 \int_{{\BZ}} \frac{d^2 k}{(2\pi)^2} \left[ f_+(\bsl{k}) \Tr[g_+(\bsl{k})] +\frac{f_-(\bsl{k})}{|\Delta_\shpa(\bsl{k})||\Delta_\perp(\bsl{k})|} z_{\bsl{k}}\right]\ ,
}
where $z_{\bsl{k}}=\sum_i z_i(\bsl{k})$ and
\eq{
f_{\pm}(\bsl{k})=\frac{1}{2}\left( \frac{(|\Delta_\perp(\bsl{k})|+|\Delta_\shpa(\bsl{k})|)^2}{\sqrt{\mu^2 + (|\Delta_\perp(\bsl{k})|+|\Delta_\shpa(\bsl{k})|)^2}}\pm \frac{(|\Delta_\perp(\bsl{k})|-|\Delta_\shpa(\bsl{k})|)^2}{\sqrt{\mu^2 + (|\Delta_\perp(\bsl{k})|-|\Delta_\shpa(\bsl{k})|)^2}}\right)\ .
}
Here $\frac{f_-(\bsl{k})}{|\Delta_\shpa(\bsl{k})||\Delta_\perp(\bsl{k})|}$ is continuous even at $\bsl{k}$ with $|\Delta_\shpa(\bsl{k})||\Delta_\perp(\bsl{k})|=0$ since 
\eqa{
 \frac{f_-(\bsl{k})}{|\Delta_\shpa(\bsl{k})||\Delta_\perp(\bsl{k})|} = & \frac{2}{\sqrt{\mu^2 + (|\Delta_\perp(\bsl{k})|+|\Delta_\shpa(\bsl{k})|)^2}+\sqrt{\mu^2 + (|\Delta_\perp(\bsl{k})|-|\Delta_\shpa(\bsl{k})|)^2}}\times \\
&\left( 1 + \frac{\mu^2 }{\sqrt{\mu^2 + (|\Delta_\perp(\bsl{k})|+|\Delta_\shpa(\bsl{k})|)^2} \sqrt{\mu^2 + (|\Delta_\perp(\bsl{k})|-|\Delta_\shpa(\bsl{k})|)^2}}\right)\ .
}
Although we derive \eqnref{eq:D_SF_sim} in the Chern gauge, \eqnref{eq:D_SF_sim} holds for all gauges since both sides are gauge-invariant.

Note that for any $\bsl{k}$ such that $|\Delta_{\shpa}(\bsl{k})| \neq 0$,
\eqa{
\label{eq:ineq_z_1}
\left|\Tr[P_{\shpa}(\bsl{k})C_{2z} \partial_{k_i}P_{+}(\bsl{k}) \partial_{k_i}P_{+}(\bsl{k})]/|\Delta_{\shpa}(\bsl{k})|\right|\leq g_{+,ii}(\bsl{k})\ .
}
Then, we have
\eq{
\left|\frac{f_-(\bsl{k})}{|\Delta_\shpa(\bsl{k})||\Delta_\perp(\bsl{k})|} z_{\bsl{k}}\right| \leq |f_{-}(\bsl{k})|\Tr[g_{+}(\bsl{k})]
}
for all $\bsl{k}$ since it trivially holds for $|\Delta_\shpa(\bsl{k})||\Delta_\perp(\bsl{k})|=0 \Rightarrow z_{\bsl{k}}=0$ and follows from \eqnref{eq:ineq_z_1} for $|\Delta_\shpa(\bsl{k})||\Delta_\perp(\bsl{k})|\neq 0$.
As a result, we have 
\eqa{
\label{eq:D_SF_bound}
\Tr[D_{SF}] & = 4 e^2 \int_{{\BZ}} \frac{d^2 k}{(2\pi)^2} \left[ f_+(\bsl{k}) \Tr[g_+(\bsl{k})] +\frac{f_-(\bsl{k})}{|\Delta_\shpa(\bsl{k})||\Delta_\perp(\bsl{k})|} z_{\bsl{k}}\right] \\
& \geq 4 e^2\int_{{\BZ}} \frac{d^2 k}{(2\pi)^2} \left[ f_+(\bsl{k}) \Tr[g_+(\bsl{k})] -\left|\frac{f_-(\bsl{k})}{|\Delta_\shpa(\bsl{k})||\Delta_\perp(\bsl{k})|} z_{\bsl{k}}\right|\right] \\
&  \geq 4 e^2\int_{{\BZ}} \frac{d^2 k}{(2\pi)^2} \left[ f_+(\bsl{k}) \Tr[g_+(\bsl{k})] -|f_{-}(\bsl{k})|\Tr[g_{+}(\bsl{k})] \right]\\
& \geq 4 e^2\int_{{\BZ}} \frac{d^2 k}{(2\pi)^2} \frac{\left[|\Delta_{\perp}(\bsl{k})|-|\Delta_{\shpa}(\bsl{k})|\right]^2}{\sqrt{\left[|\Delta_{\perp}(\bsl{k})|-|\Delta_{\shpa}(\bsl{k})|\right]^2+\mu^2}} \Tr[g_+(\bsl{k})]
\ ,
}
where we have used $|f_{-}(\bsl{k})|=f_{-}(\bsl{k})$.
Then, combined with 
\eq{
\int_{\text{{\BZ}}} \frac{d^2 k}{(2\pi)^2} \Tr[g_+(\bsl{k})] \geq \frac{\N_+}{\pi} > 0
}
derived in \refcite{Xie2020TopologyBoundSCTBG} for any isolated set of two bands with nonzero Euler number $\N_+$,
we eventually get
\eq{
\label{eq:Tr_DSF_bound}
\Tr[D_{SF}]\geq \Tr[D_{SF}^{bound}] = \left\langle \frac{\left[|\Delta_{\perp}(\bsl{k})|-|\Delta_{\shpa}(\bsl{k})|\right]^2}{\sqrt{\left[|\Delta_{\perp}(\bsl{k})|-|\Delta_{\shpa}(\bsl{k})|\right]^2+\mu^2}} \right\rangle_{g_+} \frac{4 e^2}{\pi}\N_+\ ,
}
where 
\eq{
\left\langle x(\bsl{k}) \right\rangle_{g_+} = \frac{\int_{\text{{\BZ}}} d^2 k\ x(\bsl{k}) \Tr[g_+(\bsl{k})] } { \int_{\text{{\BZ}}} d^2 k  \Tr[g_+(\bsl{k})] } \ .
}
Similar to \eqnref{eq:D_SF_sim}, \eqnref{eq:Tr_DSF_bound} also holds for all gauges since both sides are gauge-invariant.

If we choose the time-reversal-invariant uniform pairing for the flat bands in MATBG in \refcite{Xie2020TopologyBoundSCTBG}, we would have $|\Delta_\perp(\bsl{k})|=|\Delta|$ is momentum-independent and $|\Delta_{\shpa}(\bsl{k})|=0$.
Then, \eqnref{eq:Tr_DSF_bound} reproduces the bound for the time-reversal-invariant uniform pairing in \refcite{Xie2020TopologyBoundSCTBG} as
\eqa{
\Tr[D_{SF}]\geq \Tr[D_{SF}^{bound}] = \frac{4 e^2}{\pi}\N_+ 2 \sqrt{\nu(1-\nu)}|\Delta|\ ,
}
where $\nu$ is the filling ratio of the normal-state flat bands with $\nu=0$ meaning that the entire normal-state flat bands are empty.

\section{Euler Obstructed Cooper Pairing in MATBG at Zero Temperature}
\label{app:EOCP_MATBG}

In this section, we provide more details on the Euler obstructed Cooper pairing in MATBG.
We start with the BM model that captures the normal state, then discuss the possible Euler obstructed Cooper pairing in MATBG and its relation to nematic nodal superconductivity, and finally address the bounded zero-temperature superfluid weight.
All discussions in this section are at zero temperature, unless specified otherwise.

\subsection{BM Model}

In this part, we review the BM model~\cite{Bistritzer2011BMModel} that describes the normal state of MATBG.
If we rotate the top (bottom) layer by $-\theta/2$ ($\theta/2$) about the out of plane axis, the BM model reads
\eq{
\label{eq:BM_model}
H_{BM}=\widetilde{H}_+ + \widetilde{H}_- 
}
with
\eq{
\widetilde{H}_+=\int d^2 r\ \psi_{+,\bsl{r}}^\dagger 
\mat{
 -\ii v_0 \nabla_{\bsl{r}}\cdot\bsl{\sigma}_{\theta/2} & T(\bsl{r}) \\
  T^\dagger(\bsl{r}) & -\ii v_0 \nabla_{\bsl{r}}\cdot\bsl{\sigma}_{-\theta/2}
}\otimes s_0
\psi_{+,\bsl{r}}
}
and $\widetilde{H}_-=\TR  \widetilde{H}_+ \TR^{-1}$.
Here $\psi_{\pm,\bsl{r}}^\dagger=(\psi_{\pm,\bsl{r},t,A}^\dagger,\psi_{\pm,\bsl{r},t,B}^\dagger,\psi_{\pm,\bsl{r},b,A}^\dagger,\psi_{\pm,\bsl{r},b,B}^\dagger)$, $A/B$ labels the graphene sublattice (not confused with the patch index $A$ for patchwise-smooth gauges in \appref{app:general_formalism}), $t/b$ labels the top and bottom layers (not confused with the $b=\perp,\shpa$ for the channel splitting in \eqnref{eq:Delta_split_app}), $\psi_{\pm,\bsl{r},t/b,A/B}^\dagger$ has two spin components, and $\sigma$'s are Pauli matrices for the sublattice index.
$\sigma_{\theta}=e^{-\ii \theta \sigma_z /2 } (\sigma_x,\sigma_y)e^{\ii \theta \sigma_z /2 }$.
\eq{
T(\bsl{r})=\sum_{j=1}^3 e^{-\ii \bsl{r}\cdot\bsl{q}_{j}} T_j\ ,
}
$\bsl{q}_1=k_D (0,-1)^T$, $\bsl{q}_2=k_D (\sqrt{3}/2,1/2)^T$, $\bsl{q}_3=k_D (-\sqrt{3}/2,1/2)^T$, 
$k_D=\frac{4\pi}{3 a_0} 2 \sin(\theta/2)$
and 
\eq{
T_j=w_0 \sigma_0 + w_1 \left[ \sigma_x \cos\left( \frac{2\pi (j-1)}{3} \right) + \sigma_y  \sin\left( \frac{2\pi (j-1)}{3} \right)\right]\ .
}
All parameters $v_0$, $a_0$, $w_0$ and $w_1$ are real.

The symmetry group of the BM model \eqnref{eq:BM_model} is spanned by the spin-charge $\U(2)$ rotation in each valley (in total $\U(2)\times\U(2)$), the Moir\'e lattice translation $T_{\bsl{a}_{M,i}}$ with $i=1,2$ and ${\bsl{a}_{M,1}}=\frac{4\pi}{3 k_D}(\frac{\sqrt{3}}{2},\frac{1}{2})$ and ${\bsl{a}_{M,2}}=\frac{4\pi}{3 k_D}(-\frac{\sqrt{3}}{2},\frac{1}{2})$, $\overline{C}_{2z}\TR$, $\overline{C}_{3z}$, $\overline{C}_{2x}$, and $\TR$.
Although we mention the $\overline{C}_{2x}$ symmetry, we will not use $\overline{C}_{2x}$ in the discussion of nodal superconductivity.
Specifically, 
\eqa{
& T_{\bsl{x}} \psi_{\pm,\bsl{r}}^\dagger T_{\bsl{x}}^{-1}= \psi_{\pm,\bsl{r}+{\bsl{x}}}^\dagger
\mat{ e^{\mp\ii K_{-\theta/2}\cdot \bsl{x}} \sigma_0 & \\ & e^{\mp\ii K_{\theta/2}\cdot \bsl{x}} \sigma_0 }\otimes s_0\\
& \overline{C}_{3z} \psi_{\pm,\bsl{r}}^\dagger \overline{C}_{3z}^{-1} =  \psi_{\pm,C_{3z}\bsl{r}}^\dagger  
\mat{ e^{\pm\ii \sigma_z \pi/3}  & \\ &  e^{\pm\ii \sigma_z \pi/3}  }\otimes e^{-\ii \frac{s_z}{2}\frac{2\pi}{3}} \\
& \overline{C}_{2z}\TR \psi_{\pm,\bsl{r}}^\dagger (\overline{C}_{2z}\TR)^{-1} =  \psi_{\pm,-\bsl{r}}^\dagger  
\mat{ \sigma_x  & \\ &   \sigma_x  }\otimes (-\ii s_x) \\
& \overline{C}_{2x} \psi_{\pm,\bsl{r}}^\dagger (\overline{C}_{2x})^{-1} =  \psi_{\pm,C_{2x}\bsl{r}}^\dagger  
\mat{ &  -\sigma_x  \\  -\sigma_x &   }\otimes (-\ii s_x) \\
& \TR \psi_{\pm,\bsl{r}}^\dagger (\TR)^{-1} =  \psi_{\mp,\bsl{r}}^\dagger  
\mat{ \sigma_0 &    \\   &  \sigma_0}\otimes \ii s_y \ ,
}
where $K_{\theta}=\frac{4\pi}{3 a_0}(\cos(\theta),\sin(\theta))^T$.
The Moir\'e reciprocal lattice vectors spanned by the linear combiantions of $\bsl{b}_{M,1}=\sqrt{3}k_D(\frac{1}{2},\frac{\sqrt{3}}{2})$ and $\bsl{b}_{M,2}=\sqrt{3}k_D(-\frac{1}{2},\frac{\sqrt{3}}{2})$.
The Moir\'e Brillouin zone ({\MBZ}) is centered at $\Gamma_M = 0$ and has two inequivalent corners $K_M=-\bsl{q}_3$ and $K_M'=\bsl{q}_2$.

Throughout the work, the realistic parameter values for the MATBG~\cite{Bistritzer2011BMModel,Song2019TBGFragile} are chosen as
\eq{
\label{eq:BM_realistic_paraval}
\theta\in[1.05^{\circ}, 1.15 ^{\circ}],\ w_0/w_1 = 0.8,\ v_0= 5817 \text{meV}\cdot\AA,\ w_1=110\text{meV},\ a_0=2.46 \AA\ .
}
$w_0=0$ is called the chiral limit~\cite{Tarnopolsky2019MagicAngleChiralLimit}, which is of great theoretical interests, and thus we will sometimes will extend the range of $w_0/w_1$ and
\eq{
\label{eq:BM_more general_paraval}
\theta\in[1.05^{\circ}, 1.15 ^{\circ}],\ w_0/w_1\in[0,0.8],\ v_0= 5817 \text{meV}\cdot\AA,\ w_1=110\text{meV},\ a_0=2.46 \AA\ .
}
The BM model can be numerically solved by choosing cutoff in the reciprocal lattice.
Specifically, we will adopt the scheme in \refcite{Song2019TBGFragile}, meaning that $\widetilde{H}_\pm$ can be expressed in the matrix representation as
\eq{
\label{eq:BM_mat}
\widetilde{H}_\pm = \sum_{\bsl{k}\in {\MBZ}}\sum_{\bsl{Q},\bsl{Q}',\sigma,\sigma',s,s'} \psi^\dagger_{\pm,\bsl{k},\bsl{Q}} \left[\widetilde{h}_{\pm}(\bsl{k})\right]_{\bsl{Q}\bsl{Q}',\sigma\sigma'} [s_0]_{ss'} \psi_{\pm,\bsl{k},\bsl{Q}',\sigma',s'} 
}
with 
\eq{
\psi^\dagger_{\pm,\bsl{k},\bsl{Q},\sigma,s} =\frac{1}{\sqrt{\mathcal{V}}} \int d^2 r\ e^{ \ii (\bsl{k}+\bsl{Q})\cdot \bsl{r} } \psi^\dagger_{\pm,l_{\bsl{Q}}^{\pm},\bsl{r},\sigma,s}\ ,
}
where $\bsl{Q}\in Q_{+,t}\cup Q_{+,b} = Q_{-,t}\cup Q_{-,b}$, $Q_{+,t} = Q_{-,b} =  \bsl{q}_3+ \{ \bsl{G}_M \}$, $Q_{+,b} = Q_{-,t} =  -\bsl{q}_2  + \{ \bsl{G}_M \}$, $l_{\bsl{Q}}^{\pm}=t$ for $\bsl{Q}\in Q_{\pm,t}$, $l_{\bsl{Q}}^{\pm}=b$ for $\bsl{Q}\in Q_{\pm,b}$, and $\bsl{G}_M$ is the Moir\'e reciprocal lattice vector.
For numerical simplicity, we choose $|\bsl{Q}|\leq 2\sqrt{7} k_D$ for the current work.

\begin{figure}
    \centering
    \includegraphics[width=\columnwidth]{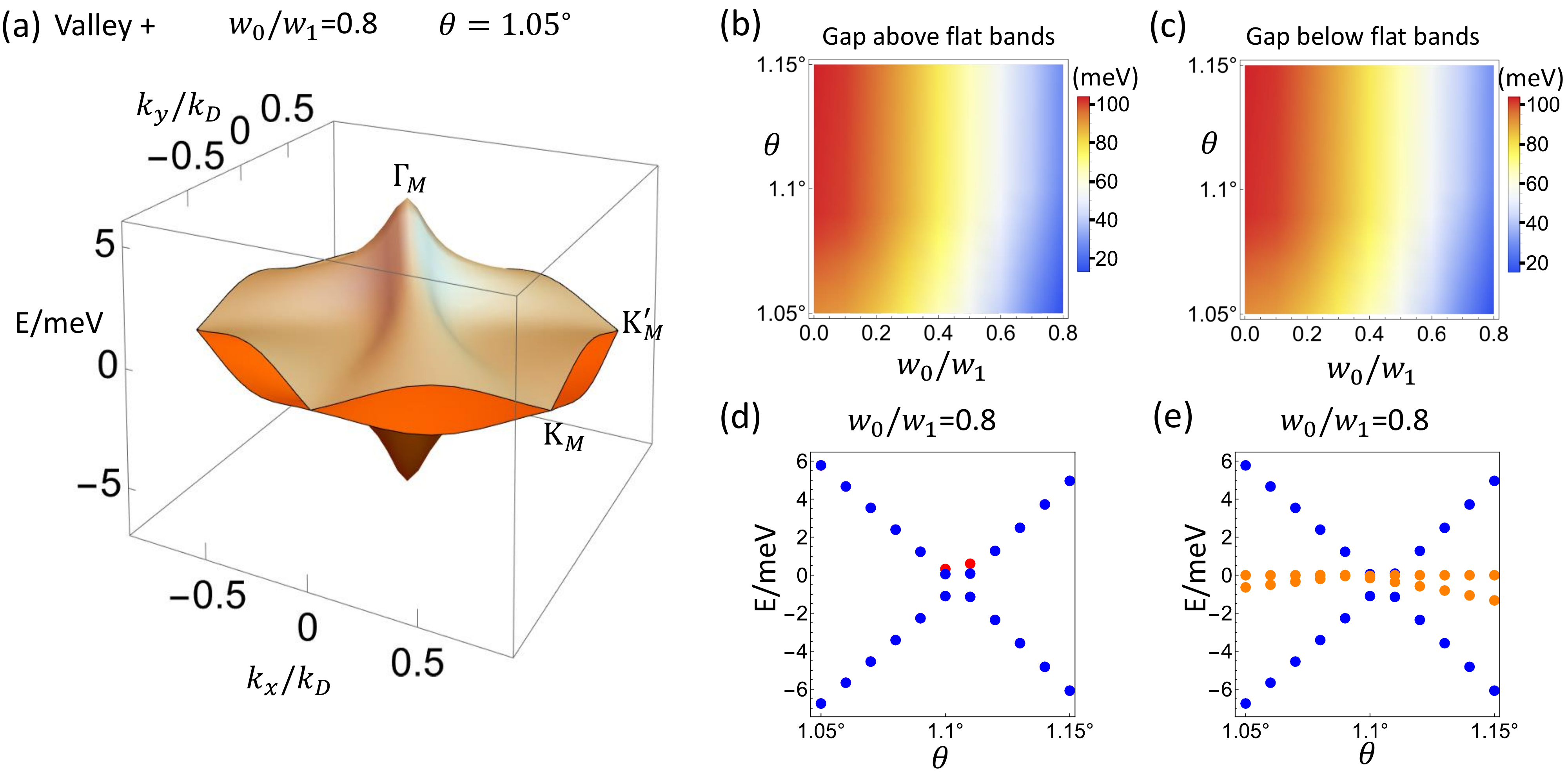}
    \caption{
    Numerical calculation for the BM model with \eqnref{eq:BM_more general_paraval}.
    (a) show the isolated set of two nearly flat bands in valley $+$ for the BM model for $\theta=1.05^\circ$ and $w_0/w_1=0.8$.
    (b) and (c) show the minimum of the direct gaps above and below the isolated set of nearly flat bands, respectively.
    In (d), we show the top and bottom of the isolated set of near flat bands as red dots, and show the energies of the two nearly flat bands at $\Gamma_M$ at blue dots. 
    In most cases, the red dots and blue dots coincide and the blue dots exactly cover the red dots, meaning that the top and bottom of the isolated set of near flat bands are typically at $\Gamma_M$.
    We can see that the band width of the isolated set of two nearly flat bands is about $2\sim 15$ meV.
    In (e), we show the energies of the two nearly flat bands at $\Gamma_M$ at blue dots, and show the normal-state chemical potentials for 2 and 3 holes per Moir\'e unit cell as the orange dots.
    We can see when there are 2 $\sim$ 3 holes per Moir\'e unit cell, the normal-state chemical potential lies in the energy range bounded by the energies of the two nearly flat bands at $\Gamma_M$.
    We caution that given a fixed filling, the chemical potential in the superconducting phase can be different from that in the normal state due to the correction brought by the pairing order parameter.
    }
    \label{fig:BM_Model_App}
\end{figure}

As exemplified in \figref{fig:BM_Model_App}(a), the BM model has two doubly degenerate nearly flat bands in each valley.
The projected effective models that capture the normal-state nearly flat bands are
\eq{
\label{eq:BM_projected}
H_{\pm}= \sum_{\bsl{k}\in\MBZ} c^\dagger_{\pm,\bsl{k}} h_{\pm}(\bsl{k})\otimes s_0 c_{\pm,\bsl{k}}\ ,
}
where $c^{\dagger}_{\pm,\bsl{k}}=(...,c^{\dagger}_{\pm,\bsl{k},a,s},...)$,  $c^\dagger_{\pm,\bsl{k},a,s}= \sum_{\bsl{Q},\sigma} \psi^\dagger_{\pm,\bsl{k},\bsl{Q},\sigma,s} [V_{\pm,a}(\bsl{k})]_{\bsl{Q},\sigma} $ stands for the creation operator for the Bloch basis of the nearly flat bands in valley $\pm$, $V_{\pm}(\bsl{k})=(V_{\pm,1}(\bsl{k}), V_{\pm,2}(\bsl{k}))$ are othonormal linear combination of eigenvectors of $\widetilde{h}_{\pm}(\bsl{k})$ (\eqnref{eq:BM_mat}) for the nearly flat bands, and \eq{
h_{\pm}(\bsl{k})= V_{\pm}^\dagger(\bsl{k}) \widetilde{h}_{\pm}(\bsl{k}) V_{\pm}(\bsl{k})\ .
}
The set of two nearly flat bands are isolated from other bands from other bands by a gap at least 15meV as shown in \figref{fig:BM_Model_App}(b-c), and it has Euler number equalling to 1.
Based on the expression of the BM model and the numerical results, we verify that the normal state satisfies the \asmref{asm:normal_state_C2T}-\ref{asm:N_pm_nonzero} for \eqnref{eq:BM_more general_paraval}, setting the stage for the study of the Euler obstructed pairing.

\subsection{Euler Obstructed Cooper Pairing in MATBG}

For the study of superconductivity, we will assume that the superconductivity in MATBG is given by the mean-field Cooper pairs of normal-state electrons.
Under this assumption, we will include the nearly flat bands over the entire {\MBZ}, instead of just the modes at the Fermi energy, for pairing operator.
The reasoning is the following.
In the standard mean-field theory of superconductivity, there is a superconductivity cutoff $\epsilon_c$, and only normal-state electrons with energies in $[\mu-\epsilon_c,\mu+\epsilon_c]$ experience effectively attractive interaction, while normal-state electrons with energies outside the range cannot form Cooper pairs that condensate at zero temperature~\cite{Frigeri2004NCSC}.
In other words, only the normal-state electrons with energies in $[\mu-\epsilon_c,\mu+\epsilon_c]$ are allowed to have nonzero matrix elements in the pairing operator, and need to be included in the study of superconductivity in general.

In the normal state, we know $\mu$ lies in the nearly flat bands, but the value of $\epsilon_c$ varies with underlying mechanism that accounts for the pairing.
Nevertheless, owing to the large gap above and below the nearly flat bands, we will assume $\epsilon_c$ is small enough such that modes outside the nearly flat bands can be neglected.
Then, if $\epsilon_c$ is larger than the small bandwidth of the nearly flat bands, we should include the nearly flat bands over the entire {\MBZ} in the pairing operator.
Even if $\epsilon_c$ is smaller than the bandwidth of the nearly flat bands, we can still define the pairing operator for the nearly flat bands over the entire {\MBZ} by just keeping the pairing matrix vanishing for the electrons beyond $\epsilon_c$.
In doing so, we should avoid discontinuous structures (like step function) for electrons at $\epsilon_c$ and try to use the corresponding smooth versions (like $\tanh$).

The discussion of the Euler obstructed Cooper pairing is done at zero temperature.
The pairing that we consider is the intervalley $C_{2z}\TR$-invariant mean-field pairing operator with form 
\eq{
\label{eq:H_pairing_MATBG}
H_{pairing} = \sum_{\bsl{k}\in {\MBZ}} c^\dagger_{+,\bsl{k}} \Delta(\bsl{k})\otimes\Pi c^{\dagger}_{-.-\bsl{k}}+h.c.\ ,
}
where {\MBZ} stands for the Moir\'e Brillouin zone, $c^{\dagger}_{\pm,\bsl{k}}=(...,c^{\dagger}_{\pm,\bsl{k},a,s},...)$ stands for the creation operator for the Bloch basis of the nearly flat bands in valley $\pm$, $s=\uparrow,\downarrow$ labels the spin index, $a=1,2$ since there are two spin-doubly-degenerate nearly flat bands in each valley, $\Delta(\bsl{k})$ is the spinless part of the pairing with $\Delta(\bsl{k}+\bsl{G}_M)=\Delta(\bsl{k})$ and $\bsl{G}_M$ the Moire reciprocal lattice vector, and $\Pi$ is the spin part of the pairing.
The global charge $\U(1)$ in $\U(2)\times\U(2)$ of the normal state is spontaneously broken.
Owing to the global spin $\SU(2)$ symmetry in $\U(2)\times\U(2)$ of the normal state, we can consider the spin-singlet and spin-triplet channels separately, and choose the spin part in each channel to be momentum independent as 
\eqa{
\label{eq:spin_part_pairing}
& \Pi=\ii s_y\text{ for spin-singlet}\\
& \Pi=s_0\text{ for spin-triplet}\ .
}
The remaining global $\U(2)$ symmetry in $\U(2)\times\U(2)$ of the normal state makes the spin-singlet and spin-triplet to have the same superconducting critical temperature~\cite{Wu2018PhononSCMATBG}; we assume that the remaining $\U(2)$ is spontaneously broken at zero temperature, and therefore, we will still study the spin-singlet and spin-triplet channels separately. 
Furthermore, \eqnref{eq:H_pairing_MATBG} can always be viewed as the projection of the pairing in the original basis $\psi$ in \eqnref{eq:BM_model}, and since the pairing in basis $\psi$ is always smooth, we know $\sum_{a,a'}\ket{u_{+,\bsl{k},a}}\ket{u_{-,-\bsl{k},a'}}[\Delta(\bsl{k})]_{aa'}$ is always smooth.
Therefore, \eqnref{eq:H_pairing_MATBG} satisfies \asmref{asm:pairing}, meaning that $\Delta(\bsl{k})$ in \eqnref{eq:H_pairing_MATBG} can be split into two channels, a trivial one and a Euler obstructed one as shown in \eqnref{eq:Delta_split_app}, and the pairing gap function of the Euler obstructed channel always has zeros  as shown in \eqnref{eq:W_b_i_N_app}.
In the remaining of this section, we will always imply that the pairing has the form in \eqnref{eq:H_pairing_MATBG}.

\subsection{Symmetry Classification of the Pairing in \eqnref{eq:H_pairing_MATBG}}

To prepare for the study of the nodal superconductivity, we need to consider more symmetry properties of the pairing.
As we have used the $\U(2)\times\U(2)$, we need to consider the rest of the symmetries.
Since we have chosen the pairing to be $C_{2z}\TR$-symmetric and invariant under the Moir\'e lattice translations and the spin part of all rotation symmetries have been included in $\U(2)\times\U(2)$, we only need to care about the remaining $C_{2z}$ and $C_{3z}$, where
\eqa{
& C_{2z} =  \overline{C}_{2z} [\overline{C}_{2z}^{spin}]^{-1}\text{ with } C_{2z} \psi_{\pm,\bsl{r}}^\dagger (C_{2z})^{-1} =  \psi_{\mp,-\bsl{r}}^\dagger  
\mat{ \sigma_x  & \\ &   \sigma_x  }\otimes s_0 \\
& C_{3z} =  \overline{C}_{3z} [\overline{C}_{3z}^{spin}]^{-1}\text{ with } C_{3z} \psi_{\pm,\bsl{r}}^\dagger C_{3z}^{-1} =  \psi_{\pm,C_{3z}\bsl{r}}^\dagger
\mat{ e^{\pm\ii \sigma_z \pi/3}  & \\ &  e^{\pm\ii \sigma_z \pi/3}  }\otimes s_0\\
}
which form the point group $C_{6}$.
Owing to the $C_{2z}\TR$-invariant of the pairing, we can only consider the real irreducible representations (irreps) of $C_6$, which are (i) $C_{2z}$-even $A$ irrep, (ii) $C_{2z}$-odd $A$ irrep, (iii) $C_{2z}$-even $E$ irrep, and (iv) $C_{2z}$-odd $E$ irrep.
Here $A$ means the irrep is $C_{3z}$-invariant, while $E$ is a 2D real irrep under $C_{3z}$.

We will split the pairing according to the real irreps of $C_{6}$, and study them separately.
Since the symmetry irrep, $|\Delta_b|$ and the BdG nodes are gauge-invariant properties, we will choose a Chern gauge for the normal-state basis that satisfies \asmref{asm:Chern_gauge} and
\eqa{
\label{eq:MATBG_Ch_Gauge}
& C_{2z}\ket{u^{Ch}_{+,\bsl{k}}}=\ket{u^{Ch}_{-,-\bsl{k}}} \\
& C_{3z}\ket{u^{Ch}_{\pm,K_M}}=\ket{u^{Ch}_{\pm,C_{3z}K_M}}e^{\ii\tau_z\frac{2\pi}{3}}\\
& C_{3z}\ket{u^{Ch}_{\pm,K_M'}}=\ket{u^{Ch}_{\pm,C_{3z}K_M'}}e^{\ii\tau_z\frac{2\pi}{3}}\\
& C_{3z}\ket{u^{Ch}_{\pm,\Gamma_M}}=\ket{u^{Ch}_{\pm,\Gamma_M}}\ ,
}
and we have numerically checked that the such a Chern gauge is allowed for the BM model with \eqnref{eq:BM_realistic_paraval}.
Then, for $C_{2z}$, we have  
\eqa{
\label{eq:C_2_pairing}
& [\Delta^{Ch}(\bsl{k})]^T = \Delta^{Ch}(\bsl{k}) \text{ if the pairing is parity-even}\\
& [\Delta^{Ch}(\bsl{k})]^T = -\Delta^{Ch}(\bsl{k}) \text{ if the pairing is parity-odd}\ ,
}
where the $C_{2z}$ eigenvalue is the same as (opposite to) the parity for spin-singlet (spin-triplet) pairing according to the expression of chosen $\Pi$ in \asmref{asm:pairing}. 
Then combined with \eqnref{eq:Delta_split_Ch_app}, we know that parity-odd pairing has $|\Delta_\shpa(\bsl{k})|=0$ for all $\bsl{k}\in\dsR^2$.

For parity-even pairing, we know $|\Delta_\shpa(\bsl{k})|$ can be non-vanishing at certain momenta and is required to have zeros.
The distribution of zeros of $|\Delta_\shpa(\bsl{k})|$ is influenced by the $A/E$ irrep of the parity-even pairing.
In the case of $A$ irrep, we combine \eqnref{eq:MATBG_Ch_Gauge} with \eqnref{eq:Delta_split_Ch_app} and get
\eqa{
& d_{\shpa}(K_M/K_M') e^{-\ii 4\pi/3} = d_{\shpa}(K_M/K_M') \Rightarrow |\Delta_{\shpa}(K_M/K_M')|=0\ .
}
Then, the simplest nodal structure of $|\Delta_{\shpa}(\bsl{k})|$ in $A$ irrep contains and only contains two zeros at $K_M$ and $K_M'$ with winding number 1.

In the case of $E$ irrep, the pairing would be the linear combination of the two component of $E$, which spontaneously breaks $C_{3z}$ symmetry and thus is called spontaneously nematic pairing.
Specifically,
\eqa{
& H_{pairing}= a_1 H_{pairing,1} + a_2 H_{pairing,2} \\
& C_{3z} (H_{pairing,1}\ H_{pairing,2}) C_{3z}^{-1} = (H_{pairing,1}\  H_{pairing,2}) e^{-\ii \tau_y \frac{2\pi}{3}} \ .
}
where $a_1,a_2\in\dsR$ and $H_{pairing,1} $ and $H_{pairing,2}$ also satisfy \asmref{asm:pairing}.
Then, combined with \eqnref{eq:MATBG_Ch_Gauge} and \eqnref{eq:Delta_split_Ch_app}, we get
\eq{
\mat{ d_{\shpa,1}(\Gamma_M) & d_{\shpa,2}(\Gamma_M)}  =  \mat{ d_{\shpa,1}(\Gamma_M) & d_{\shpa,2}(\Gamma_M)} e^{-\ii \tau_y \frac{2\pi}{3}} \Rightarrow  d_{\shpa,1}(\Gamma_M) = d_{\shpa,2}(\Gamma_M) =0\ ,
}
which means $|\Delta_{\shpa}(\Gamma_{M})|=0$.
Then, the simplest nodal structure of $|\Delta_{\shpa}(\bsl{k})|$ is to have two zeros with winding number 1 and at least one of them is at $\Gamma_M$.

Here, when we count the number of zeros of $\Delta_\shpa$, we always mean the smallest number of zeros with winding $\pm 1$. 
So, if we have a winding-2 zero, then we would say we have two winding-1 zeros at the same momentum.

\subsection{General Discussion on the Possible Nodal Superconductivity}
\label{app:nodal_SC_gen}
In this part, we will present a general discussion on the possible nodal superconductivity.

First of all, the mean-field Hamiltonian reads
\eq{
H_{MF} = H_{BdG,+} + H_{BdG,-} + const.\ ,
}
where 
\eq{
H_{BdG,\pm}=\frac{1}{2}\sum_{\bsl{k}\in{\MBZ}}\Psi_{\pm,\bsl{k}}^\dagger 
h_{BdG,\pm}(\bsl{k})
\Psi_{\pm,\bsl{k}}\ ,
}
\eq{
h_{BdG,+}(\bsl{k})=\mat{ 
[h_+(\bsl{k})-\mu] \otimes s_0 & \Delta(\bsl{k})\otimes \Pi\\
 \Delta^\dagger(\bsl{k})\otimes \Pi^\dagger & -[h_-(-\bsl{k})-\mu]^T \otimes s_0
} \ , 
}
\eq{
h_{BdG,-}(\bsl{k})=\mat{ 
[h_-(\bsl{k})-\mu] \otimes s_0 & -\Delta^T(-\bsl{k})\otimes \Pi^T\\
 -\Delta^*(-\bsl{k})\otimes \Pi^* & -[h_+(-\bsl{k})-\mu]^T \otimes s_0
} \ ,
}
$\Psi_{+,\bsl{k}}^\dagger = (c^\dagger_{+,\bsl{k}}, c^T_{-,-\bsl{k}})$, $\Psi_{-,\bsl{k}}^\dagger = (c^\dagger_{-,\bsl{k}}, c^T_{+,-\bsl{k}})$, and $\mu$ is the chemical potential.
$h_{BdG,-}(\bsl{k})$ is related to $h_{BdG,+}(\bsl{k})$ by the particle-hole symmetry as
\eq{
-h_{BdG,-}(\bsl{k}) = \rho_x\tau_0\s_0 h_{BdG,+}^*(-\bsl{k}) \rho_x\tau_0\s_0\ ,
}
which means the gapless nodes of $h_{BdG,-}(\bsl{k})$ are completely determined by those of $h_{BdG,+}(\bsl{k})$ and vice versa.

For spin-singlet pairing, we have 
\eq{
h_{BdG,+}(\bsl{k}) = 
\mat{
h_+(\bsl{k})-\mu  & & & \Delta(\bsl{k}) \\
 & h_+(\bsl{k})-\mu & -\Delta(\bsl{k}) & \\
 & -\Delta^\dagger(\bsl{k}) & -[h_-(-\bsl{k})-\mu]^T & \\
\Delta^\dagger(\bsl{k}) & & & -[h_-(-\bsl{k})-\mu]^T
}
\sim 
\mat{
\H(\bsl{k})   & \\
 & \rho_z\tau_0 \H(\bsl{k}) \rho_z\tau_0
}\ ,
}
where $\sim$ means differing by a $\bsl{k}$-independent unitary transformation and
\eq{
\label{eq:h_BdG_+_up}
\H(\bsl{k}) = 
\mat{
h_+(\bsl{k})-\mu  & \Delta(\bsl{k}) \\
\Delta^\dagger(\bsl{k}) & -[h_-(-\bsl{k})-\mu]^T 
}\ .
}
For spin-triplet pairing, we have 
\eqa{
& h_{BdG,+}(\bsl{k}) = 
\mat{
h_+(\bsl{k})-\mu  & & \Delta(\bsl{k}) & \\
 & h_+(\bsl{k})-\mu &  & \Delta(\bsl{k}) \\
   \Delta^\dagger(\bsl{k}) &   & -[h_-(-\bsl{k})-\mu]^T & \\
& \Delta^\dagger(\bsl{k})  & & -[h_-(-\bsl{k})-\mu]^T
} \sim 
\mat{
 \H(\bsl{k})  & \\
 &  \H(\bsl{k}) 
}\ .
}
where we used \eqnref{eq:spin_part_pairing}.
We can see for both spin-singlet and spin triplet pairings,  $h_{BdG,+}(\bsl{k})$ can be decomposed into two blocks, where each block is either equal to or similar to $\H(\bsl{k})$ in \eqnref{eq:h_BdG_+_up}, meaning that the dispersion of $h_{BdG,+}(\bsl{k})$ is just the double copy of that of $\H(\bsl{k})$.
Therefore,  the gapless nodes of $h_{BdG,+}(\bsl{k})$ are completely determined by those of $\H(\bsl{k})$.
So we only need to study the gapless nodes of $\H(\bsl{k})$ in the following.

Since we care about the nodal superconductivity related to the Euler obstructed Cooper pairing, we will only study the parity-even pairing.
Therefore, in the remaining of this part, we will always adopt the following condition unless specified otherwise.
\begin{assumption}
\label{asm:Existence_Delta_para}
The pairing is parity-even, and $|\Delta_{\shpa}(\bsl{k})|$ is not glabally vanishing (\ie, $|\Delta_{\shpa}(\bsl{k})|\neq 0$ holds for at least one $\bsl{k}$ points).
\end{assumption}

Since the band structure of $\H(\bsl{k})$ is gauge invariant, we can choose any gauge to study it.
Let us first choose a generic gauge for $c^\dagger_{\pm,\bsl{k}}$ such that
\eq{
\label{eq:C_2_gauge}
C_{2z} c^\dagger_{+,\bsl{k}} C_{2z}^{-1} = c^\dagger_{-,-\bsl{k}}\ ,
}
In this gauge, \eqnref{eq:h_BdG_+_up} becomes
\eq{
\label{eq:h_BdG_+_up_smooth}
\H(\bsl{k}) = 
\mat{
h_+(\bsl{k})-\mu  & \Delta_\perp(\bsl{k}) + \Delta_\shpa(\bsl{k}) \\
\Delta_\perp^\dagger(\bsl{k}) + \Delta_\shpa^\dagger(\bsl{k}) & -[h_+(\bsl{k})-\mu]^T 
}\ .
}
$\H(\bsl{k})$ has the spinless $C_{2z}\TR$ symmetry as
\eq{
U_{BdG}(\bsl{k}) \H^*(\bsl{k}) U_{BdG}^\dagger(\bsl{k}) = \H(\bsl{k})\ ,
}
where 
\eq{
U_{BdG}(\bsl{k})=\mat{U_+(\bsl{k}) & \\ & U_+^*(\bsl{k})}\ ,
}
and $U_+(\bsl{k})=\bra{u_{\bsl{k},+}}C_{2z}\TR \ket{{u_{\bsl{k},+}}}$.
Moreover, since the pairing is parity-even, $\H(\bsl{k})$ has a chiral symmetry as 
\eq{
C(\bsl{k}) \H(\bsl{k}) C^\dagger(\bsl{k}) = -\H(\bsl{k})
}
with 
\eq{
C(\bsl{k}) = \mat{  & U_+(\bsl{k}) \\ -U_+^*(\bsl{k}) & }\ .
}

We can diagonalize $C(\bsl{k})$ as
\eq{
U_C^\dagger(\bsl{k}) C(\bsl{k}) U_C(\bsl{k}) = 
\mat{ 
\ii & & & \\
& \ii & & \\
& & -\ii & \\
& & & -\ii
}
}
with 
\eq{
U_C(\bsl{k}) = \frac{1}{\sqrt{2}} \mat{1 & 1 \\ \ii U_{+,\bsl{k}}^* & -\ii U_{+,\bsl{k}}^* } \ .
}
Then, $U_C(\bsl{k})$ can make $\H(\bsl{k})$ offdiagonal as
\eq{
U_C^\dagger(\bsl{k}) \H(\bsl{k}) U_C(\bsl{k}) = \mat{ & h_+(\bsl{k})-\mu-\ii \Delta(\bsl{k})U_{+,\bsl{k}}^* \\ h.c. & }\ . 
}
In particular, the complex $\det[h_+(\bsl{k})-\mu-\ii \Delta(\bsl{k})U_{+,\bsl{k}}^*]$ is independent of the gauge choices, as long as the gauges satisfy \eqnref{eq:C_2_gauge}.
This quantity is the key to the stable nodal superconductivity, as discussed in the following.

To show this, we can choose a guage such that \eqnref{eq:C_2_gauge} holds and $\ket{u_{\pm,\bsl{k}}}$ is globally-smooth, where the existence of such gauge is guaranteed by the zero total Chern number of $\ket{u_{\alpha,\bsl{k}}}$~\cite{Brouder2007Wannier}.
Then, $\det[h_+(\bsl{k})-\mu-\ii \Delta(\bsl{k})U_{+,\bsl{k}}^*]$ is globally smooth in this gauge, and thus has interger $\U(1)$ winding along any closed loop $\gamma$ if $\det[h_+(\bsl{k})-\mu-\ii \Delta(\bsl{k})U_{+,\bsl{k}}^*]\neq 0\ \forall \bsl{k}\in\gamma$.
This winding number is called the chiral symmetry protected winding number~\cite{Bzdusek2017AZInversionNodal}, or in short chirality in this work.
If the chirality is nonzero, then $\det[h_+(\bsl{k})-\mu-\ii \Delta(\bsl{k})U_{+,\bsl{k}}^*]$ must have zero(s) inside $\gamma$, and  thus $\H(\bsl{k})$ must have zero-energy gapless node(s) inside $\gamma$.
Indeed, since $[C(\bsl{k})]^2=-1$ and $U_{BdG}(\bsl{k}) C^*(\bsl{k}) = C(\bsl{k}) U_{BdG}(\bsl{k})$, $\H(\bsl{k})$ in this smooth gauge belongs to the CI nodal class proposed in \refcite{Bzdusek2017AZInversionNodal}, which can support stable zero-energy gapless points.
The stable zero-energy gapless nodes are indeed protected by the nonzero chiralities.
Since the gapless nodes of $\H(\bsl{k})$ are always at zero energy, so does $H_{BdG}$; $H_{BdG}$ is gapless (or equivalently the superconductivity is nodal) if and only if $\H(\bsl{k})$ has zero-energy gapless nodes.
In particular, if we smoothly vary 
\eq{
\label{eq:smooth_parts}
\mu,\ P_{h,+}(\bsl{k})=\ket{u_{+,\bsl{k}}}h_+(\bsl{k}) \bra{u_{+,\bsl{k}}},\ P_\shpa(\bsl{k})\ ,P_\perp(\bsl{k})\ ,
}
$\H(\bsl{k})$ would change smoothly, and the zero-energy gapless nodes of $\H(\bsl{k})$ would change continuously.
Note that the chirality can be evaluated in any gauge that satisfies \eqnref{eq:C_2_gauge}, since $\det[h_+(\bsl{k})-\mu-\ii \Delta(\bsl{k})U_{+,\bsl{k}}^*]$  is independent of the gauge choices as long as the gauges satisfy \eqnref{eq:C_2_gauge}.

\subsection{Nodal Superconductivity Guaranteed by Sufficiently-Dominant $\Delta_{\shpa}$ in MATBG}

In the last part, we have shown that $\H$ or $H_{BdG}$ can be stably nodal in certain cases. 
The remaining question is what the relation between a gapless $\H(\bsl{k})$ and the dominance of the Euler obstructed $\Delta_\shpa$ is .
In this part, we will study the nodal superconductivity guaranteed by sufficiently-dominant $\Delta_{\shpa}$ in MATBG under \asmref{asm:normal_state_C2T}-\ref{asm:pairing} and \asmref{asm:Existence_Delta_para}.

First, nodal superconductivity enforced by a sufficiently-dominant $\Delta_\shpa$ is defined as the following.
\begin{definition}
\label{def:nodal_SC_dom_EOCP}
For any choice of $(\mu, P_{h,+}(\bsl{k}), P_{\shpa}(\bsl{k}))$, nodal superconductivity is enforced by a sufficiently-dominant $\Delta_\shpa$ if and only if there exists $\lambda>0$ such that $\H(\bsl{k})$ has gapless nodes for all choices of symmetry-preserving $P_{\perp}(\bsl{k})$ with $\max(|\Delta_{\perp}(\bsl{k})|)<\lambda$.
\end{definition}

To look for the nodal superconductivity enforced by a sufficiently-dominant $\Delta_\shpa$, we will use the Chern gauge that satisfies \asmref{asm:Chern_gauge} and \eqnref{eq:MATBG_Ch_Gauge}, since $|\Delta_{\shpa}(\bsl{k})|$ and the gapless nodes of $\H(\bsl{k})$ are gauge invariant.
Then, $\H(\bsl{k})$ becomes
\eq{
\label{eq:h_BdG_Ch_Gauge_app}
\H(\bsl{k})=\mat{ h_{Ch,+}(\bsl{k})-\mu & \Delta_\perp^{Ch}(\bsl{k}) + \Delta_{\parallel}^{Ch}(\bsl{k}) \\  [\Delta_\perp^{Ch}(\bsl{k}) + \Delta_{\parallel}^{Ch}(\bsl{k})]^\dagger & -h^T_{Ch,+}(\bsl{k})+\mu}\ ,
}
where $h_{Ch,+}(\bsl{k}) = \epsilon(\bsl{k}) + Re[f (\bsl{k})]\tau_x + Im[f (\bsl{k})]\tau_y$.

Let us first study the case where $|\Delta_{\perp}(\bsl{k})|=0$, and label $\H(\bsl{k})$ with $|\Delta_{\perp}(\bsl{k})|=0$ as $\H^{(0)}(\bsl{k})$.
We care about $\H^{(0)}(\bsl{k})$ because of the following proposition.
\begin{proposition}
\label{prop:nodal_SC_EOCP_gen_app}
Under\asmref{asm:normal_state_C2T}-\ref{asm:pairing} and \asmref{asm:Existence_Delta_para}, for any choice of $(\mu, P_{h,+}(\bsl{k}), P_{\shpa}(\bsl{k}))$, nodal superconductivity is enforced by a sufficiently-dominant $\Delta_\shpa$ if $\H^{(0)}(\bsl{k})$ has at least one isolated gapless node with nonzero chirality.
\end{proposition}
The reasoning for the above proposition is the following.
The bands of $\H(\bsl{k})$ are 
\eq{
\pm\sqrt{|d_{\shpa}|^2 + (\epsilon-\mu)^2 + |f|^2 +|d_\perp(\bsl{k})|^2 \pm 2 \sqrt{|d_\perp|^2 |d_{\shpa}|^2+ 2 d_\perp (\epsilon-\mu) Re(d_{\shpa}f^*)+ |Im(fd_{\shpa}^*)|^2 + |f|^2(\epsilon-\mu)^2 }}\ ,
}
where $\bsl{k}$-dependence of $f$, $d_{\shpa}$, $d_{\perp}$ and $\epsilon$ is implicit and we have used that $d_\perp$ is real for parity-even pairing.
Then, $\H(\bsl{k})$ is gapless at $\bsl{k}$ iff
\eq{
a_0(\bsl{k})+a_1(\bsl{k})d_\perp(\bsl{k})+a_2(\bsl{k})d_\perp^2(\bsl{k})+d_\perp^4(\bsl{k})=0\ ,
}
where 
\eqa{
& a_0(\bsl{k}) = \left[|d_{\shpa}(\bsl{k})|^2 + (\epsilon_{\bsl{k}}-\mu)^2 + |f(\bsl{k})|^2\right]^2 - 4 \left[ |Im(f(\bsl{k})d_{\shpa}^*(\bsl{k}))|^2 + |f(\bsl{k})|^2(\epsilon_{\bsl{k}}-\mu)^2 \right]\\
& a_1(\bsl{k})=-8 Re(d_{\shpa}^*(\bsl{k})f(\bsl{k}))(\epsilon_{\bsl{k}}-\mu)\\
& a_2(\bsl{k})=2 (-|d_{\shpa}(\bsl{k})|^2 + |f(\bsl{k})|^2 + (\epsilon_{\bsl{k}}-\mu)^2)
}
are all smooth in $\dsR^2$.

Consider that $\bsl{k}_0$ is an isolated gapless node of $\H^{(0)}(\bsl{k})$ with nonzero chirality.
There exists a circle $\gamma_0$ surrounding $\bsl{k}_0$ such that $\H^{(0)}(\bsl{k})$ is gapped on $\gamma_0$ and has nonzero chirality along $\gamma_0$.
According to \eqnref{eq:bands_hBdG_0}, $a_0(\bsl{k})>0$ for all $\bsl{k}\in\gamma_0$.
Since $a_{0,1,2}(\bsl{k})$ is smooth and $\gamma_0$ is closed, we have $\bar{a}_0=\min_{\bsl{k}\in\gamma_0}[a_0(\bsl{k})]>0$, and $\bar{a}_{1,2}=\max_{\bsl{k}\in\gamma_0}[a_{1,2}(\bsl{k})]$ is finite.
Then, we define 
\eq{
\lambda=\min\left[(\frac{\bar{a}_0}{6})^{1/4}, (\frac{\bar{a}_0}{6 \bar{a}_2})^{1/2}, \frac{\bar{a}_0}{6 \bar{a}_1} \right]>0
}
where we choose $1/\bar{a}_{1,2}=+\infty$ if $\bar{a}_{1,2}=0$.
As a result, for all $\bsl{k}\in\gamma_0$, $|d_{\perp}(\bsl{k})|<\lambda$ infers $a_0(\bsl{k})+a_1(\bsl{k})d_\perp(\bsl{k})+a_2(\bsl{k})d_\perp^2(\bsl{k})+d_\perp^4(\bsl{k})>\bar{a}_0/2>0$, meaning that $\H(\bsl{k})$ is gapped at $\bsl{k}$.

Then, for any symmetry-preserving $P_{\perp}(\bsl{k})$ such that $\max(|\Delta_{\perp}(\bsl{k})|)<\lambda$, we have $\H(\bsl{k})$ gapped on $\gamma_0$.
Since any $P_{\perp}(\bsl{k})$ such that $\max(|\Delta_{\perp}(\bsl{k})|)<\lambda$ can be smoothly deformed to $0$ while keeping $\max(|\Delta_{\perp}(\bsl{k})|)<\lambda$, $\H(\bsl{k})$ can be smoothly deformed to $\H^{(0)}(\bsl{k})$ in the globally smooth gauge while staying gapped on $\gamma_0$.
Therefore, $\H(\bsl{k})$ must have nonzero chirality along $\gamma_0$ and thus must be gapless for any symmetry-preserving $P_{\perp}(\bsl{k})$ such that $\max(|\Delta_{\perp}(\bsl{k})|)<\lambda$.
So \propref{prop:nodal_SC_EOCP_gen_app} holds.

Now we can see that \propref{prop:nodal_SC_EOCP_gen_app} based on $\H^{(0)}$ gives us a sufficient condition for nodal superconductivity enforced by sufficiently-dominant $\Delta_\shpa$.
Furthermore, for any choice of $(\mu, P_{h,+}(\bsl{k}), P_{\shpa}(\bsl{k}))$, if $\H^{(0)}$ is fully gapped, then nodal superconductivity cannot be enforced by a sufficiently-dominant $\Delta_\shpa$, since $\H$ is gapped for $P_\perp = 0$.
Therefore, the nodal structure of $\H^{(0)}$ is crucial in our consideration. 

To facilitate later discussions, let us present some general results on the possible nodal structure of  $\H^{(0)}$.
$\H^{(0)}(\bsl{k})$ has the four bands, which read
\eq{
\label{eq:bands_hBdG_0}
\pm\sqrt{|d_{\shpa}(\bsl{k})|^2 + (\epsilon_{\bsl{k}}-\mu)^2 + |f(\bsl{k})|^2 \pm 2 \sqrt{|Im(f(\bsl{k})d_{\shpa}^*(\bsl{k}))|^2 + |f(\bsl{k})|^2(\epsilon_{\bsl{k}}-\mu)^2 }}\ ,
}
meaning that a gapless node appears at $\bsl{k}$ if and only if 
\eqa{
Re(f^*(\bsl{k})d_{\shpa}(\bsl{k}))=0 \ \text{and}\ \mu = \epsilon(\bsl{k})\pm\sqrt{|f(\bsl{k})|^2 - |d_{\shpa}(\bsl{k})|^2} \ \text{and}\ |f(\bsl{k})|\geq |d_{\shpa}(\bsl{k})|
}
Since $f^*(\bsl{k})d_{\shpa}(\bsl{k})$ and $|f(\bsl{k})|^2$ are independent of the choice of the Chern gauge that satisfies the requirement, they are smooth in $\dsR^2$.
$|d_{\shpa}(\bsl{k})|^2 = |\Delta_{\shpa}(\bsl{k})|^2$ is also smooth in $\dsR^2$.
Then, the solution set of $Re(f^*(\bsl{k})d_{\shpa}(\bsl{k}))=0 $ and $|f(\bsl{k})|\geq |d_{\shpa}(\bsl{k})|$ consists of isolated points or lines without fine tuning. 
Combined with the fact that $d_{\shpa}(\bsl{k}+\bsl{G}_M)=d_{\shpa}(\bsl{k})$ and $f(\bsl{k}+\bsl{G}_M)=f(\bsl{k})$, we only need to care about the solution set of $Re(f^*(\bsl{k})d_{\shpa}(\bsl{k}))=0 $ and $|f(\bsl{k})|\geq |d_{\shpa}(\bsl{k})|$ in {\MBZ}, labeled as $\Sigma$.
$\Sigma$ is not empty since the zeros of $\Delta_{\shpa}(\bsl{k})$ are definitely in it.
We further define 
\eq{
E(\Sigma)= \left\{ \left.\epsilon(\bsl{k})+\sqrt{|f(\bsl{k})|^2 - |d_{\shpa}(\bsl{k})|^2} \right| \bsl{k}\in\Sigma \right\} \cup   \left\{ \left. \epsilon(\bsl{k})-\sqrt{|f(\bsl{k})|^2 - |d_{\shpa}(\bsl{k})|^2} \right| \bsl{k}\in\Sigma \right\}\ ,
}
and we know $\mu \in E(\Sigma)$ is equivalent to that $\H^{(0)}(\bsl{k})$ has zero-energy gapless nodes.

An alternative expression of $\Sigma$ is useful for the following discussions. 
The useful alternative expression is based on the fact that $|d_{\shpa}(\bsl{k})\pm \ii s f(\bsl{k}) |^2$ with $s\in[0,1]$ is smooth in $\dsR^2$, and we label the solution set of  $|d_{\shpa}(\bsl{k})\pm \ii s f(\bsl{k}) |^2=0$ as $\Sigma_{\pm}(s)\subset{\MBZ}$.
Then, we have 
\eq{
\label{eq:alt_Sigma_zero_path}
\Sigma=\bigcup_{s\in[0,1]} [\Sigma_{+}(s) \cup \Sigma_{-}(s)]
}
since  $|d_{\shpa}(\bsl{k}) + \ii s f(\bsl{k}) |^2=0$ or  $|d_{\shpa}(\bsl{k}) - \ii s f(\bsl{k}) |^2=0$ is equivalent to $Re(f^*(\bsl{k}) d_{\shpa}(\bsl{k}))=0$ and $|d_{\shpa}(\bsl{k})|=s |f(\bsl{k})|$.
Since $|d_{\shpa}(\bsl{k}) \pm \ii s f(\bsl{k}) |^2$ are smooth functions of $(s,\bsl{k})$, we know the zeros of $|d_{\shpa}(\bsl{k}) \pm \ii s f(\bsl{k}) |^2$ are moving continuously in {\MBZ} as $s$ continuously varies if we view {\MBZ} as a torus.
Then, $\Sigma$ is nothing but the paths of zeros of $|d_{\shpa}(\bsl{k})\pm \ii s f(\bsl{k}) |^2$ as $s$ increases from 0 to 1 continuously.
%
Crucially, unless invoking fine tuning, $d_{\shpa}(\bsl{k}) \pm \ii s f(\bsl{k})$ always have zeros in {\MBZ} for $s\in[0,1]$ and the total winding number of the zeros is independent of $s$ and is equal to $2$, where the winding number for a zero of $d_{\shpa}(\bsl{k}) \pm \ii s f(\bsl{k})$ is defined by the winding of $\arg[d_{\shpa}(\bsl{k}) \pm \ii s f(\bsl{k})]$ along a small circle around that zero.

The ultimate goal is to specify whether nodal superconductivity guaranteed by a sufficiently-dominant $\Delta_\shpa$ can exist for $C_{3z}$-invariant and spontaneously nematic pairings.
To do so, we will impose the following assumption in this part
\begin{assumption}
\label{asm:nodal SC}
$\mu\in[E_1(\Gamma_M), E_2(\Gamma_M)]$, where $E_{1}(\bsl{k})$ and $E_{2}(\bsl{k})$ are the lower and upper nearly flat bands of $h_{+}(\bsl{k})$, respectively.
$\Delta_{\shpa}$ only has two zeros with winding $1$ in {\MBZ}.
\end{assumption}
Using the terms in \eqnref{eq:h_BdG_Ch_Gauge_app}, we have $E_1(\bsl{k})=\epsilon_{\bsl{k}}-|f(\bsl{k})|$ and $E_2(\bsl{k})=\epsilon_{\bsl{k}}+|f(\bsl{k})|$.
In the normal state, $\mu\in[E_1(\Gamma_M), E_2(\Gamma_M)]$ is typically true for $2\sim 3$ holes per Moir\'e unit cell, since the top and bottom of the set of nearly flat bands are typically at $\Gamma_M$, as shown in \figref{fig:BM_Model_App}(d-e).
However, we caution that given a fixed filling, the chemical potential in the superconducting phase can be different from that in the normal state due to the correction brought by the pairing order parameter.
Thus, since we care about the zero-temperature superconducting phase, $\mu\in[E_1(\Gamma_M), E_2(\Gamma_M)]$ in \asmref{asm:nodal SC} should be tested when applying to specific superconducting phase with specific pairing.
Note that we do \emph{not} require the pairing order parameter to be much smaller than the gap between two nearly flat bands at the Fermi surfaces.

On the other hand, recall that when we count the zeros of $\Delta_{\shpa}$, we would treat a zero with winding $\pm n$ as $n$ zeros with winding $\pm 1$ at the same momentum.
The reason for requiring $\Delta_{\shpa}$ to only have two zeros with winding $1$ in {\MBZ} in \asmref{asm:nodal SC} is that the least number of zeros for $\Delta_{\shpa}$ tends to be physically favored since more zeros of $\Delta_{\shpa}$ typically require a more complex structure of the pairing, which tends to be physically suppressed.

In the following, we will discuss the nodal superconductivity guaranteed by sufficiently-dominant $\Delta_{\shpa}$ for both $C_{3z}$-invariant and spontaneously nematic pairings.
We will use \propref{prop:nodal_SC_EOCP_gen_app}, \eqnref{eq:alt_Sigma_zero_path}, and \asmref{asm:nodal SC}.

\subsubsection{$C_{3z}$-Invariant}

Recall that we work under \asmref{asm:normal_state_C2T}-\ref{asm:pairing} and \asmref{asm:Existence_Delta_para}-\ref{asm:nodal SC}. 
The main result for the nodal superconductivity for the $C_{3z}$-invariant pairing is the following.

\begin{proposition}
\label{prop:C3inv_nodalSC}
Under \asmref{asm:normal_state_C2T}-\ref{asm:pairing} and \asmref{asm:Existence_Delta_para}-\ref{asm:nodal SC}, for $C_{3z}$-invariant pairing ($A$ irrep), nodal superconductivity is not always guaranteed by sufficiently dominant $\Delta_{\shpa}$, even if fine-tuned cases are ruled out.
\end{proposition}
The proposition is true as long as we can find a codimension-0 subregion of choices of $(\mu, P_{h}, P_{\shpa})$ with $C_{3z}$-invariant not-globally-vanishing $P_{\shpa}$, in which nodal superconductivity is not guaranteed by sufficiently dominant $\Delta_{\shpa}$.
To find the codimension-0 region, we will use the Chern gauge that satisfies \asmref{asm:Chern_gauge} and \eqnref{eq:MATBG_Ch_Gauge}.

Consider a special $(P_{h,+}, P_{\shpa})=(\widetilde{P}_{h,+}, \widetilde{P}_{\shpa})$ such that $f(\bsl{k})=10 d_{\shpa}(\bsl{k})$,  $\epsilon_{\Gamma_M}+|f(\Gamma_M)|>1$meV and $\epsilon_{K_M/K_M'}=0$.
For this special choice, $\Lambda$ only contains $K_M$ and $K_M'$, and $E(\Lambda)=\{0\}$.
Since any infinitesimal derivation of $(P_{h,+}, P_{\shpa})$ from  $(\widetilde{P}_{h,+}, \widetilde{P}_{\shpa})$ can only change $E(\Lambda)$ and $\epsilon_{\Gamma_M}+|f(\Gamma_M)|$ infinitesimally, there exists a codimension-0 region $X$ of $(P_{h,+}, P_{\shpa})$ that contains  $(\widetilde{P}_{h,+}, \widetilde{P}_{\shpa})$ such that $E(\Lambda)\subset [-0.5 \text{meV}, 0.5 \text{meV}]$ and $\epsilon_{\Gamma_M}+|f(\Gamma_M)|>1$meV for all $(P_{h,+}, P_{\shpa})\in X$.
Then, for any $(\mu,P_{h,+}, P_{\shpa})$ in the codimension-0 $(0.5\text{meV}, 1.0 \text{meV})\times X$, $\mu\notin E(\Lambda)\Rightarrow \H^{(0)}$ is gapped.
In the codimension-0 $(0.5\text{meV}, 1.0 \text{meV})\times X$, nodal superconductivity is certainly not guaranteed by the sufficiently dominant $\Delta_{\shpa}$ since $\H^{(0)}(\bsl{k})$ is gapped everywhere.
So \propref{prop:C3inv_nodalSC} holds.

\subsubsection{Spontaneously Nematic Pairing}
\label{app:nodal_SC_nematic}

Now we turn to the spontaneously nematic pairing, which spontaneously breaks the $C_{3z}$ symmetry.
It means that the pairing is the linear combination of two components of the $E$ irrep, or in short belongs to the $E$ irrep.
Recall that we work under \asmref{asm:normal_state_C2T}-\ref{asm:pairing} and \asmref{asm:Existence_Delta_para}-\ref{asm:nodal SC}. 
The main result for the nodal superconductivity for the $E$ pairing is the following.

\begin{proposition}
\label{prop:nematic_EOCP_nodalSC}
Under \asmref{asm:normal_state_C2T}-\ref{asm:pairing} and \asmref{asm:Existence_Delta_para}-\ref{asm:nodal SC}, for spontaneously nematic pairing ($E$ irrep), nodal superconductivity is always guaranteed by sufficiently dominant $\Delta_{\shpa}$, if fine-tuned cases are ruled out.
\end{proposition}
The reasoning is the following.
Since zeros of $d_{\shpa}(\bsl{k}) \pm \ii s f(\bsl{k})$ can only be created in pairs with zero total winding, there exist at least two zeros of $\Delta_{\shpa}$ that persist through the continuous change of $s$ from $0$ to $1$ and becomes two zeros of $d_{\shpa}(\bsl{k}) \pm \ii s f(\bsl{k})$, in order to carry the total winding number $2$.
Here, recall that when we count the zeros of $\Delta_{\shpa}$, we would treat a zero with winding $\pm n$ as $n$ zeros with winding $\pm 1$ at the same momentum. 
The paths of the two zeros of $\Delta_{\shpa}$ definitely connects the two zeros of $\Delta_{\shpa}$ to zeros of $d_{\shpa}(\bsl{k}) \pm \ii s f(\bsl{k})$.

Since $P_{\shpa}$ belongs to the $E$ irrep, we know at least one of the two zeros of $\Delta_{\shpa}$ is at $\Gamma_M$.
Then, there exists a continuous path $\gamma\subset \Sigma$ (continuous when treating {\MBZ} as a torus) such that it connects $\Gamma_M$ to $\bsl{k}_0$ that satisfies  $d_{\shpa}(\bsl{k}_0) + \ii f(\bsl{k}_0)=0$ or $d_{\shpa}(\bsl{k}_0) - \ii f(\bsl{k}_0)=0$, unless invoking fine tuning.
Finally, we have
\eq{
\left[\epsilon_{\Gamma_M}-|f_{\Gamma_M}|, \epsilon_{\Gamma_M}+|f_{\Gamma_M}|\right]\subset E(\Sigma)\ .
}
since as $\bsl{k}$ goes from $\Gamma_M$ to $\bsl{k}_0$ through $\gamma$, $\epsilon(\bsl{k})\pm\sqrt{|f(\bsl{k})|^2 - |d_{\shpa}(\bsl{k})|^2}$ changes from $\epsilon_{\Gamma_M}\pm|f_{\Gamma_M}|$ to $\epsilon_{\bsl{k}_0}$ continuously.
It means that for all choices of $(\mu, P_{h}, P_{\shpa})$ with not-globally-vanishing $P_{\shpa}$ belonging to the $E$ irrep, $\H^{(0)}(\bsl{k})$ is always nodal, unless invoking fine tuning.
Since a gapless $\H^{(0)}(\bsl{k})$ has at least one isolated gapless node unless invoking fine tuning, \propref{prop:nodal_SC_EOCP_gen_app} suggests that the nodal superconductivity is always guaranteed by sufficiently dominant $\Delta_{\shpa}$ for all choices of $(\mu, P_{h}, P_{\shpa})$ with not-globally-vanishing $P_{\shpa}$, if the pairing is spontaneously nematic and if fine-tuned cases are ruled out.
So \propref{prop:nematic_EOCP_nodalSC} holds.

\refcite{Wu2019TSCMATBG} provides an alternative way for nematic pairing to give the nodal superconductivity.
The mechanism proposed in \refcite{Wu2019TSCMATBG} does not require a sufficiently-dominant  Euler obstructed pairing channel and thus is different from our mechanism presented above.
On the other hand, the mechanism proposed in \refcite{Wu2019TSCMATBG} requires the pairing order parameter to be much smaller than the gap between the two nearly flat bands on the Fermi ``surface"  so that the pairing order parameter can be projected to the Fermi surface and become scalar; this condition is not required here by our mechanism.
We note that for MATBG, the pairing order parameter is not always much smaller than the gap between the two nearly flat bands on the Fermi ``surface".
In \appref{app:local_int_example}, we will present such an example where the mechanism proposed in \refcite{Wu2019TSCMATBG} failed but our mechanism presented above works.

\subsection{Bounded Zero-temperature superfluid weight in MATBG}

In this part, we will discuss the applicability of the bounded zero-temperature superfluid weight in \eqnref{eq:Tr_DSF_bound} to MATBG.

As discussed before, \eqnref{eq:Tr_DSF_bound} is derived from \eqnref{eq:SFweight_gen} under \asmref{asm:normal_state_C2T}-\ref{asm:pairing} and \asmref{asm:ForSW}.
We already know that \asmref{asm:normal_state_C2T}-\ref{asm:N_pm_nonzero} are satisfied in the normal state of MATBG based on the BM model, and we have always imposed \asmref{asm:pairing}.
Then, the extra assumptions for \eqnref{eq:Tr_DSF_bound} to hold in this part would be (i) \eqnref{eq:SFweight_gen} is valid, and (ii) choosing the normal-state flat bands exactly flat and choosing $\mu\neq 0$ in \asmref{asm:ForSW}.

Let us first discuss the validity of \eqnref{eq:SFweight_gen}.
\eqnref{eq:SFweight_gen} is valid for gapped superconductors.
Owing to the $C_{2z}\TR$ symmetry, the superconductor might be nodal as discussed above.
For the 2D $C_{2z}\TR$-invariant pairing model discussed above, \eqnref{eq:SFweight_gen} will also hold for nodal superconductors unless invoking fine tuning.
It is because, unless invoking fine tuning, the nodal superconductors only have isolated gapless point nodes with linear dispersion, and then these nodes cannot contribute to the superfluid weight and can be directly neglected, since they cannot give Dirac-delta-like integrands in \eqnref{eq:SFweight_gen}.
In other words, we can directly neglect all the nodal points in the superconductors unless invoking fine tuning, which means \eqnref{eq:SFweight_gen} is valid unless invoking fine tuning.

Now let us discuss the extra assumptions in \asmref{asm:ForSW}.
$\mu\neq 0$ is true unless invoking fine-tuning.
Choosing the normal-state flat bands exactly flat was previously adopted in the study of bound of superfluid weight for time-reversal invariant uniform pairings in \refcite{Xie2020TopologyBoundSCTBG}.
It is not exactly true based on the BM model with realistic parameter values (\eqnref{eq:BM_realistic_paraval}), but it is true for the chiral limit with twist angle exactly at the magic angle~\cite{Tarnopolsky2019MagicAngleChiralLimit}.
For the nodal superconductivity discussed above, we know the normal-state dispersion is very important, since it determines the range of the zero-temperature chemical potential, in which the superconductor is enforced to be nodal by nematic Euler obstructed pairing as discussed in \asmref{asm:nodal SC} and \appref{app:nodal_SC_nematic}.
Nevertheless, \asmref{asm:ForSW} can be a good approximation for the calculation of the superfluid weight of the pairing model chosen in \asmref{asm:pairing}.
By being a good approximation, we mean that the superfluid weight of the pairing model in \asmref{asm:pairing} does not change dramatically (at least within the same order of magnitude) if we choose the normal-state bands to be exactly flat to satisfy \asmref{asm:ForSW} while keeping the pairing matrix and the filling unchanged.
Then, the remaining question is when \asmref{asm:ForSW} is a good approximation.
It turns out that \asmref{asm:ForSW} is not always a good approximation, and in general, the \asmref{asm:ForSW} becomes a better approximation if the ratio between the pairing amplitude and the normal-state bandwidth becomes large~\cite{Hu2019MATBGSW,Julku2020MATBGSW}.
As shown in the next section, the \asmref{asm:ForSW} becomes good when the twist angle is very close to $1.1^\circ$---for which the normal-state bandwidth is smallest. 
Then, we know in certain cases, \eqnref{eq:Tr_DSF_bound} can be applied to MATBG.

\section{Euler Obstructed Cooper Pairing Induced by Attractive Local Interaction in MATBG}
\label{app:local_int_example}

In this section, we will provide more details on the pairing induced by the local attractive interaction that has the similar form as the attractive interaction given by the electron-acoustic phonon coupling in MATBG proposed in \refcite{Wu2019PhononResSCMATBG}.

According to \refcite{Wu2019PhononResSCMATBG}, we choose the attractive interaction to be local in $\bsl{r}$, $\U(2)\times \U(2)$-invariant, intralayer, and inter-valley. 
With these constraints, the form of the interaction is 
\eq{
\label{eq:H_int_app}
\widetilde{H}_{int}=  - 4 \sum_{\sigma_1,\sigma_2,\sigma_3,\sigma_4, s, s',l}g_{l}(\sigma_1\sigma_2\sigma_3\sigma_4)\int d^2 r \psi^\dagger_{+,\bsl{r},l,\sigma_1,s}\psi^\dagger_{-,\bsl{r},l,\sigma_2,s'} \psi_{-,\bsl{r},l,\sigma_3,s'} \psi^\dagger_{+,\bsl{r},l,\sigma_4,s}\ ,
}
where $\psi^\dagger_{\pm,\bsl{r},l,\sigma,s}$ are the basis of the BM model in \eqnref{eq:BM_model}, $l,l'\in\{t,b\}$,  $\sigma,\sigma'\in\{A,B\}$, and $s,s'\in\{ \uparrow, \downarrow \}$.
The interaction should be Hermitian and is assumed to satisfy $\overline{C}_{2z}\TR$, $\overline{C}_{3z}$, $\overline{C}_{2x}$ and $\TR$ symmetries, resulting in 
\eqa{
& h.c.:\ g_{l}^*(\sigma_1\sigma_2\sigma_3\sigma_4)=g_{l}(\sigma_4\sigma_3\sigma_2\sigma_1)\\
& \overline{C}_{2z}\TR:\ g_{l}^*(\sigma_1'\sigma_2'\sigma_3'\sigma_4')(\sigma_{x})_{\sigma_1\sigma_1'}(\sigma_{x})_{\sigma_2\sigma_2'}(\sigma_{x})_{\sigma_3\sigma_3'}(\sigma_{x})_{\sigma_4\sigma_4'}=g_{l}(\sigma_1\sigma_2\sigma_3\sigma_4)\\
& \overline{C}_{3z}:\ \sum_{\sigma_1'\sigma_2'\sigma_3'\sigma_4'}g_{l}(\sigma_1'\sigma_2'\sigma_3'\sigma_4')
(e^{\ii \sigma_z \frac{2\pi}{3}})_{\sigma_1\sigma_1'}
(e^{-\ii \sigma_z \frac{2\pi}{3}})_{\sigma_2\sigma_2'}
(e^{\ii \sigma_z \frac{2\pi}{3}})_{\sigma_3\sigma_3'}
(e^{-\ii \sigma_z \frac{2\pi}{3}})_{\sigma_4\sigma_4'}=g_{l}(\sigma_1\sigma_2\sigma_3\sigma_4)\\
& \overline{C}_{2x}:\ \sum_{\sigma_1'\sigma_2'\sigma_3'\sigma_4'}g_{t}(\sigma_1'\sigma_2'\sigma_3'\sigma_4')(\sigma_{x})_{\sigma_1\sigma_1'}(\sigma_{x})_{\sigma_2\sigma_2'}(\sigma_{x})_{\sigma_3\sigma_3'}(\sigma_{x})_{\sigma_4\sigma_4'}=g_{b}(\sigma_1\sigma_2\sigma_3\sigma_4)\\
& \TR:\ g_{l}^*(\sigma_2\sigma_1\sigma_4\sigma_3)=g_{l}(\sigma_1\sigma_2\sigma_3\sigma_4)\ .
}
Here we use the $\overline{C}_{2x}$ to simplify the interaction in this example, but the symmetry is not essential for the general discussions in \appref{app:EOCP_MATBG}.
We can use the above relations to simplify \eqnref{eq:H_int_app}.
To do so, we can define
\eq{
\widetilde{O}_{l,j,j_s}(\bsl{r})=\psi^\dagger_{+,\bsl{r},l} \widetilde{\sigma}_j \otimes (\widetilde{s}_{j_s}\ii s_y) (\psi^\dagger_{-,\bsl{r},l})^T
}
with $\widetilde{s}_{0}=s_{0}$, $\widetilde{s}_{x,y,z}=\ii s_{x,y,z}$, $\widetilde{\sigma}_{0,x,y}=\sigma_{0,x,y}$, and $\widetilde{\sigma}_{z}=\ii \sigma_{z}$.
Then, we get
\eq{
\label{eq:H_int_sim_app}
\widetilde{H}_{int}= - \sum_{l,j,j_s} g_{j} \int d^2 r \widetilde{O}_{l,j,j_s}(\bsl{r}) \widetilde{O}_{l,j,j_s}^\dagger(\bsl{r})\ ,
}
where 
\eq{
g_x=g_y=g_1\ .
}
If we choose $g_0=g_1=g_z$, \eqnref{eq:H_int_sim_app} becomes the effective attractive interaction mediated by the acoustic phonon derived in \refcite{Wu2019PhononResSCMATBG}.
As shown in \refcite{Wu2019PhononResSCMATBG,Wu2019TSCMATBG}, the pairing given by the attractive interaction in general would contain intrasublattice and intersublattice channels simultaneously, but the mixing between them is typically small, meaning that we may study the intrasublattice and intersublattice channels separately.
To do so, we will impose 
\eq{
g_z=g_0
}
but allow $g_1\neq g_0$ just to study the intra-sublattice and inter-sublattice pairings independently.

However, the interaction is \eqnref{eq:H_int_sim_app} has a physical problem---the attractive interaction should happen only to low-energy electrons instead of all electrons with all possible energies, since we need to integrate out high-energy electrons to renomralize the Coulomb interaction.
To fix this issue, a more physically reasonable attractive interaction should be obtained by projecting \eqnref{eq:H_int_sim_app} to the nearly-flat bands, which gives us
\eq{
\label{eq:H_int_sim_proj_app}
H_{int}= - \sum_{l,j,j_s,\bsl{G}_M} g_{j} \int_{{\MBZ}} \frac{d^2 q}{(2\pi)^2} O_{l,j,j_s,\bsl{G}_M}(\bsl{q}) O_{l,j,j_s,\bsl{G}_M}^\dagger(\bsl{q})\ ,
}
where  $c^{\dagger}_{\pm,\bsl{k}}=(...,c^{\dagger}_{\pm,\bsl{k},a,s},...)$ stands for the creation operator for the Bloch basis of the nearly flat bands in valley $\pm$, 
\eq{
O_{l,j,j_s,\bsl{G}_M}(\bsl{q})=\sum_{\bsl{k}\in\MBZ} c^{\dagger}_{+,\bsl{k}} \phi_{l,j,\bsl{G}_M}(\bsl{k},\bsl{q})\otimes (\widetilde{s}_{j_s} \ii s_y) (c^{\dagger}_{-,-\bsl{k}+\bsl{q}})^T\ ,
}
\eq{
\phi_{l,j,\bsl{G}_M}(\bsl{k},\bsl{q})=V_{+,\bsl{k}}^\dagger \overline{m}_{l,j,\bsl{G}_M} V_{-,-\bsl{k}+\bsl{q}}^*\ ,
}
$V_{\pm}(\bsl{k})=(V_{\pm,1}(\bsl{k}), V_{\pm,2}(\bsl{k}))$ are othonormal linear combination of eigenvectors of $\widetilde{h}_{\pm}(\bsl{k})$ (\eqnref{eq:BM_mat}) for the nearly flat bands, 
\eqa{
& [\overline{m}_{l,j,\bsl{G}_M}]_{\bsl{Q} \bsl{Q}'} = \widetilde{\sigma}_j \delta_{\bsl{Q}+\bsl{Q}',+\bsl{G}_M} \delta_{\bsl{Q}\in Q_{+,l}}\ ,
}
$Q_{+,t}$ and $Q_{+,b}$ are defined  below \eqnref{eq:BM_mat}, and  $\bsl{Q},\bsl{Q}$ takes values in $Q_{+,b}\cup Q_{+,t}$ according to the convention in \refcite{Song2019TBGFragile}.
The total Hamiltonian is 
\eq{
H=H_+ + H_- + H_{int}\ ,
}
where $H_\pm$ are the projections of $\widetilde{H}_\pm$ onto the nearly flat bands in \eqnref{eq:BM_projected}.

We will adopt the mean-field approximation to derive the superconductivity from the interaction by defining the superconductivity mean-field order parameter as 
\eq{
\widetilde{\Delta}_{l,j,j_s,\bsl{G}_M}(\bsl{q}) = -  g_j  \Tr[ O_{l,j,j_s,\bsl{G}_M}(\bsl{q}) e^{-\beta (H-\mu N)} ]/ \Tr[ e^{-\beta (H-\mu N)} ]\ ,
}
where $N$ is the electron number operator, $\beta=1/(k_B T)$, and $T$ is the temperature.
Since we care about the order parameter that preserves the Moir\'e lattice translation, we requires 
\eq{
\widetilde{\Delta}_{l,j,j_s,\bsl{G}_M}(\bsl{q})=\widetilde{\Delta}_{l,j,j_s,\bsl{G}_M} (2\pi)^2 \delta(\bsl{q})\ ,
}
resulting in the mean-field pairing operator as
\eq{
H_{pairing}= \sum _{l,j,j_s,\bsl{G}_M} \widetilde{\Delta}_{l,j,j_s,\bsl{G}_M}  O_{l,j,j_s,\bsl{G}_M}(0) + h.c.
}
and the mean-field Hamiltonian as
\eq{
H_{MF}= H_+ + H_-  - \mu N + H_{pairing}\ .
}
Then, the self-consistent equation for the order parameter reads
\eqa{
\label{eq:self_consist}
& \widetilde{\Delta}_{l,j,j_s,\bsl{G}_M}= -  \frac{g_j}{\mathcal{V}}  \Tr[ O_{l,j,j_s,\bsl{G}_M}^\dagger(\bsl{0})  e^{-\beta H_{MF}} ]/ \Tr[ e^{-\beta H_{MF}} ] \\
& =- \frac{g_j}{\mathcal{V}}\sum_{\bsl{k}\in\MBZ} \Tr\left[ \mat{ 0 & 0 \\ [\phi_{l,j,\bsl{G}_M}(\bsl{k},0)\otimes \widetilde{s}_{j_s}\ii s_y]^\dagger & 0 } \frac{1}{\exp[\beta h_{BdG,+}(\bsl{k})]+1} \right]\ ,
}
where $\mathcal{V}$ is the volume of the system, 
\eq{
h_{BdG,+}(\bsl{k})
=
\mat{ 
(h_+(\bsl{k})-\mu)\otimes s_0 & M(\bsl{k}) \\ 
M^\dagger(\bsl{k}) & -(h_-(-\bsl{k})-\mu)^T\otimes s_0
}\ ,
}
and 
\eq{
M(\bsl{k})=\sum_{l,j,j_s,\bsl{G}_M} \widetilde{\Delta}_{l,j,j_s,\bsl{G}_M} \phi_{l,j,\bsl{G}_M}(\bsl{k},0)\otimes \widetilde{s}_{j_s}\ii s_y\ .
}

Nontrivial solutions (\ie, $\widetilde{\Delta}_{l,j,j_s,\bsl{G}_M}$ has nonzero components) of \eqnref{eq:self_consist} are pairing order parameters of the pairing. 
In particular, there always exists solutions (might be trivial) to \eqnref{eq:self_consist} such that 
\eq{
\label{eq:Delta_real_y}
\widetilde{\Delta}_{l,j,j_s,\bsl{G}_M} = -\widetilde{\Delta}_{l,j,\bsl{G}_M} \delta_{j_s y }
}
with real $\widetilde{\Delta}_{l,j,\bsl{G}_M}$, and if these solutions are nontrivial, they will naturally lead to the $C_{2z}\TR$-invariant pairing in \eqnref{eq:H_pairing_MATBG} that satisfied \asmref{asm:pairing} with
\eqa{
& \Delta(\bsl{k})= \sum_{l,j,\bsl{G}_M} \widetilde{\Delta}_{l,j,\bsl{G}_M}  \phi_{l,j,\bsl{G}_M}(\bsl{k},0) \\
& \Pi=s_0\ .
}
In the following, we will focus on solutions that satisfy \eqnref{eq:Delta_real_y}.
For those solutions, \eqnref{eq:self_consist} can be simplified into
\eqa{
\label{eq:self_consist_sim}
\widetilde{\Delta}_{l,j,\bsl{G}_M}= -  2 g_j  \int_{\MBZ} \frac{d^2 k}{(2\pi)^2 } \Tr\left\{ U^\dagger(\bsl{k}) \mat{ 0 & 0 \\ \phi_{l,j,\bsl{G}_M}^{\dagger}(\bsl{k},0) & 0 } U(\bsl{k})
\mat{n_F(E_1(\bsl{k})) & & & \\ 
& n_F(E_2(\bsl{k})) & & \\
& & n_F(E_3(\bsl{k})) & \\
& & & n_F(E_4(\bsl{k})) } \right\}\ ,
}
where $n_F(x)=[e^{\beta x}+1]^{-1}$, we choose the gauge such that $C_{2z}c^\dagger_{+,\bsl{k}}C_{2z}^{-1} = c^\dagger_{-,-\bsl{k}}$, and $U(\bsl{k})$ is unitary matrix such that 
\eq{
U^\dagger(\bsl{k}) \mat{ 
h_+(\bsl{k})-\mu &  \Delta(\bsl{k}) \\ 
 \Delta^\dagger(\bsl{k}) & -(h_+(\bsl{k})-\mu)^T\ 
} U(\bsl{k}) = \mat{E_1(\bsl{k}) & & & \\ 
& E_2(\bsl{k}) & & \\
& & E_3(\bsl{k}) & \\
& & & E_4(\bsl{k}) } \ .
}

We solved \eqnref{eq:self_consist_sim} with 
\eqa{
\label{eq:SCE_paraval}
& g_0 = 40 \text{ meV (nm)$^2$ and } g_1 = 0 \text{ for intrasublattice pairing}\\
& g_0 = 0 \text{ and } g_1 = 70 \text{ meV (nm)$^2$} \text{ for intersublattice pairing}\\
& \theta\in\{1.05^{\circ},1.06^{\circ}, ..., 1.15 ^{\circ}\},\ w_0/w_1 = 0.8,\ v_0= 5817 \text{meV}\cdot\AA,\ w_1=110\text{meV},\ a_0=2.46 \AA\ ,\\
&\text{filling = 2.5 holes per Moir\'e unit cell}\ ,\\
& k_B T\in 10^{-5}+\{ 0,0.02, 0.04, ... , 0.98, 1\}\text{meV, where we approximate 0 as $10^{-5}$}\ ,\\
&|\bsl{Q}|\leq 2\sqrt{7}k_D \text{ for the index $\bsl{Q}$}\ ,\\
& |\bsl{G}_M|\leq 2\sqrt{7}k_D\text{ for $\bsl{G}_M$ in $\widetilde{\Delta}_{l,j,\bsl{G}_M}$}\ .
}
We find nontrivial real solutions to \eqnref{eq:self_consist_sim} for all twisted angles in \eqnref{eq:BM_realistic_paraval} and for both intrasublattice and intersublattice pairing.
According to the terminology in \refcite{Wu2019PhononResSCMATBG}, the intrasublattice pairing is f-wave spin-triplet, and the intrasublattice pairing is d-wave spin-singlet.
Indeed, the intrasublattice pairing that we numerically got is $C_{2z}\TR$-invariant parity-even $C_{3z}$-invariant, while the intersublattice pairing that we numerically got is $C_{2z}\TR$-invariant parity-even and belongs to $E$ irrep.

\begin{figure}
    \centering
    \includegraphics[width=0.9\columnwidth]{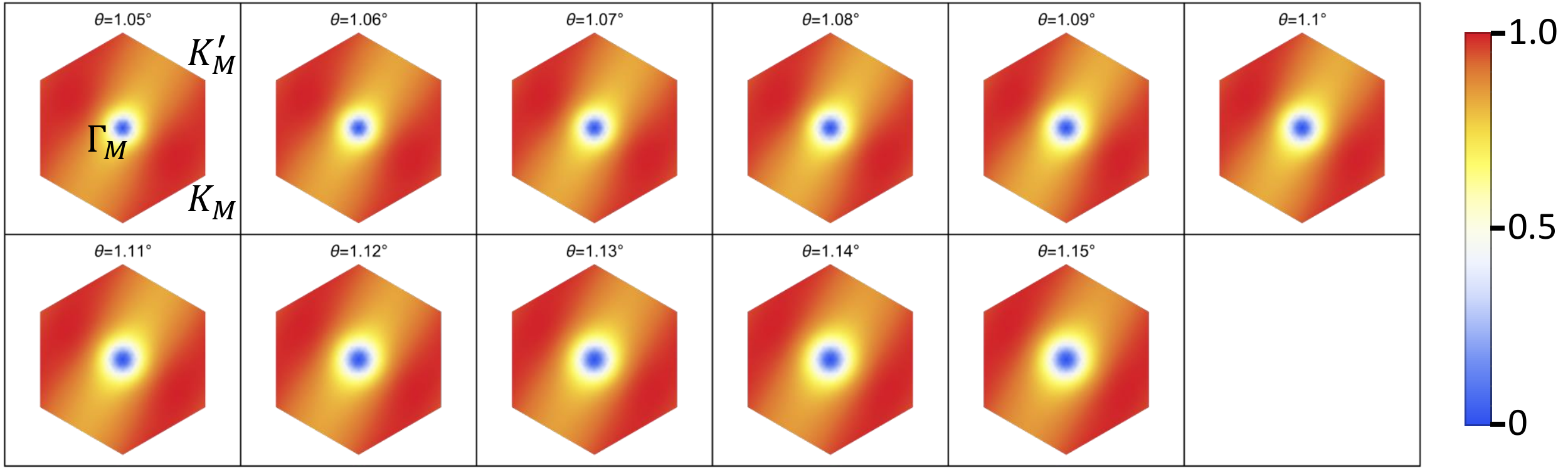}
    \caption{The plot of $\frac{|\Delta_{\shpa}(\bsl{k})|}{\max[|\Delta_{\shpa}(\bsl{k})|]}$ for the intersublattice pairing at zero temperature.
    The parameter values in \eqnref{eq:SCE_paraval} are adopted.
    }
    \label{fig:EOCP_Delta_Para_App}
\end{figure}

\begin{figure}
    \centering
    \includegraphics[width=0.6\columnwidth]{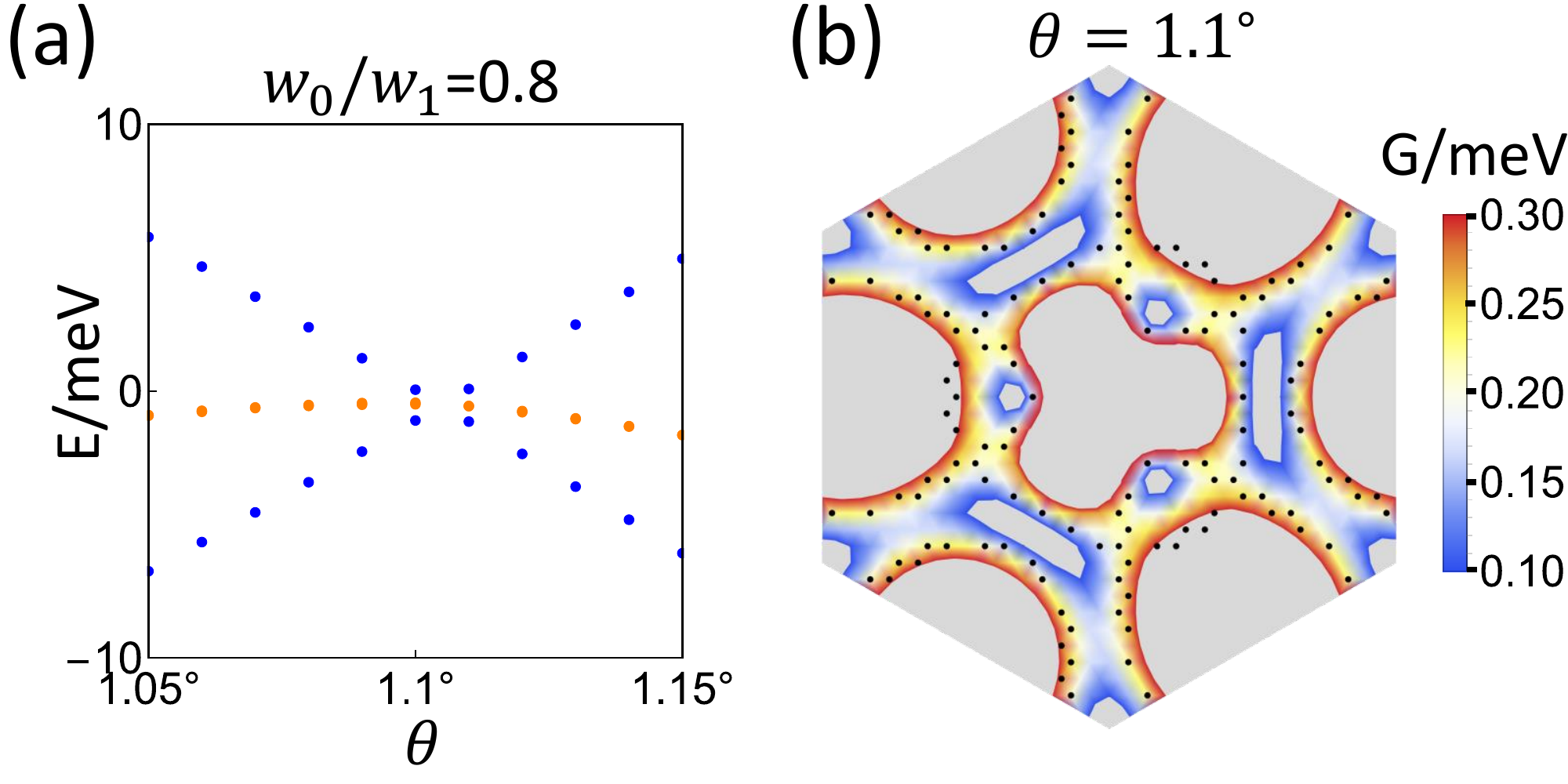}
    \caption{In (a), we show the energies of the normal-state two nearly flat bands at $\Gamma_M$ as blue dots, and show the zero-temperature chemical potentials for both intrasublattice and intersublattice pairings with \eqnref{eq:SCE_paraval} as the orange dots.
    The chemical potential roughly coincide for both intrasublattice and intersublattice pairings at each value of the twist angle.
    We can see that zero-temperature chemical potential lie in the energy range bounded by the normal-state energies of the two nearly flat bands at $\Gamma_M$.
    In (b), we show the area of the {\MBZ} in which the normal-state gap (G) between two nearly flat bands for $\theta=1.1^\circ$ is in $[0.1,0.3]$meV by the color map, and show the Fermi ``surface" as the black dots.
    We can see the normal-state gap between two nearly flat bands stays roughly in the range $[0.1,0.3]$meV at the Fermi ``surface".
    }
    \label{fig:EOCP_Delta_mu_eg_App}
\end{figure}

Let us first focus on the zero temperature.
In \figref{fig:EOCP_Delta_Para_App}, we show the $|\Delta_{\shpa}(\bsl{k})|$ for the intersublattice pairing at zero temperature, and we can see they all only have zeros at $\Gamma_M$.
We numerically check that the total winding number along a circle surrounding $\Gamma_M$ is 2, and therefore we know there are two zeros with winding number 1 coinciding at $\Gamma_M$.
In \figref{fig:EOCP_Delta_mu_eg_App}(a), we show that the zero-temperature chemical potential (after including the correction due to the pairing) in \eqnref{eq:h_BdG_+_up} always lies in the range $[\epsilon_{\Gamma_M}-|f(\Gamma_M)|,\epsilon_{\Gamma_M}+|f(\Gamma_M)|]$ for both intersublattice and intrasublattice pairings.
Thus, we know \asmref{asm:nodal SC} is satisfied.
Therefore, \propref{prop:nematic_EOCP_nodalSC} suggests that the nodal superconductivity should be expected for the intersublattice pairing for all the $\theta$ values in \eqnref{eq:SCE_paraval} at zero temperature, as long as its trivial channel is small enough, which is numerically verified as discussed in the main text.
In particular, the zero BdG gap for the intersublattice pairing shown in the main text is checked by finding closed loops with nonzero chiralities of the BdG Hamiltonian, which must include zero-energy gapless ndoes as discussed in \appref{app:nodal_SC_gen}.
On the other hand, as shown in \figref{fig:EOCP_Delta_mu_eg_App}(b) for $\theta=1.1^\circ$, the normal-state gap between two nearly flat bands at the FS is roughly $0.1\sim 0.3$meV, which is smaller than the average zero-temperature pairing amplitude ($\sim 0.5$meV, shown in the main text), meaning that the mechanism for nodal superconductivity arising from nematic pairing proposed in \refcite{Wu2019TSCMATBG} does not always work in this case, while \propref{prop:nematic_EOCP_nodalSC} works. 

\begin{figure}[H]
    \centering
    \includegraphics[width=0.6\columnwidth]{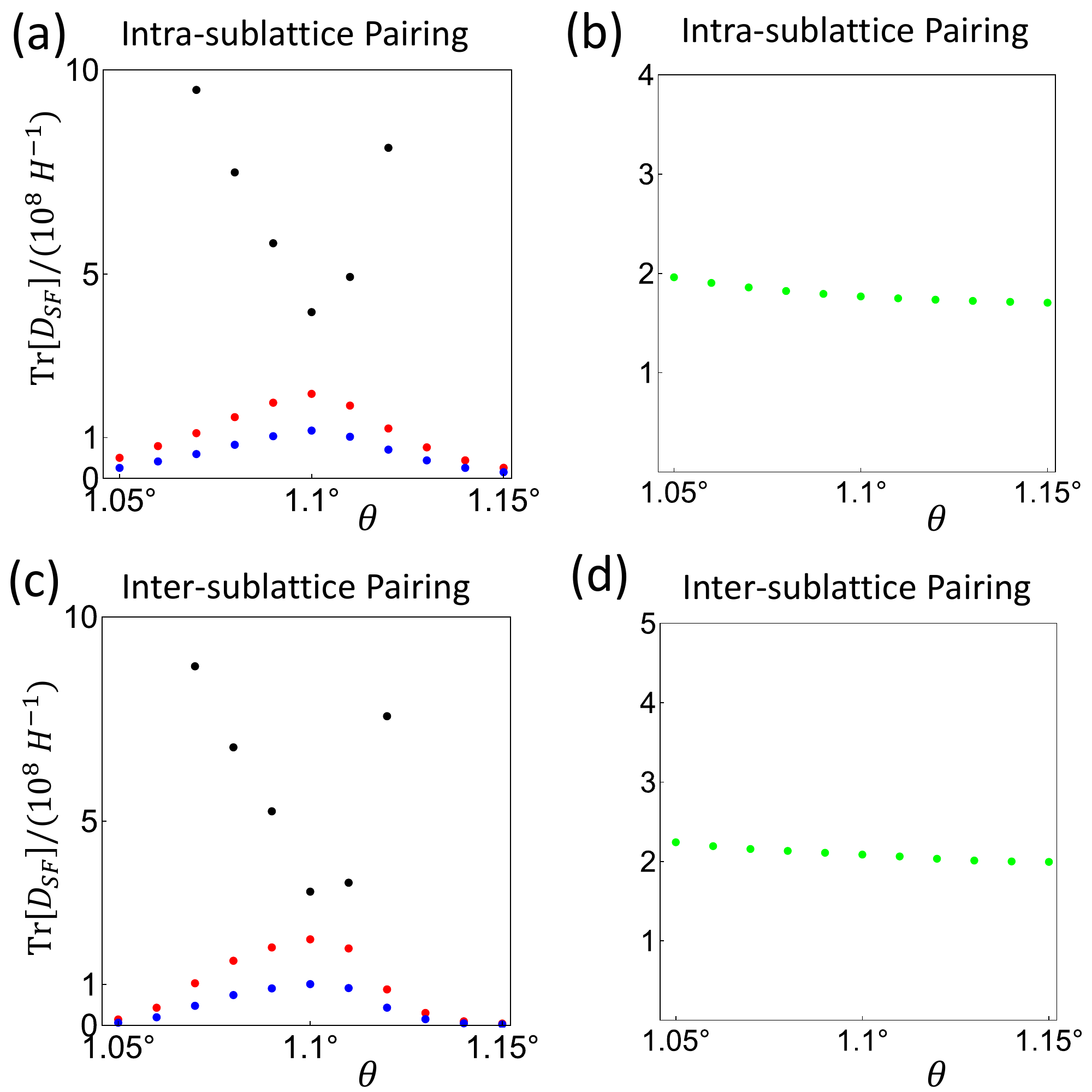}
    \caption{
    The superfluid weight has been converted to SI unit in this figure, and is in the zero temperature.
    (a-b) are for the intrasublattice pairings derived from \eqnref{eq:self_consist_sim} for MATBG, and (c-d) are for the intersublattice pairings derived from \eqnref{eq:self_consist_sim} for MATBG.
    The black dots stand for the trace of the superfluid weight derived from \eqnref{eq:SFweight_gen} for \eqnref{eq:SCE_paraval}, which neglects all high-energy bands.
    For $\theta$ values at which the black dots are missing, the trace of the superfluid weight derived from \eqnref{eq:SFweight_gen} is larger than $10^9$ H$^{-1}$.
    The red, blue and green dots are obtained by artificially limiting the dispersion of the normal-state nearly flat bands to zero, while keeping the normal-state eigenvectors, the pairing matrix and the filling unchanged.
    Specifically, the red, blue and green dots stand for the trace of the superfluid weight $\Tr[D_{SF}]$ derived from \eqnref{eq:D_SF_efb_sim}, the values of the lower bound $\Tr[D_{SF}^{bound}]$ in \eqnref{eq:Tr_DSF_bound}, and $\Tr[D_{SF}]/\Tr[D_{SF}^{bound}]$, respectively.
    }
    \label{fig:EOCP_ZeroT_SFW_App}
\end{figure}

We further check what are the cases for the MATBG where \asmref{asm:ForSW} is a good approximation for the study of zero-temperature superfluid weight.
As shown in \figref{fig:EOCP_ZeroT_SFW_App}(a,c), \asmref{asm:ForSW} is a good approximation for the study of zero-temperature superfluid weight for the intrasublattice and intersubalattice pairings derived from \eqnref{eq:self_consist_sim}, when the twist angle is very close to $1.1^\circ$---for which the normal-state bandwidth is smallest. 
Therefore, there are cases for the MATBG where \asmref{asm:ForSW} is a good approximation.
As shown in \figref{fig:EOCP_ZeroT_SFW_App}(b,d), $\Tr[D_{SF}^{bound}]$ is of the same order of magnitude as $\Tr[D_{SF}]$ evaluated under \asmref{asm:ForSW} for both intrasublattice and intersublattice pairings derived in \eqnref{eq:self_consist_sim} for MATBG.

\begin{figure}[H]
    \centering
    \includegraphics{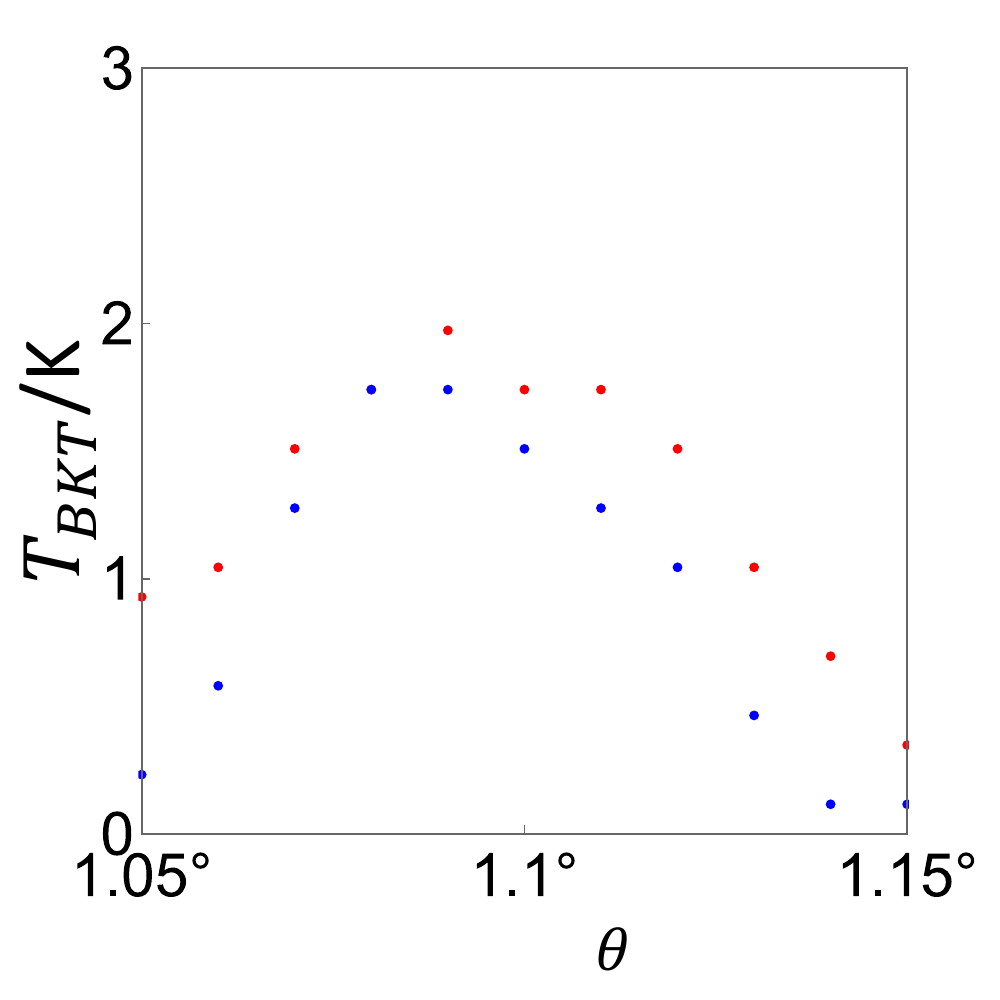}
    \caption{We show the $T_{BKT}$ estimated from \eqnref{eq:TBKT_DSF} with \eqnref{eq:SCE_paraval} for both intrasublattice (blue) and intersublattice  (red) pairings.
    }
    \label{fig:TBKT_App}
\end{figure}

As temperature increases, the pairing would eventually reach zero at the mean-field critical temperature.
However, the mean-field critical temperature is typically not what is measured in MATBG experiments; it is the Berezinskii–Kosterlitz–Thouless (BKT) temperature that is typically measured since the superconductivity transition in most 2D systems should be the BKT transition~\cite{Xie2020TopologyBoundSCTBG}.
So, we should estimate the BKT temperature from the pairing in order to compare with the experiments.
With $\mu$ and $\Delta(\bsl{k})$ derived from \eqnref{eq:self_consist_sim}, we can get $D_{SF}(T)$ using \eqnref{eq:SFweight_gen}, and then the BKT temperature can be estimated by \eqnref{eq:TBKT_DSF} (be aware of different unit systems in \eqnref{eq:SFweight_gen} and  \eqnref{eq:TBKT_DSF}).
In \figref{fig:TBKT_App}, we plot the $T_{BKT}$ estmated from \eqnref{eq:TBKT_DSF} for both intersublattice and intrasublattice.
It shows that the $T_{BKT}\in [1,2]$K for most of the twist angles, roughly coinciding with the experimentally observed values~\cite{Balents2020TBGSCReview}.

\end{document}